\documentclass[11pt,a4paper]{article}
\usepackage{jheppub,epstopdf}

\usepackage{amsmath,amssymb,epsfig,graphicx,epstopdf}

\makeatletter
\newcommand{\fmslash}[2][0mu]{
\mathchoice
    {\fmsl@sh\displaystyle{#1}{#2}}%
        {\fmsl@sh\textstyle{#1}{#2}}%
    {\fmsl@sh\scriptstyle{#1}{#2}}%

    {\fmsl@sh\scriptscriptstyle{#1}{#2}}}
\newcommand{\fmsl@sh}[3]{%
\m@th\ooalign{$\hfil#1\mkern#2/\hfil$\crcr$#1#3$}}

\newcommand{\tr}{\hbox{tr}}

\title{\center{Super Yang-Mills and $\theta$-exact Seiberg-Witten map: Absence of quadratic noncommutative IR divergences}}

\author[a,1]{Carmelo P. Martin
\note{carmelop@fis.ucm.es}}
\affiliation[a]{Departamento de Física Teórica I, Facultad de Ciencias F\'{\i}sicas,
Universidad Complutense de Madrid, 28040-Madrid, Spain}
\author[b,2,3]{Josip Trampetic
\note{trampeti@mpp.mpg.de}
\note{On leave of absence from the Rudjer Bo\v skovi\' c Institute, Zagreb, Croatia}}
\affiliation[b]{Max-Planck-Institut f\"ur Physik, (Werner-Heisenberg-Institut),
F\"ohringer Ring 6, D-80805 M\"unchen, Germany}
\author[c,4]{ and  Jiangyang You
\note{youjiangyang@gmail.com}}
\affiliation[c]{Rudjer Bo\v skovi\' c Institute, Divisions of Theoretical Physics, P.O.Box 180, HR-10002 Zagreb, Croatia}

\abstract{We compute  the one-loop 1PI  contributions to all the  propagators of the noncommutative ${\cal N}=$1, 2, 4 super Yang-Mills (SYM) U(1) theories defined by the means of the $\theta$-exact Seiberg-Witten (SW) map in the Wess-Zumino gauge. Then we extract the UV divergent contributions and the noncommutative IR divergences. We show that all the quadratic noncommutative IR divergences add up to zero in each propagator.}


\begin{document}

\maketitle

\section{Introduction}

The one-loop UV/IR mixing structure of noncommutative (NC) ${\cal N}$=1 super Yang-Mills theory defined in terms of the noncommutative fields was studied some years ago in a number of papers~\cite{Zanon:2000nq, Ruiz:2000hu, Ferrari:2003vs, AlvarezGaume:2003mb,Ferrari:2004ex}. The outcome was the famous quadratic noncommutative IR divergences which occur in the one-loop gauge field propagator of the non-supersymmetric version of the theory cancel here due to Supersymmetry. The one-loop gauge field propagator still carries a logarithmic UV divergence -a simple pole in Dimensional Regularization- and the dual logarithmic noncommutative IR divergence $\ln(p^2(\theta p)^2)$ as a result of the UV/IR mixing being at work. By increasing the number of supersymmetries of the noncommutative Yang-Mills theory one makes the UV  behaviour of the theory softer and eventually finite for ${\cal N}=4$~\cite{Jack:2001cr}, at which point noncommutative IR divergences cease to exist~\cite{Santambrogio:2000rs, Pernici:2000va}. In the ${\cal N}=2$ super Yang-Mills case, there still remain logarithmic UV divergences at one-loop in the two-point function which give rise via UV/IR mixing to the corresponding IR divergences~\cite{Zanon:2000nq, Buchbinder:2001at}. That noncommutative ${\cal N}=4$ super Yang-Mills has a smooth commutative limit has been shown in Ref.~\cite{Hanada:2014ima}.

It is known that classically  noncommutative gauge field theories admit a dual formulation in terms of ordinary fields, a formulation that is obtained by using the celebrated Seiberg-Witten map~\cite{Seiberg:1999vs}. However we still do not know whether this duality holds at the quantum level, i.e., whether the quantum theory defined in terms of noncommutative fields is the same as the ordinary quantum theory --called the dual ordinary theory-- whose classical action is obtained from the noncommutative action by using the Seiberg-Witten map. The existence of the UV/IR mixing effects in noncommutative field theory defined in terms of the noncommutative fields is thought to be a characteristic feature of those field theories. It is thus sensible to think such effects should also occur in the ordinary dual theory obtained, as previously explained, by using the Seiberg-Witten map. That these UV/IR mixing effects actually occur in the propagator of the gauge field of the dual ordinary theory was first shown in Ref.~\cite{Schupp:2008fs} by using the $\theta$-exact Seiberg-Witten map expansion~\cite{Mehen:2000vs,Jurco:2001rq}. 
The analysis of the properties and phenomenological implications of the UV/IR mixing effects that occur in noncommutative gauge theories defined by means of the $\theta$-exact Seiberg-Witten map has been pursued in Refs.~\cite{Trampetic:2015zma, Horvat:2013rga, Trampetic:2014dea, Horvat:2015aca}.

Up to the best of our knowledge no analysis of the UV and the noncommutative IR  structures of the noncommutative super Yang-Mills theory defined by means of the $\theta$-exact Seiberg-Witten map has been carried out as yet. In particular, it is not known whether the cancellation of the quadratic noncommutative IR divergences of the gauge-field propagator that occurs, as we mentioned above, in  noncommutative super Yang-Mills theory  defined in terms of noncommutative fields also  works in its dual ordinary theory. Answer to this question is far from obvious since Supersymmetry is linearly realized in terms of the noncommutative fields --and thus there exists a superfield formalism-- but is non-linearly realized --see Ref.~\cite{Martin:2008xa}-- in terms of the ordinary fields of the dual ordinary theory defined by means of the Seiberg-Witten map. It has long been known that the proper definition  of theories with non-linearly realized symmetries is a highly non-trivial process.

The purpose of this paper is to work out  all the one-loop 1PI two-point functions, and analyze the UV  and noncommutative IR  structures of those functions, in noncommutative U(1) ${\cal N}$=1,2 and 4 super Yang-Mills theories in the Wess-Zumino gauge,  when those theories are defined in terms of ordinary fields  by means of the $\theta$-exact Seiberg-Witten  maps. To analyze the gauge dependence of the UV and noncommutative IR of the gauge field two-point functions we shall consider two types of gauge-fixing terms for the ordinary gauge field: the standard ordinary Feynman gauge-fixing term  and the noncommutative Feynman gauge-fixing term.

The layout of this paper is as follows. Section 2 is devoted to the computation of the one-loop contributions to the photon and photino propagators in ${\cal N}=1$ super Yang-Mills U(1) theory in the ordinary Feynman gauge. In section 3 we discuss, for later use, the construction by using the $\theta$-exact Seiberg-Witten map of a noncommutative U(1) theory  with a noncommutative scalar field transforming under the adjoint representation. The one-loop propagators of the ordinary fields of noncommutative ${\cal N}$=2 and 4 super Yang-Mills U(1) theories defined by using the $\theta$-exact Seiberg-Witten map are worked out in sections 4 and 5 in the ordinary Feynman gauge. In sections 6 and 7 we use a noncommutative Feynman gauge to quantize ${\cal N}=1$ super Yang-Mills U(1) theory and compute the one-loop photon propagator. Sections 6 and 7 are introduced to analyze the dependence on the gauge-fixing term of the UV and noncommutative IR contributions found in previous sections. The overall discussion of our results is carried out in Section 8. We also include several appendices which are needed to complete properly the analysis and computations carried out in the body of the paper.

\section{Noncommutative $\cal N$=1 SYM U(1) theory and the $\theta$-exact SW map}

The noncommutative field content of the noncommutative U(1) super Yang-Mills theory in the Wess-Zumino gauge is the noncommutative gauge field $A_\mu$, its supersymmetric fermion partner $\Lambda_{\alpha}$ and the noncommutative SUSY-auxiliary field $D^{(nc)}$. The ordinary/commutative counterparts of $A_\mu$, $\Lambda_{\alpha}$ and $D^{(nc)}$ will be denoted by $a_\mu$ (photon), $\lambda_\alpha$ (photino) and $D$, respectively. Regarding dotted and undotted fermions and $\sigma^{\alpha\dot{\alpha}}$ traces, we shall follow the
conventions of~\cite{Dreiner:2008tw}.

In terms of the noncommutative fields and in the Wess-Zumino gauge the action of U(1) super Yang-Mills theory reads
\begin{equation}
S_{{\cal N} =1}\,=\,\int\, -\frac{1}{4}\,F_{\mu\nu}\star F^{\mu\nu} + i \bar{\Lambda}_{\dot{\alpha}}\bar{\sigma}^{\mu\,
\dot{\alpha}\alpha}{\cal D}_\mu[A]\Lambda_{\alpha}+\frac{1}{2} D^{(nc)}D^{(nc)},
\label{2.1}
\end{equation}
where $F_{\mu\nu}=\partial_{\mu}A_{\nu}-\partial_{\nu}A_{\mu}-i[A_{\mu}\stackrel{\star}{,} A_{\nu}]$ and ${\cal D}_{\mu}[A]\Lambda_{\alpha}=\partial_{\mu}\Lambda_{\alpha}-i[A_{\mu}\stackrel{\star}{,} \Lambda_{\alpha}]$.

The above action $\rm S_{{\cal N} =1}$ (\ref{2.1}), is invariant under the following noncommutative supersymmetry transformations
\begin{equation}
\begin{split}
&\delta_{\xi}\Lambda_{\alpha}=-iD^{(nc)}\xi_{\alpha}-e^{-1}(\sigma^{\mu}\bar{\sigma}^{\nu})_{\alpha}^{\phantom{\alpha}\beta}\xi_\beta F_{\mu\nu},
\\&\delta_{\xi}A^{\mu}=ie(\xi\sigma^{\mu}\bar{\Lambda}-\Lambda\sigma^{\mu}\bar{\xi}),
\\&\delta_{\xi}D^{(nc)}=(\xi\sigma^{\mu}{\cal D}_{\mu}\bar{\Lambda}-{\cal D}_{\mu}\Lambda\sigma^{\mu}\bar{\xi}).
\end{split}
\label{2.2}
\end{equation}
These supersymmetry transformations close on translations modulo noncommutative gauge transformations and tell us that supersymmetry is linearly realized on the noncommutative fields---see~\cite{Martin:2008xa} for further discussion. The action in \eqref{2.1}, is also invariant under noncommutative U(1) transformations, which in the noncommutative BRST form read
\begin{equation}
s_{\rm NC}\Lambda_\alpha=-i[\Lambda_\alpha\stackrel{\star}{,} \Omega],\quad s_{\rm NC}A_{\mu}=\partial_{\mu}\Omega-i[A_\mu\stackrel{\star}{,} \Omega],\quad s_{\rm NC}\Omega=i\Omega\star\Omega,
\label{2.3}
\end{equation}
with $\Omega$ being the noncommutative U(1) ghost field. The above action $S_{{\cal N} =1}$ can be expressed in terms of ordinary fields, $a_\mu$, $\lambda_{\alpha}$ and $D$, by means of the SW map. The resulting functional is invariant under ordinary U(1) BRST transformations:
\begin{equation}
s\lambda_\alpha=-i[\lambda_\alpha,\omega]\,,\quad s a_{\mu}=\partial_{\mu}\omega\,,\quad s\omega=0,
\label{2.4}
\end{equation}
where $\omega$ is the ordinary U(1) ghost field. Indeed, the SW map maps ordinary BRST orbits into the noncommutative BRST orbits.

The $\theta$-exact SW map for $F_{\mu\nu}$ has been worked out in~\cite{Trampetic:2015zma}  up to the three ordinary U(1) gauge fields $a_\mu$. It reads
\begin{equation}
F_{\mu\nu}\left(e\cdot a_\mu,\theta^{\mu\nu}\right)=e f_{\mu\nu}+F^{e^2}_{\mu\nu}+F^{e^3}_{\mu\nu}+\mathcal O\left(e^4\right),
\label{2.5}
\end{equation}
where, up to the $e^2$ order, the gauge field strength SW map $F^{e^2}_{\mu\nu}$ expansion is fairly universal
\cite{Horvat:2013rga,Trampetic:2014dea,Trampetic:2015zma}
\begin{equation}
F^{e^2}_{\mu\nu}=e^2\theta^{ij}
\Big( f_{\mu i}\star_{2}f_{\nu j}-a_i\star_2\partial_j
f_{\mu\nu}\Big).
\label{2.6}
\end{equation}
The $e^3$ order SW map for the gauge field strength from \cite{Trampetic:2015zma} in that case reads
\begin{equation}
\begin{split}
F^{e^3}_{\mu\nu}(x)=
\frac{e^3}{2}\theta^{ij}\theta^{kl}&\bigg(\left[f_{\mu k}f_{\nu i} f_{l j}\right]_{\star_{3'}}+\left[f_{\nu l}f_{\mu i}f_{kj}\right]_{\star_{3'}}-\left[f_{\nu l}a_i\partial_j f_{\mu k}\right]_{\star_{3'}}
-\left[f_{\mu k}a_i\partial_j f_{\nu l}\right]_{\star_{3'}}
\\&-\left[a_k\partial_l\left(f_{\mu i}f_{\nu j}\right)\right]_{\star_{3'}}
+\left[a_i\partial_j a_k \partial_l f_{\mu\nu}\right]_{\star_{3'}}
+\left[\partial_l f_{\mu\nu}a_i\partial_j a_k\right]_{\star_{3'}}
\\&+\left[a_k a_i \partial_l\partial_j f_{\mu\nu}\right]_{\star_{3'}}
-\frac{1}{2}\Big(\left[a_i\partial_k a_j\partial_l f_{\mu\nu}\right]_{\star_{3'}}
+\left[\partial_l f_{\mu\nu}a_i\partial_k a_j\right]_{\star_{3'}}\Big)\bigg).
\label{2.7}
\end{split}
\end{equation}
The generalized star products relevant for this work are defined as follows~\cite{Trampetic:2014dea,Horvat:2015aca}
\begin{gather}
\begin{split}
(f\star_2 g)(x)&=\int\,e^{-i(p+q)x}\tilde f(p)\tilde g(q)f_{\star_2}\left(p,q\right),
\\
[f g h]_{\star_{3'}}(x)&=\int\,e^{-i(p+q+k)x}\tilde f(p)\tilde g(q)\tilde h(k)f_{\star_3}\left(p,q,k\right),\;\;
\\
[f g h]_{\mathcal M_{\rm (I)}}(x)&=\int\,e^{-i(p+q+k)x}\tilde f(p)\tilde g(q)\tilde h(k)f_{\rm (I)}\left(p,q,k\right),
\label{2.71}
\end{split}
\end{gather}
with
\begin{gather}
\begin{split}
f_{\star_2}(p,q)&=\frac{\sin\frac{p\theta q}{2}}{\frac{p\theta q}{2}},
\\
f_{\star_{3'}}\left(p,q,k\right)&=\frac{\cos(\frac{p\theta q}{2}+\frac{p\theta k}{2}-\frac{q\theta k}{2})-1}{(\frac{p\theta q}{2}+\frac{p\theta k}{2}-\frac{q\theta k}{2})\frac{q\theta k}{2}}-\frac{\cos(\frac{p\theta q}{2}+\frac{p\theta k}{2}+\frac{q\theta k}{2})-1}{(\frac{p\theta q}{2}+\frac{p\theta k}{2}+\frac{q\theta k}{2})\frac{q\theta k}{2}},
\\
f_{\rm (I)}\left(p,q,k\right)&=\frac{1}{p\theta q}\left(f_{\star_{3'}}[p,q,-(p+q+k)]-f_{\star_{3'}}[p,q,k]\right).
\end{split}
\label{f23'}
\end{gather}

The $\theta$-exact SW map for ${\cal D}_{\mu}[A]\Lambda_\alpha$ up to the two ordinary fields $a_\mu$ can be retrieved from the expression for $F_{\mu\nu}$ in \eqref{2.5}, \eqref{2.6} and \eqref{2.7} as explained in the appendix A:
\begin{equation}
{\cal D}_{\mu}[A]\Lambda_\alpha=\partial_{\mu}\lambda_\alpha
+{\cal D}^{e}_{\mu\alpha}[e\cdot a,\lambda]+{\cal D}^{e^2}_{\mu\alpha}[e\cdot a,\lambda]+\mathcal O\left(e^3\right),
\label{2.8}
\end{equation}
where
\begin{equation}
{\cal D}^{e}_{\mu\alpha}[e\cdot a,\lambda]=-e\theta^{ij}
\Big(f_{\mu i}\star_{2}\partial_j\lambda_\alpha+a_i\star_2\partial_j
\partial_{\mu}\lambda_\alpha\Big),
\label{2.9}
\end{equation}
and
\begin{equation}
\begin{split}
&{\cal D}^{e^2}_{\mu\alpha}[e\cdot a,\lambda]=
\frac{e^2}{2}\theta^{ij}\theta^{kl}\bigg(-\left[f_{\mu k}\partial_ i\lambda_\alpha f_{l j}\right]_{\star_{3'}}-
\left[\partial_l\lambda_\alpha f_{\mu i}f_{kj}\right]_{\star_{3'}}
+\left[\partial_l\lambda_{\alpha}a_i\partial_j f_{\mu k}\right]_{\star_{3'}}
\\&+\left[f_{\mu k}a_i\partial_j \partial_l\lambda_\alpha\right]_{\star_{3'}}+\left[a_k\partial_l\left(f_{\mu i}\partial_j\lambda_\alpha\right)\right]_{\star_{3'}}
+\left[a_i\partial_j a_k \partial_l \partial_{\mu}\lambda_\alpha\right]_{\star_{3'}}
+\left[\partial_l \partial_\mu\lambda_\alpha a_i\partial_j a_k\right]_{\star_{3'}}
\\&+\left[a_k a_i \partial_l\partial_j \partial_\mu\lambda_\alpha\right]_{\star_{3'}}
-\frac{1}{2}\Big(\left[a_i\partial_k a_j\partial_l \partial_\mu\lambda_\alpha\right]_{\star_{3'}}
+\left[\partial_l \partial_\mu\lambda_\alpha a_i\partial_k a_j\right]_{\star_{3'}}\Big)\bigg).
\label{2.10}
\end{split}
\end{equation}

To compute the full one-loop photon two-point function, one also needs the SW maps for the $\bar{\Lambda}_{\dot{\alpha}}$ and $D$ fields. They read
\begin{gather}
\begin{split}
\bar{\Lambda}_{\dot{\alpha}}=&\bar{\lambda}_{\dot{\alpha}}-e\theta^{ij}a_i\star_2\partial_j \bar{\lambda}_{\dot{\alpha}}+\frac{e^2}{4}\theta^{ij}\theta^{kl}\bigg(\Big[a_i\partial_j\big(a_k\partial_l \bar{\lambda}_{\dot{\alpha}}\big)\Big]_{\star_{3'}}
\\&-\Big[a_i(f_{jk}\partial_l\bar{\lambda}_{\dot{\alpha}}-a_k\partial_l \partial_j\bar{\lambda}_{\dot{\alpha}})\Big]_{\star_{3'}}
+\Big[\partial_j \bar{\lambda}_{\dot{\alpha}}a_k(\partial_l a_i+f_{li})\Big]_{\star_{3'}}\bigg)+\mathcal O\left(e^3\right),
\end{split}
\label{2.11}\\
\begin{split}
D^{(nc)}=&D-e\theta^{ij}a_i\star_2\partial_j D+\frac{e^2}{4}\theta^{ij}\theta^{kl}\bigg(\Big[a_i\partial_j\big(a_k\partial_l D\big)\Big]_{\star_{3'}}
\\&-\Big[a_i(f_{jk}\partial_lD-a_k\partial_l \partial_jD)\Big]_{\star_{3'}}
+\Big[\partial_j Da_k(\partial_l a_i+f_{li})\Big]_{\star_{3'}}\bigg)+\mathcal O\left(e^3\right).
\end{split}
\label{2.12}
\end{gather}

Let us stress that the way we have constructed---by appropriate restriction of the SW map for the gauge-field---the SW map for $\Lambda_{\alpha}$ and $D^{(nc)}$ is very much in harmony with the idea that if supersymmetry and
gauge symmetry are not to clash, the superpartners must have similar behavior with respect to the gauge group.

As discussed in~\cite{Martin:2008xa} the noncommutative supersymmetric transformations in \eqref{2.2} can be understood as the push-forward under the SW map of the appropriate $\theta$-dependent supersymmetric transformations of the ordinary fields. Here we have worked out the $\theta$-exact expression for such deformed transformations up to the order $e^2$,
\begin{equation}
\begin{split}
&\delta^{(nc)}_{\xi}\lambda_{\alpha}=\delta^{(0)}_{\xi}\lambda_{\alpha}+e\delta^{(1)}_{\xi}\lambda_{\alpha}+e^2\delta^{(2)}_{\xi}\lambda_{\alpha}+\mathcal O\left(e^3\right),
\\&\delta^{(nc)}_{\xi}a_\mu=\delta^{(0)}_{\xi}a_\mu+e\delta^{(1)}_{\xi}a_\mu+e^2\delta^{(2)}_{\xi}a_\mu+\mathcal O\left(e^3\right),
\\&\delta^{(nc)}_{\xi}D=\delta^{(0)}_{\xi}D+e\delta^{(1)}_{\xi}a_\mu+e^2\delta^{(2)}_{\xi}D+\mathcal O\left(e^3\right),
\end{split}
\label{2.13}
\end{equation}
where
\begin{equation}
\begin{split}
&\delta^{(0)}_{\xi}\lambda_{\alpha}=-iD\xi_{\alpha}-\frac{1}{2}(\sigma^{\mu}\bar{\sigma}^{\nu})_{\alpha}^{\phantom{\alpha}\beta}\xi_\beta f_{\mu\nu},
\\&\delta^{(0)}_{\xi}a^{\mu}=i(\xi\sigma^{\mu}\bar{\lambda}-\lambda\sigma^{\mu}\bar{\xi}),
\\&\delta^{(0)}_{\xi}D=(\xi\sigma\partial_{\mu}\bar{\Lambda}-\partial_{\mu}\lambda\sigma^{\mu}\bar{\xi}),
\\&\delta^{(1)}_{\xi}\lambda_{\alpha}=-iD^{(1)}\xi_{\alpha}-\frac{1}{2e^2}(\sigma^{\mu}\bar{\sigma}^{\nu})_{\alpha}^{\phantom{\alpha}\beta}\xi_\beta F^{e^2}_{\mu\nu}-\delta^{(0)}_{\xi}\Lambda^{(1)}_{\alpha},
\\&\delta^{(1)}_{\xi}a^{\mu}=i(\xi\sigma\bar{\Lambda}^{(1)}-\Lambda^{(1)}\sigma^{\mu}\bar{\xi})-\delta^{(0)}_{\xi}A^{(1)}_\mu,
\\&\delta^{(1)}_{\xi}D=(\xi\sigma^{\mu}\bar{{\cal D}}_\mu^{e}-{\cal D}_\mu^{e}\sigma^{\mu}\bar{\xi})-\delta^{(0)}_{\xi}D^{(1)},
\\&\delta^{(2)}_{\xi}\lambda_{\alpha}=-iD^{(2)}\xi_{\alpha}-\frac{1}{2e^3}(\sigma^{\mu}\bar{\sigma}^{\nu})_{\alpha}^{\phantom{\alpha}\beta}\xi_\beta F^{e^3}_{\mu\nu}-\delta^{(1)}_{\xi}\Lambda^{(1)}_{\alpha}-\delta^{(0)}_{\xi}\Lambda^{(2)}_{\alpha},
\\&\delta^{(2)}_{\xi}a^{\mu}=i(\xi\sigma\bar{\Lambda}^{(2)}-\Lambda^{(2)}\sigma^{\mu}\bar{\xi})-\delta^{(1)}_{\xi}A^{(1)}_\mu-\delta^{(0)}_{\xi}A^{(2)}_\mu,
\\&\delta^{(2)}_{\xi}D=(\xi\sigma^{\mu}\bar{{\cal D}}_\mu^{e^2}-{\cal D}_\mu^{e^2}\sigma^{\mu}\bar{\xi})-\delta^{(1)}_{\xi}D^{(1)}-\delta^{(0)}_{\xi}D^{(2)}.
\end{split}
\label{2.14}
\end{equation}
The reader shall find in the appendix A the values of objects in the previous equations that have not been given yet.

The $\theta$-exact deformed supersymmetry transformations given in \eqref{2.13} and \eqref{2.14} can be rightly called supersymmetry transformations  since, as shown in~\cite{Martin:2008xa}, they close on translations modulo gauge transformations and, hence, they carry a representation of the supersymmetry algebra. However notice that these deformed supersymmetry transformations of the ordinary field do not realize the supersymmetry algebra linearly. Finally, since these supersymmetry transformations generate the noncommutative supersymmetry transformations of \eqref{2.2}, we conclude that the total $\theta$-exact action (given explicitly in the next subsection) has to be invariant up to the second order in $e$, under the deformed supersymmetry transformations in \eqref{2.13}.

\subsection{The action}

Now, substituting into \eqref{2.1}, the Seiberg-Witten maps from \eqref{2.6}, \eqref{2.8} and \eqref{2.11} and dropping any contribution of order $e^3$,  one obtains the SYM U(1) action in terms of commutative fields:
\begin{equation}
S\,=\,S^{\rm photon}+S^{\rm photino}+S^{\rm SUSY-auxiliary},
\label{2.15}
\end{equation}
where
\begin{equation}
\begin{split}
S^{\rm photon}=&\int -\frac{1}{4}f_{\mu\nu}f^{\mu\nu}-\frac{e}{2}\theta^{ij}f^{\mu\nu}\bigg( f_{\mu i}\star_2 f_{\nu j}-\frac{1}{4}f_{ij}\star_2f_{\mu\nu}\bigg)
\\&-\frac{e^2}{4}\theta^{ij}\theta^{kl}\Bigg((f_{\mu i}\star_2 f_{\nu j})(f^\mu_{\;\,\; k}\star_2 f^\nu_{\;\,\; l})-(f_{ij}\star_2 f_{\mu\nu})(f^\mu_{\;\,\; k}\star_2 f^\nu_{\;\,\; l})
\\&+2 f^{\mu\nu}\Big(a_i\star_2\partial_j(f_{\mu k}\star_2 f_{\nu l})-[f_{\mu k} a_i\partial_j f_{\nu l}]_{\star_{3'}}-[a_i f_{\mu k}\partial_j f_{\nu l}]_{\star_{3'}}
\\&+[f_{\mu i}f_{\nu k}f_{jl}]_{\star_{3'}}
-\frac{1}{8}\left[f_{\mu\nu}f_{ik}f_{jl}\right]_{\star_{3'}}\Big)
\\&+\frac{1}{8}\left(f^{\mu\nu}\star_2 f_{ij}\right)\left(f_{kl}\star_2 f_{\mu\nu}\right)
+\frac{1}{2}\theta^{pq}f_{\mu\nu}\left[\partial_i f_{jk} f_{lp}\partial_q f_{\mu\nu}\right]_{\mathcal M_{\rm (I)}}\Bigg),
\end{split}
\label{2.16}
\end{equation}
\begin{equation}
\begin{split}
&S^{\rm photino}=\int\,i\bar{\lambda}_{\dot{\alpha}}\bar{\sigma}^{\mu\,\dot{\alpha}\alpha}\partial_\mu\lambda_{\alpha}
\\&-ie\theta^{ij}\,\Big(\bar{\lambda}_{\dot{\alpha}}\bar{\sigma}^{\mu\,\dot{\alpha}\alpha}\Big[ f_{\mu i}\star_{2}\partial_j\lambda_\alpha+a_i\star_2\partial_j
\partial_{\mu}\lambda_\alpha\Big]+ a_i\star_2\partial_j \bar{\lambda}_{\dot{\alpha}}\bar{\sigma}^{\mu\,\dot{\alpha}\alpha}\partial_\mu\lambda_{\alpha}\Big)
\\&-i\frac{e^2}{2}\theta^{ij}\theta^{kl}\,\Bigg(\bar{\lambda}_{\dot{\alpha}}\bar{\sigma}^{\mu\,\dot{\alpha}\alpha}\Big(\Big[f_{\mu k}\partial_ i\lambda_\alpha f_{l j}\Big]_{\star_{3'}}+\Big[\partial_l\lambda_\alpha f_{\mu i}f_{kj}\Big]_{\star_{3'}}
-\Big[\partial_l\lambda_{\alpha}a_i\partial_j f_{\mu k}\Big]_{\star_{3'}}
\\&-\Big[f_{\mu k}a_i\partial_j \partial_l\lambda_\alpha\Big]_{\star_{3'}}-\Big[a_k\partial_l\left(f_{\mu i}\partial_j\lambda_\alpha\right)\Big]_{\star_{3'}}
-\Big[a_i\partial_j a_k \partial_l \partial_{\mu}\lambda_\alpha\Big]_{\star_{3'}}
-\Big[\partial_l \partial_\mu\lambda_\alpha a_i\partial_j a_k\Big]_{\star_{3'}}
\\&-\Big[a_k a_i \partial_l\partial_j \partial_\mu\lambda_\alpha\Big]_{\star_{3'}}
+\frac{1}{2}\Big[a_i\partial_k a_j\partial_l \partial_\mu\lambda_\alpha\Big]_{\star_{3'}}+\frac{1}{2}\Big[\partial_l \partial_\mu\lambda_\alpha a_i\partial_k a_j\Big]_{\star_{3'}}\Big)
\\&-2a_i\star_2\partial_j \bar{\lambda}_{\dot{\alpha}}\bar{\sigma}^{\mu\,\dot{\alpha}\alpha}
\Big(f_{\mu k}\star_{2}\partial_l\lambda_\alpha+a_k\star_2\partial_l
\partial_{\mu}\lambda_\alpha\Big)-\frac{1}{2}\Big(\Big[a_i\partial_j\big(a_k\partial_l \bar{\lambda}_{\dot{\alpha}}\big)\Big]_{\star_{3'}}
\\&
-\Big[a_i(f_{jk}\partial_l\bar{\lambda}_{\dot{\alpha}}-a_k\partial_l \partial_j\bar{\lambda}_{\dot{\alpha}})\Big]_{\star_{3'}}
+\Big[\partial_j \bar{\lambda}_{\dot{\alpha}}a_k(\partial_l a_i+f_{li})\Big]_{\star_{3'}}\Big)\bar{\sigma}^{\mu\,\dot{\alpha}\alpha}\partial_\mu\lambda_{\alpha}\Bigg)
+\mathcal O\left(e^3\right),
\end{split}
\label{2.17}
\end{equation}
and
\begin{equation}
\begin{split}
&S^{\rm SUSY-auxiliary}=\int\,\frac{1}{2}DD\;+\;e\theta^{ij}\,D (a_i\star_2\partial_j D)
\\&+\frac{e^2}{4}\theta^{ij}\theta^{kl}\,D\bigg(\Big[a_i\partial_j\big(a_k\partial_l D\big)\Big]_{\star_{3'}}
-\Big[a_i(f_{jk}\partial_lD-a_k\partial_l \partial_jD)\Big]_{\star_{3'}}
+\Big[\partial_j Da_k(\partial_l a_i+f_{li})\Big]_{\star_{3'}}\bigg)
\\&+\frac{e^2}{2}\theta^{ij}\theta^{kl}\,\big(a_i\star_2\partial_j D\big)\big( a_k\star_2\partial_l D\big)+\mathcal O\left(e^3\right).
\label{2.18}
\end{split}
\end{equation}

First we note that, since the Feynman rules of  the 3- and 4-photon self-couplings (\ref{2.16}), are already given in previous papers \cite{Horvat:2013rga} and \cite{Horvat:2015aca}, respectively, we shall not repeat them here. Photino-photon Feynman rules, obtained from (\ref{2.17}), are given explicitly in the appendix C.

\subsection{The photon one-loop contributions to the photon polarization tensor}

Most generally speaking, the total photon one-loop 1PI two-point function $\Pi^{\mu\nu}(p)$ in the ${\cal N} =1,2,4$ SYM theory is the sum of the following contributions
\begin{equation}
\Pi^{\mu\nu}(p)=(B^{\mu\nu}(p)+T^{\mu\nu}(p))+n_f(P^{\mu\nu}(p)_{\rm bub}+P^{\mu\nu}(p)_{\rm tad})+n_s(S^{\mu\nu}(p)_{\rm bub}+S^{\mu\nu}(p)_{\rm tad}),
\label{4.1Pi}
\end{equation}
where $B^{\mu\nu}(p)$, $T^{\mu\nu}(p)$, $P^{\mu\nu}(p)_{\rm bub}$, $P^{\mu\nu}(p)_{\rm tad}$, $S^{\mu\nu}(p)_{\rm bub}$ and $S^{\mu\nu}(p)_{\rm tad}$ refer to the contributions from the photon bubble and tadpole, the photino bubble and tadpole, and the adjoint scalar bubble and tadpole diagrams, respectively. The last two diagrams appear only in the extended SUSY, of course. We use $n_f$ for the number of photinos (Weyl fermions) and $n_s$ for the number of real adjoint scalar bosons (one complex scalar is counted as two real scalars), which are uniquely determined by ${\cal N} =1,2,4$ supersymmetry.

Explicit computation revolves that each of these diagrams can be expressed as a linear combinations of five transverse tensor structures
\begin{equation}
\begin{split}
&\Pi(B,T,P,S_{\rm bub},S_{\rm tad})^{\mu\nu}(p)=\frac{e^2}{(4\pi)^2}\bigg\{\Big[g^{\mu\nu}p^2-p^\mu p^\nu\Big]\Pi_1(B_1,T_1,P_1,S^{\rm bub}_1,S^{\rm tad}_1)(p)
\\&+(\theta p)^\mu (\theta p)^\nu \Pi_2(B_2,T_2,P_2,S^{\rm bub}_2,S^{\rm tad}_2)(p)
\\&+\Big[g^{\mu\nu}(\theta p)^2-(\theta\theta)^{\mu\nu}p^2
+ p^{\{\mu}(\theta\theta p)^{\nu\}}\Big]\Pi_3(B_3,T_3,P_3,S^{\rm bub}_3,S^{\rm tad}_3)(p)
\\&+\Big[(\theta\theta)^{\mu\nu}(\theta p)^2+(\theta\theta p)^\mu(\theta\theta p)^\nu\Big]\Pi_4(B_4,T_4,P_4,S^{\rm bub}_4,S^{\rm tad}_4)(p)
\\&+ (\theta p)^{\{\mu} (\theta\theta\theta p)^{\nu\}} \Pi_5(B_5,T_5,P_5,S^{\rm bub}_5,S^{\rm tad}_5)(p)\bigg\}.
\end{split}
\label{2.1P} 
\end{equation}
The sum \eqref{4.1Pi} can be expressed, in the language of the five tensor decomposition (\ref{2.1P}), as 
\footnote{As we will see soon, the photino tadpole diagram vanishes, so we can simply denote $P^{\mu\nu}(p)=P^{\mu\nu}(p)_{\rm bub}$ and consequently $P_i(p)=P_i(p)_{\rm bub}$.}
\begin{equation}
\Pi_i(p)=B_i(p)+T_i(p)+n_f\left(P_i^{\rm bub}(p)+P_i^{\rm tad}(p)\right)+n_s\left(S_i^{\rm bub}(p)+S_i^{\rm tad}(p)\right).
\label{2.1Pi}
\end{equation}
In the subsequent sections we are going to compute and give the coefficients $B_i(p)$, $T_i(p)$, $P^{\rm bub}_i(p)$ and $P^{\rm tad}_i(p)$, $S^{\rm bub}_i(p)$ and $S^{\rm tad}_i(p)$ in detail via equations \eqref{2.25}, \eqref{2.Ti}, \eqref{2.Pfinal}, \eqref{Sbubi} and \eqref{Stadi}, respectively.
For the ${\cal N}=1$ theory $n_f=1$, $n_s=0$, thus from (\ref{2.1Pi}) we have
\begin{equation}
\Pi_i^{{\cal N}=1}(p)=B_i(p)+T_i(p)+P^{\rm bub}_i(p)+P^{\rm tad}_i(p).
\label{2.1Pi1}
\end{equation}
In this section we are going to show that all quadratic IR divergences cancel in each of the $\Pi_i^{{\cal N}=1}$'s, then we extend our results to the ${\cal N}=2,4$ theories as well.

We choose one specific set of four (five in the sections 6 and 7) nonplanar/special function integrals $T_0$, $I_K^0$, $I_K^1$ and $I_H$ alongside the usual planar/commutative UV divergent integrals to express all loop diagrams/coefficients in this article. This decomposition enjoys the advantage that each nonplanar integral bear distinctive asymptotic behavior in the IR regime: $T_0$ carries all the quadratic IR divergence $(\theta p)^{-2}$, with a pre-factor $-2$, while $I_K^0$ and $I_K^1$ contain the dual logarithmic noncommutative IR divergence $\ln(p^2(\theta p)^2)$, with pre-factors $-1/2$ and $-1/12$, respectively. The last integral $I_H$ is finite at the IR limit. 
Further details of these integrals are given in the appendix B.

\subsubsection{The photon-bubble diagram}

\begin{figure}
\begin{center}
\includegraphics[width=6cm]{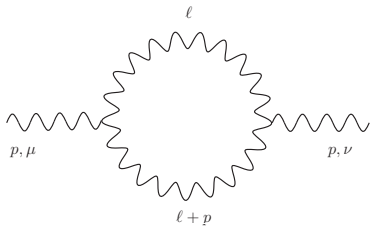}
\end{center}
\caption{Three-photon bubble contribution to the photon two-point function ${B}^{\mu\nu}(p)$.}
\label{fig:photonbubble}
\end{figure}

 From the photon bubble diagram Fig. \ref{fig:photonbubble} we obtain the following loop-coefficients~\cite{Horvat:2013rga}
\begin{equation}
\begin{split}
B&_1=\frac{2D^2-9D+8}{D-1}(4\pi\mu^2)^{2-\frac{D}{2}}(p^2)^{\frac{D}{2}-1}{\rm\Gamma}\left(2-\frac{D}{2}\right){\rm B}\left(\frac{D}{2}-1, \frac{D}{2}-1\right)\Bigg|_{D\to 4-\epsilon}
\\&-16I_K^1-4 I_H
\\&+3\tr\theta\theta\frac{p^2}{(\theta p)^2}\frac{1}{2}\bigg(2(4\pi\mu^2)^{2-\frac{D}{2}}(p^2)^{\frac{D}{2}-1}{\rm\Gamma}\left(2-\frac{D}{2}\right){\rm B}\left(\frac{D}{2}-1, \frac{D}{2}-1\right)\Bigg|_{D\to 4-\epsilon}
\\&-4I_K^0-4 I_H\bigg)
\\&+(\theta\theta p)^2\frac{p^2}{(\theta p)^4}\bigg(4(4\pi\mu^2)^{2-\frac{D}{2}}(p^2)^{\frac{D}{2}-1}{\rm\Gamma}\left(2-\frac{D}{2}\right){\rm B}\left(\frac{D}{2}-1, \frac{D}{2}-1\right)\Bigg|_{D\to 4-\epsilon}
\\&-8I_K^0-4 I_H\bigg),
\\
B&_2=\frac{1}{(\theta p)^2}\bigg(2(D-2)(4\pi\mu^2)^{2-\frac{D}{2}}(p^2)^{\frac{D}{2}-1}{\rm\Gamma}\left(2-\frac{D}{2}\right){\rm B}\left(\frac{D}{2}-1, \frac{D}{2}-1\right)\Bigg|_{D\to 4-\epsilon}
\\&-\frac{16}{3}T_0+p^2(48I_K^1-16I_K^0-4 I_H)
\\&+\frac{1}{2}\tr\theta\theta\frac{p^4}{(\theta p)^2}\Big(-4(4\pi\mu^2)^{2-\frac{D}{2}}(p^2)^{\frac{D}{2}-2}{\rm\Gamma}\left(2-\frac{D}{2}\right){\rm B}\left(\frac{D}{2}-1, \frac{D}{2}-1\right)\Bigg|_{D\to 4-\epsilon}
\\&+8I_K^0+8 I_H\Big)\bigg),
\\
B_3&=\frac{1}{(\theta p)^2}\bigg(2(4\pi\mu^2)^{2-\frac{D}{2}}(p^2)^{\frac{D}{2}-1}{\rm\Gamma}\left(2-\frac{D}{2}\right){\rm B}\left(\frac{D}{2}-1, \frac{D}{2}-1\right)\Bigg|_{D\to 4-\epsilon}
\\&+\frac{16}{3}T_0-p^2(4I_K^0+8 I_H)\bigg),
\\
B_4&=\frac{p^2}{(\theta p)^4}\bigg(-4(4\pi\mu^2)^{2-\frac{D}{2}}(p^2)^{\frac{D}{2}-2}{\rm\Gamma}\left(2-\frac{D}{2}\right){\rm B}\left(\frac{D}{2}-1, \frac{D}{2}-1\right)\Bigg|_{D\to 4-\epsilon}
\\&+\frac{16}{3}T_0+8I_K^0+4 I_H\bigg),
\\
B_5&=\frac{p^2}{(\theta p)^4}\bigg(4(4\pi\mu^2)^{2-\frac{D}{2}}(p^2)^{\frac{D}{2}-2}{\rm\Gamma}\left(2-\frac{D}{2}\right){\rm B}\left(\frac{D}{2}-1, \frac{D}{2}-1\right)\Bigg|_{D\to 4-\epsilon}
\\&-8I_K^0-4 I_H\bigg).
\label{2.25}
\end{split}
\end{equation}
Extracting the divergent parts from each of the $B_i$'s
\begin{gather}
\begin{split}
B_1(p)&=+\bigg(\frac{4}{3} +3\; \frac{p^2(\tr\theta\theta)}{(\theta p)^2}
+4\; \frac{p^2(\theta\theta p)^2}{(\theta p)^4}\bigg)
\left(\frac{2}{\epsilon} + \ln(\mu^2(\theta p)^2)\right) +{\rm finite \; terms},
\\
B_2(p)&=+2\frac{p^2}{(\theta p)^2}\bigg(2-\frac{p^2(\tr\theta\theta)}{(\theta p)^2}
\bigg)\left(\frac{2}{\epsilon} + \ln(\mu^2(\theta p)^2)\right)+\frac{32}{3}\frac{1}{(\theta p)^4}+{\rm finite \; terms},
\\
B_3(p)&=+2\frac{p^2}{(\theta p)^2}
\left(\frac{2}{\epsilon} + \ln(\mu^2(\theta p)^2)\right)-\frac{32}{3}\frac{1}{(\theta p)^4}+{\rm finite \; terms},
\\
B_4(p)&=-4\frac{p^4}{(\theta p)^4}\left(\frac{2}{\epsilon} + \ln(\mu^2(\theta p)^2)\right)-\frac{32}{3}\frac{p^2}{(\theta p)^6}+{\rm finite \; terms},
\\
B_5(p)&=+4\frac{p^4}{(\theta p)^4}
\left(\frac{2}{\epsilon} + \ln(\mu^2(\theta p)^2)\right)+{\rm finite \; terms},
\end{split}
\label{2.1Bi}
\end{gather}
we observe the presence of the UV plus logarithmic IR divergences in all of them. The logarithmic IR divergences from both planar and nonplanar sources in the bubble diagram appear to have identical coefficient and combine into a single $\ln(\mu^2(\theta p)^2)$ term, confirming the expected UV/IR mixing. The quadratic IR divergence, on the other hand, exists only in the $B_{2,3,4}$ terms.

\subsubsection{The photon-tadpole diagram}

\begin{figure}
\begin{center}
\includegraphics[width=5cm]{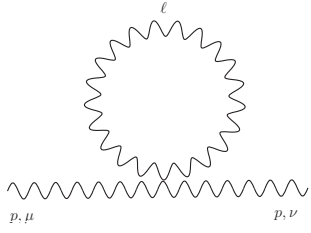}
\end{center}
\caption{Four-photon tadpole contribution to the photon two-point function $T^{\mu\nu}(p)$.}
\label{fig:photontadpole}
\end{figure}
From tadpole Fig. \ref{fig:photontadpole} we obtain the same tensor structure as from the photon bubble diagram (Fig. \ref{fig:photonbubble}) with the following loop-coefficients $T_i(p)$:
\begin{equation}
\begin{split}
T_1(p)=&T_5(p)=0,
\\
T_2(p)=&-\frac{32}{3}\frac{1}{(\theta p)^2}T_0=\frac{64}{3}\frac{1}{(\theta p)^4},\;\;
\\
T_3(p)=&-\frac{16}{3}\frac{1}{(\theta p)^2}T_0=\frac{32}{3}\frac{1}{(\theta p)^4},\;\;
\\
T_4(p)=&-\frac{16}{3}\frac{1}{(\theta p)^4}T_0=\frac{32}{3}\frac{p^2}{(\theta p)^6}.
\end{split}
\label{2.Ti}
\end{equation}
We notice immediately the absence of UV plus logarithmic divergent terms contrary to the photon-bubble diagram results (\ref{2.1Bi}). In addition, the tadpole diagram produces no finite terms either, and the quadratic IR are again present in the second, third and fourth term!

\subsection{The photino one-loop contributions to the photon polarization tensor}

\begin{figure}
\begin{center}
\includegraphics[width=5cm]{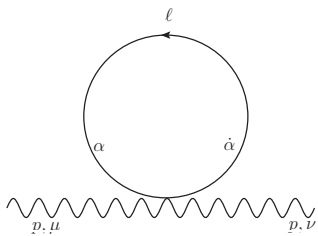}
\end{center}
\caption{Photon-2photinos tadpole contribution to the photon two-point function
 $P^{\mu\nu}_{\rm tad}(p)$.}
\label{fig:photino-tadpol}
\end{figure}

The photino sector contains two diagrams: photino tadpole Fig. \ref{fig:photino-tadpol} and photino bubble Fig. \ref{fig:photino-bubble}. We are going to see below that only the latter contributes to the photon polarization tensor.

\subsubsection{The photino-tadpole diagram}

The photino-tadpole contribution is computed using vertex (\ref{C.4}).
It produces only the quadratic IR divergent terms which cancel each other internally, thus giving vanishing contribution to the photon polarization tensor
\begin{equation}
\begin{split}
P_{\rm tad}^{\mu\nu}(p)
&=-\mu^{4-D}\int\,\frac{d^D  \ell}{(2\pi)^D}
\frac{1}{ \ell^2}{V^{e^2}}_{\mu_1}^{\mu\nu}[ \ell,-p,p, \ell;\theta^{ij}] \ell_{\mu_2}
{\rm Tr}(\bar{\sigma}^{\mu_1}\sigma^{\mu_2})\Big |_{D\to4}
\\&
=-\frac{e^2}{3\pi^2}\frac{(\theta p)^\mu (\theta p)^\nu}{(\theta p)^4}(4-2-2)=0.
\end{split}
\label{2.27}
\end{equation}

\subsubsection{The photino-bubble diagram}
Taking the photino-photon Feynman rules from appendix C we obtain
\begin{figure}
\begin{center}
\includegraphics[width=6.5cm]{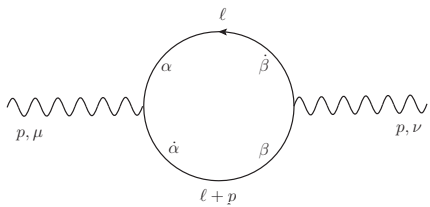}
\end{center}
\caption{Photon-photino bubble contribution to the photon two-point function:
 $P_{\rm bub}^{\mu\nu}(p)$.}
\label{fig:photino-bubble}
\end{figure}
the following photino-bubble contribution to the photon polarization tensor, $P_{\rm bub}^{\mu\nu}(p)$, from Fig. \ref{fig:photino-bubble}:
\begin{equation}
\begin{split}
P_{\rm bub}^{\mu\nu}(p)&=-\mu^{4-D}\int\,\frac{d^D \ell}{(2\pi)^D}
\frac{1}{ \ell^2( \ell+p)^2}
\\&\cdot {V^e}_{\mu_1}^{\mu}[ \ell+p,p, \ell;\theta^{ij}] \ell_{\mu_2}{V^e}_{\mu_3}^{\nu}[ \ell,-p, \ell+p;\theta^{ij}]( \ell+p)_{\mu_4}
{\rm Tr}({\bar{\sigma}^{\mu_1}}\sigma^{\mu_2}{\bar{\sigma}^{\mu_3}}\sigma^{\mu_4}).
\end{split}
\label{2.2P}
\end{equation}
Taking into account the trace
\begin{equation}
{\rm Tr}(\bar{\sigma}^{\mu_1}\sigma^{\mu_2}\bar{\sigma}^{\mu_3}\sigma^{\mu_4})=
2(\eta^{\mu_1\mu_2}\eta^{\mu_3\mu_4}-\eta^{\mu_1\mu_3}\eta^{\mu_2\mu_4}+\eta^{\mu_1\mu_4}\eta^{\mu_2\mu_3}-i\epsilon^{\mu_1\mu_2\mu_3\mu_4}),
\label{2.2P1}
\end{equation}
and that  $P^{\mu\nu}_{\rm bub}(p)$ cannot have, at the end of the day, contributions depending on $\epsilon^{\mu_1\mu_2\mu_3\mu_4}$, one arrives at
\begin{equation}
\begin{split}
P_{\rm bub}^{\mu\nu}(p)=-\mu^{4-D}\int\,&\frac{d^D  \ell}{(2\pi)^D}\frac{1}{ \ell^2( \ell+p)^2}
\\&\cdot\Big[{V^e}_{\mu_1}^{\mu}[ \ell+p,p, \ell;\theta^{ij}] \ell^{\mu_1}{V^e}_{\mu_3}^{\nu}[ \ell,-p, \ell+p;\theta^{ij}]
( \ell+p)^{\mu_3}
\\&+{V^e}_{\mu_1}^{\mu}[ \ell+p,p, \ell;\theta^{ij}]( \ell+p)^{\mu_1}{V^e}_{\mu_3}^{\nu}[ \ell,-p, \ell+p;\theta^{ij}] \ell^{\mu_3}
\\&-{V^e}_{\mu_1}^{\mu}[ \ell+p,p, \ell;\theta^{ij}]\,\eta^{\mu_1\mu_3}\,{V^e}_{\mu_3}^{\nu}[ \ell,-p, \ell+p;\theta^{ij}]\,
 \ell\cdot( \ell+p)\Big].
\end{split}
\label{2.2P2}
\end{equation}
After some amount of computations we find that only first two of the general five-terms structure (\ref{2.1P}) survive in $D=4$ dimensions:
\begin{equation}
\begin{split}
P_1(p)& = -2\frac{D-2}{D-1}(4\pi\mu^2)^{2-\frac{D}{2}}(p^2)^{\frac{D}{2}-2}{\rm\Gamma}\left(2-\frac{D}{2}\right){\rm B}\left(\frac{D}{2}-1, \frac{D}{2}-1\right)\Bigg|_{D\to 4-\epsilon}+16I_K^1,
\\
P_{2}(p)&=16T_0+8p^2(I_K^0-6I_K^1),\;P_3(p)=P_4(p)=P_5(p)=0.
\end{split}
\label{2.Pfinal}
\end{equation}
After inspecting the divergences in these two terms we also find that the first of the two, $P_1(p)$, contain the logarithmic UV/IR mixing terms, while $P_{2}(p)$ possesses only quadratic IR divergence and finite terms, as the dual NC logarithmic divergences from integrals $I_K^0$ and $I_K^1$ cancel each other.

Summing over \eqref{2.1Bi}, \eqref{2.Ti} and \eqref{2.Pfinal} one can see that the total quadratic IR divergences in all $\Pi_i$'s are zero. The total UV divergences for the ${\cal N}=1$ theory are as follows
\begin{gather}
\Pi_1(p)|_{\rm UV}=\frac{p^2}{(\theta p)^4}\Big(3(\tr\theta\theta)(\theta p)^2+4(\theta\theta p)^2\Big)\left(\frac{2}{\epsilon}+\ln(\mu^2(\theta p)^2)\right),
\label{UVterm1}\\
\Pi_2(p)|_{\rm UV}=\frac{p^2}{(\theta p)^2}\bigg(2-\frac{p^2(\tr\theta\theta)}{(\theta p)^2}
\bigg)\left(\frac{2}{\epsilon}+\ln(\mu^2(\theta p)^2)\right),
\label{UVterm2}\\
\Pi_3(p)|_{\rm UV}=\frac{p^2}{(\theta p)^2}\left(\frac{2}{\epsilon}+\ln(\mu^2(\theta p)^2)\right),
\label{UVterm3}\\
\Pi_4(p)|_{\rm UV}=-4\frac{p^4}{(\theta p)^4}\left(\frac{2}{\epsilon}+\ln(\mu^2(\theta p)^2)\right),
\label{UVterm4}\\
\Pi_5(p)|_{\rm UV}=4\frac{p^4}{(\theta p)^4}
\left(\frac{2}{\epsilon} + \ln(\mu^2(\theta p)^2)\right).
\label{UVterm5}
\end{gather}


\subsection{The one-loop SUSY-auxiliary field contributions to the photon propagator}

The   free two-point function of the SUSY-auxiliary field $D$ reads
\begin{equation}
\Big< D(x) D(y)\Big>=\delta(x-y),
\label{2.28}
\end{equation}
hence, the integrals to be computed in dimensional regularization are of the type
\begin{equation}
\int \frac{d^D  \ell}{(2\pi)^D}  \frac{e^{i( \ell\theta p)}}{( \ell\theta{p})^n}.
\label{2.29}
\end{equation}
The integrals in \eqref{2.29} vanish in dimensional regularization and hence the SUSY-auxiliary field $D$ does not contribute to the one-loop photon propagator. Indeed, following~\cite{Collins:1984xc}, we first split the $D$-dimensional $ \ell(\equiv \ell^{(D)})$ into
\begin{equation}
 \ell^{(D)}= \ell^{(D-1)}+ x \frac{\theta p}{|\theta{p}|},
 \label{2.30}
\end{equation}
where $ \ell^{(D-1)}$ belongs to $D-1$ dimensional space orthogonal to $\theta p$, and $|\theta p|>0$ is the modulus of $\theta p$ --recall that $\theta p$ is a space-like vector, since $\theta^{\mu 0}=0$. Then introduce the following definition of the dimensionally regularized integral in \eqref{2.29}:
\begin{equation}
\int \frac{d^D  \ell}{(2\pi)^D}  \frac{e^{i \ell\theta{p}}}{( \ell\theta p)^n}=\int \frac{d^{(D-1)}  \ell}{(2\pi)^{(D-1)}} \lim_{\varepsilon\rightarrow 0}\int_{-\infty}^{+\infty}\,dx\,  \frac{e^{ix}}{(x+i\varepsilon)^n}.
\label{2.31}
\end{equation}
However, in dimensional regularization -see~\cite{Collins:1984xc}-
\begin{equation}
\int \frac{d^{(D-1)}  \ell}{(2\pi)^{(D-1)}}=0,
\label{2.32}
\end{equation}
which in turn leads to the conclusion that the integral \eqref{2.29} vanishes under the dimensional regularization procedure.

It is not difficult to see that the argument above can be generalized to integrals with positive $\ell$ powers too, i.e.
\begin{equation}
\int \frac{d^D  \ell}{(2\pi)^D} \ell^{\mu_1}...\ell^{\mu_{2n}}  \frac{e^{i( \ell\theta p)}}{( \ell\theta{p})^n}=0.
\end{equation}
One can explicitly verify two special cases of the identity above
\begin{equation}
\int \frac{d^D  \ell}{(2\pi)^D}  f_{\star_2}\left(\ell,p\right)^2=\int \frac{d^D  \ell}{(2\pi)^D} \ell^{\mu}\ell^{\nu} f_{\star_2}\left(\ell,p\right)^2=0,
\label{vanishingtadpole}
\end{equation}
using a generalization of the n-nested zero regulator method~\cite{Horvat:2015aca}.

\subsection{The one-loop photino 1-PI two point function}

\begin{figure}
\begin{center}
\includegraphics[width=5cm]{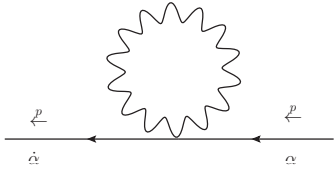}
\end{center}
\caption{$\cal N$=1 photino-photon tadpole: $\Sigma^{\dot{\alpha}\alpha}(p)_{\rm tad}$.}
\label{fig:N=2 Fermion-photon-tadpole}
\end{figure}

\begin{figure}
\begin{center}
\includegraphics[width=5cm]{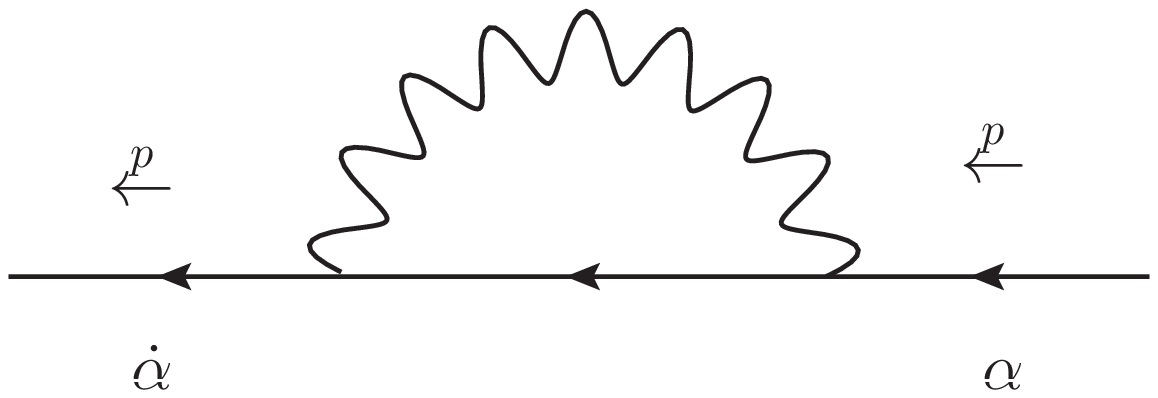}
\end{center}
\caption{$\cal N$=1 photino-photon bubble: $\Sigma^{\dot{\alpha}\alpha}(p)_{\rm bub}$.}
\label{fig:N=2 Fermion-photon-bubble}
\end{figure}

The photino self-energy consists two diagrams, a tadpole Fig. \ref{fig:N=2 Fermion-photon-tadpole} and a bubble Fig. \ref{fig:N=2 Fermion-photon-bubble}. Explicit computation shows that the tadpole diagram Fig. \ref{fig:N=2 Fermion-photon-tadpole} vanishes:
\begin{eqnarray}
\Sigma^{\dot{\alpha}\alpha}(p)_{\rm tad}&=&0.
\label{Sigmatad0}
\end{eqnarray}
The bubble diagram was computed in \cite{Horvat:2013rga}, which boils down to the following expressions
\begin{eqnarray}
\Sigma^{\dot{\alpha}\alpha}(p)_{\rm bub}&=&
-\frac{e^2}{(4\pi)^2}\sigma^{\dot{\alpha}\alpha}_\mu\bigg[ p^\mu\; N_1(p)+(\theta\theta p)^\mu\; N_2(p)\bigg],
\label{N212}
\end{eqnarray}
with
\begin{equation}
\begin{split}
N&_1(p)=-\frac{1}{2}(\theta p)^2I_H
\\&+\tr\theta\theta\frac{p^2}{(\theta p)^2}\bigg((4\pi\mu^2)^{2-\frac{D}{2}}(p^2)^{\frac{D}{2}-2}{\rm\Gamma}\left(2-\frac{D}{2}\right){\rm B}\left(\frac{D}{2}-1, \frac{D}{2}-1\right)\Bigg|_{D\to 4-\epsilon}
\\&-(2I_K^0+2 I_H)\bigg)
\\&+(\theta\theta p)^2\frac{p^2}{(\theta p)^4}\bigg(2(4\pi\mu^2)^{2-\frac{D}{2}}(p^2)^{\frac{D}{2}-2}{\rm\Gamma}\left(2-\frac{D}{2}\right){\rm B}\left(\frac{D}{2}-1, \frac{D}{2}-1\right)\Bigg|_{D\to 4-\epsilon}
\\&-(4I_K^0+2 I_H)\bigg),
\end{split}
\label{N1}
\end{equation}
and
\begin{equation}
\begin{split}
N_2(p)=&4\frac{p^2}{(\theta p)^2}I_H.
\end{split}
\label{N2}
\end{equation}
One can easily notice the absence of the quadratic IR divergent integral $T_0$. The UV divergence can be expressed as follows
\begin{equation}
\begin{split}
\Sigma^{\dot{\alpha}\alpha}(p)_{\rm bub}|_{\rm UV}=&-\frac{e^2}{(4\pi)^2}\sigma^{\dot{\alpha}\alpha}_\mu p^\mu\; N_1(p)|_{\rm UV}
\\=&-\frac{e^2}{(4\pi)^2}\sigma^{\dot{\alpha}\alpha}_\mu p^\mu\left(\tr\theta\theta\frac{p^2}{(\theta p)^2}+2(\theta\theta p)^2\frac{p^2}{(\theta p)^4}\right)\left(\frac{2}{\epsilon} + \ln(\mu^2(\theta p)^2)\right).
\end{split}
\label{2.42}
\end{equation}

\section{Minimal action of the noncommutative adjoint scalar field}

It is commonly known that extended, ${\cal N}=2,4$, super YM theories contain not only fermion (photino) but also scalar bosons in the adjoint representation. These scalar bosons couple minimally to the gauge field, and their action  for the real scalar is
\begin{equation}
S^{\rm real}=\frac{1}{2}\int {\cal D}^{\mu}\Phi {\cal D}_{\mu}\Phi,
\label{3.1}
\end{equation}
or
\begin{equation}
S^{\rm complex}=\int {\cal D}^{\mu}\Phi^\dagger {\cal D}_{\mu}\Phi,
\label{3.2}
\end{equation}
for the complex scalar. We study the minimal interacting scalar boson's contribution to 1-PI photon two point function as well as the scalar's own 1-PI two point function in this section. These results will be used for our discussion on ${\cal N}=2,4$ SYM in the subsequent sections.

It is straightforward to derive the SW map expansion of either $S^{\rm real}$ or $S^{\rm complex}$ using the method described in the appendix A\begin{equation}
\begin{split}
S^{\rm real}=&\frac{1}{2}\int {\cal D}^{\mu}\Phi {\cal D}_{\mu}\Phi=\frac{1}{2}\int \partial^{\mu}\phi \partial_{\mu}\phi-2e\theta^{ij}\partial^{\mu}\phi \Big( f_{\mu i}\star_2 \partial_j \phi+\frac{1}{4}f_{ij}\star_2\partial_{\mu}\phi\Big)+
\\&+e^2\theta^{ij}\theta^{kl}\bigg((f_{\mu i}\star_2 \partial_j \phi)(f^\mu_{\;\,\; k}\star_2 \partial_l \phi)+(\partial_\mu \phi\star_2 f_{ij})(f^\mu_{\;\,\; k}\star_2 \partial_l \phi)
\\&+\partial^\mu \phi \Big([f_{\mu i} \partial_l \phi f_{ik}]_{\star_{3'}}+[\partial_l \phi f_{\mu i}  f_{ik}]_{\star_{3'}}+ [\partial_l \phi a_i \partial_jf_{\mu k}]_{\star_{3'}}
+[f_{\mu k}a_i \partial_j\partial_l \phi  ]_{\star_{3'}}
\\&+[a_i \partial_j (f_{\mu k}\partial_l \phi)]_{\star_{3'}}-2a_i\star_2\partial_j (f_{\mu k}
\star_2\partial_l \phi)+\frac{1}{4}[\partial_\mu \phi f_{il}  f_{jk}]_{\star_{3'}}
\\&+\frac{1}{8}
f_{ij}\star_2(f _{kl}\star_2 \partial_\mu \phi)+\frac{1}{2}\theta^{pq}[\partial_i f_{jk}f_{lp}\partial_q\partial_\mu\phi]_{{\cal M_{\rm (I)}}}\Big)\bigg),
\end{split}
\label{AcSreal}
\end{equation}
\begin{equation}
\begin{split}
S^{\rm complex}=&\int {\cal D}^{\mu}\Phi^\dagger {\cal D}_{\mu}\Phi=\int\partial_\mu\phi^*\partial^\mu\phi
\\&-e\theta^{ij}\left(\partial^\mu\phi^*(f_{\mu i}\star_2\partial_j\phi)+(f_{\mu i}\star_2\partial_j\phi^*)\partial^\mu\phi+\frac{1}{2}(\partial_\mu\phi^*\star_2 f_{ij})\partial^\mu\phi\right)
\\&+e^2\theta^{ij}\theta^{kl}\Big((f_{\mu i}\star_2\partial_j\phi^*)(f^\mu_{\;\,\; k}\star_2 \partial_l \phi)+\frac{1}{2}((f_{\mu i}\star_2\partial_j\phi^*)(f_{ij}\star_2 \partial^\mu \phi)
\\&+(f_{\mu i}\star_2\partial_j\phi)(f_{ij}\star_2 \partial^\mu \phi^*))+\frac{1}{2}\partial^\mu\phi^*([f_{\mu i}\partial_l\phi f_{jk}]_{\star_{3'}}+[\partial_l\phi f_{\mu i} f_{jk}]_{\star_{3'}})
\\&+\frac{1}{2}([f_{\mu i}\partial_l\phi^* f_{jk}]_{\star_{3'}}+[\partial_l\phi^* f_{\mu i} f_{jk}]_{\star_{3'}})\partial^\mu\phi+\frac{1}{2}\partial^\mu\phi^*([\partial_l\phi a_i\partial_j f_{\mu k}]_{\star_{3'}}
\\&+[f_{\mu k} a_i\partial_j \partial_l\phi ]_{\star_{3'}}+[a_i\partial_j(f_{\mu k}  \partial_l\phi) ]_{\star_{3'}}-2a_i\star_2\partial_j(f_{\mu k}\star_2\partial_l\phi))
\\&+\frac{1}{2}([\partial_l\phi^* a_i\partial_j f_{\mu k}]_{\star_{3'}}
+[f_{\mu k} a_i\partial_j \partial_l\phi^* ]_{\star_{3'}}+[a_i\partial_j(f_{\mu k}  \partial_l\phi^*) ]_{\star_{3'}}
\\&-2a_i\star_2\partial_j(f_{\mu k}\star_2\partial_l\phi^*))\partial^\mu\phi+\frac{1}{4}\partial^\mu\phi^*[\partial_\mu\phi f_{il}f_{jk}]_{\star_{3'}}+\frac{1}{8}(\partial^\mu\phi^*\star_2 f_{ij}(\partial_\mu\phi\star_2 f_{kl})
\\&+\frac{1}{4}\theta^{pq}(\partial^\mu\phi^*[\partial_i f_{jk}f_{lp}\partial_q\partial_\mu\phi]_{{\cal M_{\rm (I)}}}-\partial_q\partial_\mu\phi^*[\partial_i f_{jk}f_{lp}\partial_\mu\phi]_{{\cal M_{\rm (I)}}})\Big),
\label{AcScompl}
\end{split}
\end{equation}
with the product $\cal{M_{\rm (I)}}$ being defined in \cite{Horvat:2015aca}.
One can show that
\begin{equation}
S^{\rm complex}\big(\phi=\frac{1}{\sqrt 2}(\varphi_1+i\varphi_2)\big)=S^{\rm real}\big(\varphi_1\big)+S^{\rm real}\big(\varphi_2\big),
\label{3.5}
\end{equation}
if we express one complex scalar in terms of two real scalars. For this reason one complex scalar contribution to the photon 1-PI two point function is twice as one real scalar, while the photon contribution to the complex scalar two point function is the same as for the real scalar two point function. Thus, we shall compute only those for the real scalar field. The scalar-photon Feynman rules are given in the appendix D.

\subsection{Scalar one-loop contributions to the photon polarization tensor}

Like the photino sector, the adjoint scalar sector contains also two diagrams that contribute to the photon polarization tensor, the scalar-bubble diagram Fig.\ref{fig:scalar-photon-bub} and scalar-tadpole diagram Fig.\ref{fig:scalar-photon-tad}. They both follow the five tensor structure decomposition~\eqref{2.1P} and stay nonzero at the $D\to 4-\epsilon$ limit.

\subsubsection{The scalar-bubble diagram}

\begin{figure}
\begin{center}
\includegraphics[width=6cm]{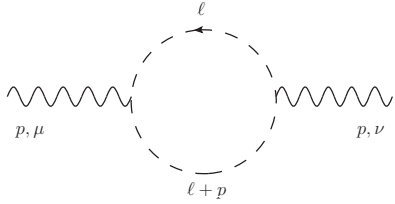}
\end{center}
\caption{Photon-scalar bubble: $S^{\mu\nu}(p)_{\rm bub}$.}
\label{fig:scalar-photon-bub}
\end{figure}

Using Feynman rule (\ref{D.1}) and employing the dimensional regularization techniques we obtain the following loop-coefficients from the photon-scalar bubble diagram Fig.\ref{fig:scalar-photon-bub}:
\begin{equation}
\begin{split}
S^{\rm bub}_{1}(p)=&-\frac{1}{D-1}(4\pi\mu^2)^{2-\frac{D}{2}}(p^2)^{\frac{D}{2}-2}{\rm\Gamma}\left(2-\frac{D}{2}\right){\rm B}\left(\frac{D}{2}-1, \frac{D}{2}-1\right)\Bigg|_{D\to 4-\epsilon}
\\&-8I_K^1+2I_K^0,
\\
S^{\rm bub}_{2}(p)=&-8\frac{1}{(\theta p)^2}T_0=\frac{16}{(\theta p)^4},\;
S^{\rm bub}_{3}(p=4\frac{1}{(\theta p)^2}T_0=-\frac{8}{(\theta p)^4},\,
\\S^{\rm bub}_{4}(p)=&\frac{16}{3}\frac{p^2}{(\theta p)^4}T_0=-\frac{32}{3}\frac{p^2}{(\theta p)^6},\;S^{\rm bub}_{5}(p) =0.
\end{split}
\label{Sbubi}
\end{equation}

\subsubsection{The scalar-tadpole diagram}

Next, with Feynman rule (\ref{D.2}) we compute the photon-scalar tadpole diagram in Fig. \ref{fig:scalar-photon-tad},
\begin{gather}
\begin{split}
S^{\mu\nu}(p)_{\rm tad}=\sum_{k'=1}^{4}S^{\mu\nu}_{k'}(p)_{\rm tad}&
=\frac{1}{2}\int\,\frac{d^D  \ell}{(2\pi)^D}
\frac{i}{ \ell^2}\sum_{k'=1}^{4}S^{\mu\nu}_{k'}(\ell,p,-p,-\ell).
\end{split}
\label{scalphottad}
\end{gather}
Starting with the first integral under dimensional regularization:
\begin{gather}
\begin{split}
&S^{\mu\nu}_{1'}(p)_{\rm tad}=-e^2\int\,\frac{d^D  \ell}{(2\pi)^D}
\frac{f_{\star_2}(\ell,p)}{ \ell^2} \bigg((p\theta \ell)\Big((p\theta \ell) g^{\mu\nu}-p^\mu(\theta \ell)^\nu -(\theta \ell)^\mu p^\nu\Big)+p^2(\theta \ell)^\mu(\theta \ell)^\nu\bigg)\bigg |_{D\to4}
\\&=\frac{e^2}{(4\pi)^2}\frac{8}{3(\theta p)^4}\Big[3\Big(g^{\mu\nu}(\theta p)^2-(\theta\theta)^{\mu\nu}p^2+ p^{\{\mu}(\theta\theta p)^{\nu\}}\Big)+\frac{4p^2}{(\theta p)^2}\Big((\theta\theta)^{\mu\nu}(\theta p)^2+(\theta\theta p)^\mu(\theta\theta p)^\nu\Big)\Big],
\end{split}
\label{scalphottad1}
\end{gather}
we obtained the IR result. To evaluate $S^{\mu\nu}_{2'}(p)_{\rm tad}$, $S^{\mu\nu}_{3'}(p)_{\rm tad}$ and $S^{\mu\nu}_{4'}(p)_{\rm tad}$ we first need to establish the following identities:
\begin{gather}
\begin{split}
&f_{\star_{3'}}(\ell,p,-p)=f_{\star_{3'}}(\ell,-p,p)=f_{\star_{3'}}(-\ell,p,-p)=f_{\star_{3'}}(-\ell,-p,p)= 1,
\\&
f_{\star_{3'}}(p,-p,-\ell)=f_{\star_{3'}}(p,-p,\ell)=f_{\star_{3'}}(-p,p,-\ell)=f_{\star_{3'}}(-p,p,\ell)= f^2_{\star_2}(\ell, p),
\\&
f_{\star_{3'}}(-\ell,p,-p)+f_{\star_{3'}}(p,-p,-\ell)-2f^2_{\star_2}(\ell, p) \sim -f^2_{\star_2}(\ell, p),
\\&
f_{\star_{3'}}(p,-p,-\ell)+f_{\star_{3'}}(-p,p,-\ell)-2f^2_{\star_2}(\ell, p)=0,
\\&
f_{\star_{3'}}(\ell,p,-p)+f_{\star_{3'}}(p,-p,\ell)-2f^2_{\star_2}(\ell, p) \sim -f^2_{\star_2}(\ell, p),
\\&
f_{\star_{3'}}(p,-p,\ell)+f_{\star_{3'}}(-p,p,\ell)-2f^2_{\star_2}(\ell, p)=0.
\end{split}
\label{f3'2}
\end{gather}
We then find the following pure IR divergent terms:
\begin{gather}
\begin{split}
&S^{\mu\nu}_{4'}(p)_{\rm tad}=-2S^{\mu\nu}_{2'}(p)_{\rm tad}=-2S^{\mu\nu}_{3'}(p)_{\rm tad}
=-S^{\mu\nu}_{2'}(p)_{\rm tad}-S^{\mu\nu}_{3'}(p)_{\rm tad}
\\&
=e^2\int\,\frac{d^D  \ell}{(2\pi)^D}
\frac{f_{\star_2}(\ell,p)}{ \ell^2} \bigg((p \ell)\Big((\theta p)^\mu(\theta \ell)^\nu +(\theta \ell)^\mu (\theta p)^\nu\Big)
-(p\theta \ell)\Big(\ell^\mu(\theta p)^\nu+(\theta p)^\mu\ell^\nu\Big)\bigg)\bigg |_{D\to4}
\\&
=\frac{e^2}{(4\pi)^2}\frac{32}{3}\frac{(\theta p)^\mu(\theta p)^\nu}{(\theta p)^4},
\end{split}
\label{scalphottad4}
\end{gather}
\begin{equation}
S^{\mu\nu}(p)_{\rm tad}=S^{\mu\nu}_{1'}(p)_{\rm tad}.
\label{tad1=tad}
\end{equation}
Using (\ref{scalphottad1}) and (\ref{tad1=tad}) and by comparing with general tensor structure (\ref{2.1P}) we have found that from scalar-photon tadpole diagram only two terms survive:
\begin{figure}
\begin{center}
\includegraphics[width=6cm]{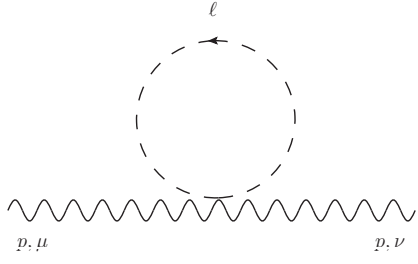}
\end{center}
\caption{Photon-2scalars tadpole: $S^{\mu\nu}(p)_{\rm tad}$.}
\label{fig:scalar-photon-tad}
\end{figure}
\begin{equation}
\begin{split}
S^{\rm tad}_{1}(p)=&S^{\rm tad}_{2}(p)=S_{5}^{\rm tad}(p)=0,\;
\\S^{\rm tad}_{3}(p)&=-4\frac{1}{(\theta p)^2}T_0=\frac{8}{(\theta p)^4},\;
S^{\rm tad}_{4}(p)=-\frac{16}{3}\frac{p^2}{(\theta p)^4}T_0=\frac{32}{3}\frac{p^2}{(\theta p)^6}.
\end{split}
\label{Stadi}
\end{equation}
Finally summing up the IR parts of bubble (\ref{Sbubi}) and tadpole (\ref{Stadi}) contributions we get:
\begin{equation}
\begin{split}
\big[S^{\mu\nu}(p)_{\rm bub}+S^{\mu\nu}(p)_{\rm tad}\big]_{\rm IR}
=&\frac{e^2}{(4\pi)^2} (\theta p)^\mu (\theta p)^\nu \;S_{2}(p)_{\rm bub}|_{\rm IR}
\\=&-\frac{e^2}{(4\pi)^2} \frac{(\theta p)^\mu (\theta p)^\nu}{(\theta p)^2}8T_0=\frac{e^2}{(4\pi)^2} (\theta p)^\mu (\theta p)^\nu\frac{16}{(\theta p)^4} ,
\end{split}
\label{scalphotbub+tad}
\end{equation}
where all IR terms from both diagrams, except the one arising from the bubble, cancels out. Interesting enough is that within this noncommutaive scalar-photon action in the adjont (\ref{AcSreal}) we are facing the exact cancelations of all divergences of the higher order terms of noncommutative tensor-parameter $\theta^{\mu\nu}$, showing thus the consistency of our computations.

\subsection{The photon one-loop contribution to scalar 1-PI two point function}

\begin{figure}
\begin{center}
\includegraphics[width=6cm]{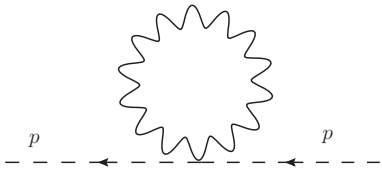}
\end{center}
\caption{The scalar-photon tadpole: $\Sigma_{(\phi)}(p)_{\rm tad}$.}
\label{fig:N=2 scalar-photon-tadpole}
\end{figure}

\begin{figure}
\begin{center}
\includegraphics[width=5.5cm]{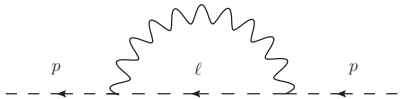}
\end{center}
\caption{The scalar-photon bubble: $\Sigma_{(\phi)}(p)_{\rm bub}$.}
\label{fig:N=2 scalar-photon-bubble}
\end{figure}

The one-loop adjoint scalar 1-PI two point function in the minimal coupled model consists the tadpole diagram Fig.\ref{fig:N=2 scalar-photon-tadpole} and the bubble diagram Fig.\ref{fig:N=2 scalar-photon-bubble}.
The evaluation is straightforward. We obtain in the end
\begin{equation}
\Sigma_{(\phi)}(p)_{\rm tad}=\frac{e^2}{(4\pi)^2}8T_0,
\label{3.14}
\end{equation}
and
\begin{equation}
\begin{split}
&\Sigma_{(\phi)}(p)_{\rm bub}=
\\&-\frac{e^2}{(4\pi)^2}\bigg[\tr\theta\theta\frac{p^4}{(\theta p)^2}\bigg(2(4\pi)^{-\frac{D}{2}}(p^2)^{\frac{D}{2}-2}{\rm\Gamma}\left(2-\frac{D}{2}\right){\rm B}\left(\frac{D}{2}-1, \frac{D}{2}-1\right)\Bigg|_{D\to 4-\epsilon}
\\&
-4I_K^0-4 I_H\bigg)
\\&
+(\theta\theta p)^2\frac{p^4}{(\theta p)^4}\bigg(4(4\pi)^{-\frac{D}{2}}(p^2)^{\frac{D}{2}-2}{\rm\Gamma}\left(2-\frac{D}{2}\right){\rm B}\left(\frac{D}{2}-1, \frac{D}{2}-1\right)\Bigg|_{D\to 4-\epsilon}
\\&
-8I_K^0-4 I_H\bigg)
\\&
+p^2\bigg(\left(2D-4\right)(4\pi)^{-\frac{D}{2}}(p^2)^{\frac{D}{2}-2}{\rm\Gamma}\left(2-\frac{D}{2}\right){\rm B}\left(\frac{D}{2}-1, \frac{D}{2}-1\right)\Bigg|_{D\to 4-\epsilon}
\\&
-8I_K^0-12 I_H\bigg)-4T_0\bigg].
\end{split}
\label{3.15}
\end{equation}
The total quadratic IR divergence reads
\begin{equation}
\big[\Sigma_{(\phi)}(p)_{\rm tad}+\Sigma_{(\phi)}(p)_{\rm bub}\big]_{\rm IR}=\frac{e^2}{(4\pi)^2}12T_0=-\frac{e^2}{(4\pi)^2}\frac{24}{(\theta p)^2}.
\label{3.16}
\end{equation}
We will soon see in the next section that this divergence is exactly canceled by the contributions from scalar-photino and scalar self-interaction diagrams.

\section{Noncommutative $\cal N$=2 SYM  U(1) theory and the $\theta$-exact SW map}

The noncommutative U(1) $\cal N$=2  super Yang-Mills theory has the following action
\begin{equation}
\begin{split}
S_{{\cal N}=2}=&\int -\frac{1}{4}F_{\mu\nu}\star F^{\mu\nu}+({\cal D}_\mu[A]\Phi)^{\dagger} {\cal D}^\mu[A] \Phi-\
\frac{e^2}{2}[\Phi^{\dagger}\stackrel{\star}{,}\Phi]^2
\\
+&i  \bar{\Lambda}\bar{\sigma}^{\mu}{\cal D}_\mu[A]\Lambda +i  \bar{\Psi}\bar{\sigma}^{\mu}{\cal D}_\mu[A]\Psi+ie\sqrt{2}\Psi[\Lambda\stackrel{\star}{,}\Phi^{\dagger}]
+ ie\sqrt{2}\bar{\Psi}[\bar{\Lambda}\stackrel{\star}{,}\Phi],
\end{split}
\label{sn2action}
\end{equation}
in the Wess-Zumino gauge. The noncommutative fields in the previous action constitute the noncommutative U(1) supermultiplet $(A_\mu,\Lambda_{\alpha},\Psi_{\alpha},\Phi)$. $\Lambda_{\alpha}$ and $\Psi_{\alpha}$ are Weyl fermion fields and $\Phi$ is a complex scalar field. Each field $\Lambda_{\alpha}$, $\Psi_{\alpha}$ and $\Phi$ transforms under the adjoint action of the NC U(1), so that the NC covariant derivative is ${\cal D}_{\mu}[A]=\partial_{\mu}\phantom{\Phi}-i[A_\mu\stackrel{\star}{,}\phantom{\Phi}]$.

By replacing the noncommutative fields of the action in (\ref{sn2action}) with the  $\theta$-exact Seiberg-Witten maps --see appendix A-- $A_\mu[a_\rho;\theta]$, $\Lambda_{\alpha}[a_\rho,\lambda_\alpha;\theta]$, $\Psi_{\alpha}[a_\rho,\psi_\alpha;\theta]$, $\Phi[a_\rho,\phi;\theta]$, the action $S_{{\cal N}=2}$ is turned
into the action of a theory which is an interacting deformation of the  free ordinary U(1) supersymmetric theory for the U(1) supermultiplet $(a_\mu,\lambda_{\alpha},\psi_{\alpha},\phi)$. This deformation is supersymmetric although  supersymmetry --${\cal N}=2$-- is nonlinearly realized on the ordinary multiplet $(a_\mu,\lambda_{\alpha},\psi_{\alpha},\phi)$; a feature we have already seen in the ${\cal N}=1$ SYM case.

The contributions to the action in (\ref{sn2action}) that are needed to compute one-loop 1PI two-point function of each field in $(a_\mu,\lambda_{\alpha},\psi_{\alpha},\phi)$ can be readily obtained by using (\ref{2.13}), ({\ref{2.14}), (\ref{AcSreal}) and
\begin{equation}
\int\; ie\sqrt{2}\psi[\lambda\stackrel{\star}{,}\phi^{\dagger}]+ ie\sqrt{2}\bar{\psi}[\bar{\lambda}\stackrel{\star}{,}\phi]-\frac{e^2}{2}([\Phi^{\dagger}\stackrel{\star}{,}\Phi])^2.
\label{N=2Act}
\end{equation}
The terms in (\ref{N=2Act}) yields the scalar-fermion $\cal N$=2 Feynman rules (\ref{E.1}) given in appendix E.
Now we are ready to display the value of each  one-loop Feynman diagram contributing to the two-point functions of the ordinary fields of the theory.

\subsection{The one-loop 1PI two-point function for photon field $a_\mu$}

For ${\cal N} =2$ theory one has $n_f=n_s=2$. Substituting these numbers as well as the scalar bubble and tadpole results to \eqref{4.1Pi}, and then restricting to quadratic IR divergence only, gives
\begin{equation}
\Pi^{\mu\nu}(p)|_{\rm IR}=\frac{e^2}{(4\pi)^2}\frac{(\theta p)^\mu(\theta p)^\nu}{(\theta p)^2}(-16T_0+16 n_f T_0-8n_s T_0)|_{n_f=n_s=2}=0,
\label{IRcancel}
\end{equation}
i.e. clean quadratic IR divergence cancellation. Remaining UV divergences can be expressed using the five-term notation in \eqref{2.1P} as follows
\begin{gather}
\Pi_1(p)|_{\rm UV}=\bigg(\frac{4}{3}-\frac{4}{3}{n_f}-\frac{1}{3}{n_s}+\frac{p^2}{(\theta p)^4}(3\tr\theta\theta(\theta p)^2+4(\theta\theta p)^2)\bigg)\left(\frac{2}{\epsilon}+\ln(\mu^2(\theta p)^2)\right)\bigg|_{n_f=n_s=2},
\label{UVterm1S}\\
\Pi_2(p)|_{\rm UV}=\frac{p^2}{(\theta p)^2}\bigg(2-\frac{p^2(\tr\theta\theta)}{(\theta p)^2}
\bigg)\left(\frac{2}{\epsilon}+\ln(\mu^2(\theta p)^2)\right),
\label{UVterm2S}\\
\Pi_3(p)|_{\rm UV}=\frac{p^2}{(\theta p)^2}\left(\frac{2}{\epsilon}+\ln(\mu^2(\theta p)^2)\right),
\label{UVterm3S}\\
\Pi_4(p)|_{\rm UV}=-4\frac{p^4}{(\theta p)^4}\left(\frac{2}{\epsilon}+\ln(\mu^2(\theta p)^2)\right),
\label{UVterm4S}\\
\Pi_5(p)|_{\rm UV}=4\frac{p^4}{(\theta p)^4}
\left(\frac{2}{\epsilon} + \ln(\mu^2(\theta p)^2)\right).
\label{UVterm5S}
\end{gather}

\subsection{The one-loop 1PI two-point function for the scalar $\phi$}

The one-loop 1PI two-point function, $\Sigma_{(\phi)}(p)$, of field $\phi$ is the sum of the four diagrams (Fig.\ref{fig:N=2 scalar-photon-tadpole}, Fig.\ref{fig:N=2 scalar-photon-bubble}, Fig.\ref{fig:N=2 scalar-scalar-tadpole} and Fig.\ref{fig:N=2 scalar-fermion bubble}). The first two are already given as equations \eqref{3.14} and \eqref{3.15} in the last section. The values of the third and fourth diagrams read
\begin{figure}
\begin{center}
\includegraphics[width=5cm]{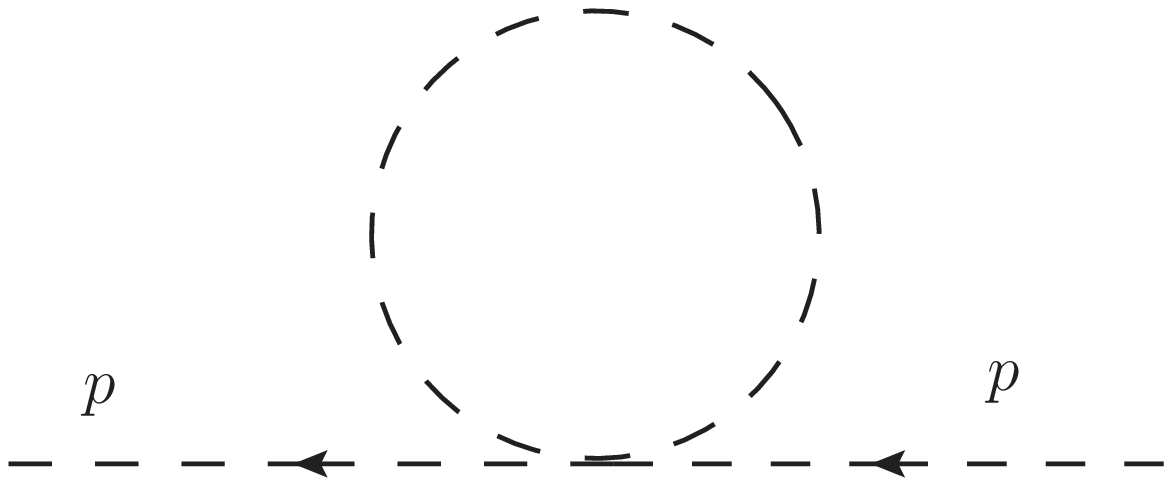}
\end{center}
\caption{$\cal N$=2 four-scalar tadpole: $\Sigma_{(\phi)}(p)_{\rm 4sc-tad}$.}
\label{fig:N=2 scalar-scalar-tadpole}
\end{figure}
\begin{equation}
\Sigma_{(\phi)}(p)_{\rm 4sc-tad}= -\frac{e^2}{(4\pi)^2} 4T_0,
\label{you23-9}
\end{equation}
\begin{equation}
\begin{split}
\Sigma&_{(\phi)}(p)_{\rm f-bub}=8e^2 \mu^{4-D}\int\; \frac{d^D\ell}{(2\pi)^D} \frac{p\ell-\ell^2}{\ell^2(\ell-p)^2}\sin^2\frac{\ell\theta p}{2}
\\
=&
-4e^2  \mu^{4-D}\int\; \frac{d^D\ell}{(2\pi)^D} \frac{1}{\ell^2}\sin^2\frac{\ell\theta p}{2}\left(2- \frac{p^2}{(\ell-p)^2}\right)
\\
=&\frac{e^2}{(4\pi)^2}\bigg[-8T_0+2p^2\bigg((4\pi\mu^2)^{2-\frac{D}{2}}(p^2)^{\frac{D}{2}-2}{\rm\Gamma}\left(2-\frac{D}{2}\right){\rm B}\left(\frac{D}{2}-1, \frac{D}{2}-1\right)\Bigg|_{D\to 4-\epsilon}
\\&-2I_K^0\bigg)\bigg].
\end{split}
\label{you23-10}
\end{equation}
\begin{figure}
\begin{center}
\includegraphics[width=5cm]{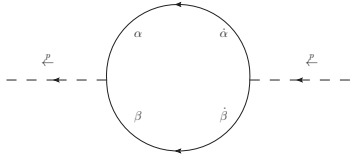}
\end{center}
\caption{$\cal N$=2 scalar-photino bubble: $\Sigma_{(\phi)}(p)_{\rm f-bub}$.}
\label{fig:N=2 scalar-fermion bubble}
\end{figure}
Hence, one gets the following full scalar two-point function:
\begin{equation}
\Sigma_{(\phi)}(p)=\Sigma_{(\phi)}(p)_{\rm tad}+\Sigma_{(\phi)}(p)_{\rm bub}+\Sigma_{(\phi)}(p)_{\rm 4sc-tad}+\Sigma_{(\phi)}(p)_{\rm f-bub},
\label{FullSc2point}
\end{equation}
which is again quadratic IR divergence free, as
\begin{equation}
\begin{split}
\Sigma_{(\phi)}(p)|_{\rm IR}=&(\Sigma_{(\phi)}(p)_{\rm tad}+\Sigma_{(\phi)}(p)_{\rm bub}+\Sigma_{(\phi)}(p)_{\rm 4sc-tad}+\Sigma_{(\phi)}(p)_{\rm f-bub})|_{\rm IR}
\\=&\frac{e^2}{(4\pi)^2}(12T_0-8T_0-4T_0)=0.
\end{split}
\label{FullSc2pointIR}
\end{equation}
The UV divergence reads
\begin{equation}
\begin{split}
\Sigma_{(\phi)}(p)|_{\rm UV}=&\left(\Sigma_{(\phi)}(p)_{\rm bub}+\Sigma_{(\phi)}(p)_{\rm f-bub}\right)|_{\rm UV}
\\=&-2\frac{e^2}{(4\pi)^2} p^2\left(1+\tr\theta\theta\frac{p^2}{(\theta p)^2}+2(\theta\theta p)^2\frac{p^2}{(\theta p)^4}\right)\left(\frac{2}{\epsilon}+\ln(\mu^2(\theta p)^2)\right).
\end{split}
\label{4.14}
\end{equation}

\subsection{The one-loop 1PI two-point function for photinos $\lambda_{\alpha}$ and $\psi_{\alpha}$}

In the $\cal N$=2 theory there is a photino-scalar loop (Fig. \ref{fig:N=2Fermion-scalar-bubble}) alongside the photino-photon loop contribution which is identical to the ${\cal N}=1$ theory value \eqref{N212} for each of the two photinos.
\begin{figure}
\begin{center}
\includegraphics[width=5.5cm]{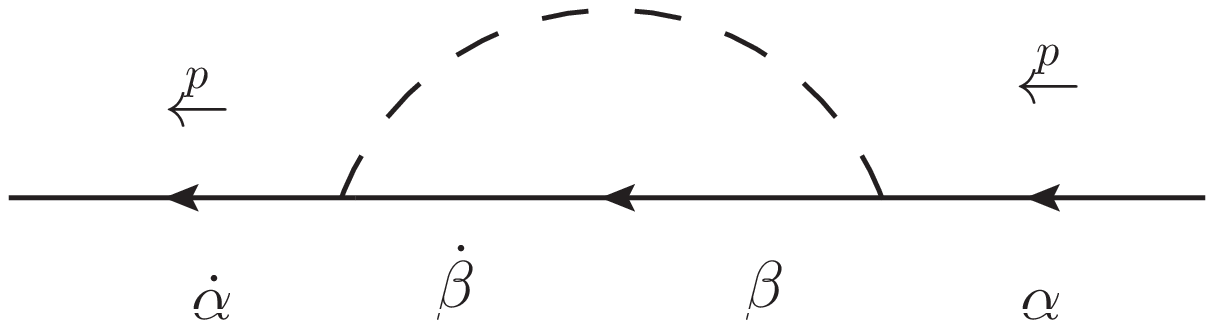}
\end{center}
\caption{$\cal N$=2 photino-scalar bubble: $\Sigma^{\dot{\alpha}\alpha}(p)_{\rm scal}$.}
\label{fig:N=2Fermion-scalar-bubble}
\end{figure}

The photino-scalar loop-integral gives the following contribution
\begin{equation}
\begin{split}
\Sigma^{\dot{\alpha}\alpha}(p)_{\rm scal}=&8e^2\mu^{4-D}\int\frac{d^D\ell}{(4\pi)^D}\sin^2\frac{\ell\theta p}{2}\frac{\ell^\mu-p^\mu}{\ell^2(\ell-p)^2}\bar\sigma_\mu^{\dot\alpha\alpha}
\\=&-\frac{e^2}{(4\pi)^2} p^\mu\bar\sigma_\mu^{\dot\alpha\alpha}2\bigg((4\pi\mu^2)^{2-\frac{D}{2}}(p^2)^{\frac{D}{2}-2}{\rm\Gamma}\left(2-\frac{D}{2}\right){\rm B}\left(\frac{D}{2}-1, \frac{D}{2}-1\right)\Bigg|_{D\to 4-\epsilon}
\\&-2I_K^0\bigg).
\end{split}
\label{Sigmascal}
\end{equation}
Total photino two-point function is finally given as a sum:
\begin{equation}
\Sigma_{(\lambda,\psi)}^{\dot{\alpha}\alpha}(p)=\Sigma^{\dot{\alpha}\alpha}(p)_{\rm tad}+ \Sigma^{\dot{\alpha}\alpha}(p)_{\rm bub}+\Sigma^{\dot{\alpha}\alpha}(p)_{\rm scal}.
\label{SigmaN2tot}
\end{equation}
It is quadratic IR divergence free and it has the following UV divergence
\begin{equation}
\Sigma_{(\lambda,\psi)}^{\dot{\alpha}\alpha}(p)|_{\rm UV}=-\frac{e^2}{(4\pi)^2}\sigma^{\dot{\alpha}\alpha}_\mu p^\mu\left(2+\tr\theta\theta\frac{p^2}{(\theta p)^2}+2(\theta\theta p)^2\frac{p^2}{(\theta p)^4}\right)\left(\frac{2}{\epsilon} + \ln(\mu^2(\theta p)^2)\right).
\label{4.17}
\end{equation}

\section{Noncommutative $\cal N$=4 SYM  U(1) theory and the $\theta$-exact SW map}

 Let $(A_\mu,\Lambda_{\alpha\,i},\Phi_m)$, $i=1,...,4$, $m=1,...,6$ define the noncommutative U(1) ${\cal N}$=4 supermultiplet; then, the action of the noncommutative U(1) $\cal N$=4  super Yang-Mills theory reads
\begin{equation}
\begin{split}
S_{{\cal N}=4}=&\int -\frac{1}{4}F_{\mu\nu}\star F^{\mu\nu}+i  \bar{\Lambda}^{i}\bar{\sigma}^{\mu}{\cal D}_\mu[A]\Lambda_{i}+
\frac{1}{2}{\cal D}_\mu[A] \Phi_m{\cal D}^\mu[A]\Phi_m+\Big(\frac{e}{2}[\Phi_m\stackrel{\star}{,}\Phi_n]\Big)^2
\\
+&i\frac{e}{2}(\tilde{\sigma}^{-1})^{ij}\Lambda_i[\Lambda_j\stackrel{\star}{,}\Phi_m]
-i\frac{e}{2}(\tilde{\sigma})_{ij}\bar{\Lambda}^i[\bar{\Lambda}^j\stackrel{\star}{,}\Phi_m].
\end{split}
\label{sn4action}
\end{equation}
The matrices $4\times 4$, $(\tilde{\sigma})_{ij}$ and $(\tilde{\sigma}^{-1})^{ij}$ give rise to the IRREP of the  Dirac matrices in 8 Euclidean dimensions; further details can be found in \cite{Sohnius:1985qm}. Let us recall that $A_\mu$ is the noncomutative gauge field, that $\Lambda_{\alpha\,i}$ is a noncommutative Weyl field and that $\Phi_m$ is a noncommutative real scalar  field. The noncommutative U(1) acts by the adjoint action on $\Lambda_{\alpha\,i}$ and  $\Phi_m$, and hence ${\cal D}_{\mu}[A]=\partial_{\mu}\phantom{\Phi}-i[A_\mu\stackrel{\star}{,}\phantom{\Phi}]$.

 By replacing, in $S_{{\cal N}=4}$ above, the fields $A_\mu$, $\Lambda_{\alpha\,i}$ and $\Phi_m$ with the corresponding $\theta$-exact Seiberg-Witten maps --namely, $A_\mu[a_\rho;\theta]$, $\Lambda_{\alpha\,i}[a_\rho,\lambda_\alpha;\theta]$ and $\Phi_m[a_\rho,\phi;\theta]$, respectively, we obtain an action which defines an interacting  deformation of the ordinary ${\cal N}=4$ SYM theory in the Wess-Zumino gauge. This deformed action is expressed in terms of the fields of the ordinary ${\cal N}=4$ Yang-Mills supermultiplet $(a_\mu,\lambda_{\alpha\,i},\phi_m)$, $i=1,...,4$, $m=1,...,6$ and it is invariant (on-shell) under the deformed supersymmetric transformations of the ordinary supermultiplet $(a_\mu,\lambda_{\alpha\,i},\phi_m)$ which give rise to the ${\cal N}=4$ supersymmetric transformations of the fields in $(A_\mu,\Lambda_{\alpha\,i},\Phi_m)$. As in the ${\cal N}=1$ and ${\cal N}=2$ cases, the supersymmetry transformations of the ordinary fields that leave $S_{{\cal N}=4}$ in (\ref{sn4action}) invariant gives rise to on-shell nonlinear realization of ${\cal N}=4$ supersymmetry algebra.

The contributions to the action in (\ref{sn4action}) that are needed to compute one-loop 1PI two-point function of each field in $(a_\mu,\lambda_{\alpha},\psi_{\alpha},\phi)$ can easily be obtained by using (\ref{2.16}), ({\ref{2.17}), (\ref{AcSreal}) and
\begin{equation}
\int\;\Big(\frac{e}{2}[\phi_m\stackrel{\star}{,}\phi_n]\Big)^2
+\frac{ie}{2}(\tilde{\sigma}^{-1})^{ij}\lambda_i[\lambda_j\stackrel{\star}{,}\phi_m]
-\frac{ie}{2}(\tilde{\sigma})_{ij}\bar{\lambda}^i[\bar{\lambda}^j\stackrel{\star}{,}\phi_m].
\label{newfeyn4}
\end{equation}
The terms in (\ref{newfeyn4}) yields the Feynman rules given in appendix D.

Below we shall  display the value of each  one-loop Feynman diagram contributing to the two-point functions of the ordinary fields of the theory.

\subsection{The one-loop 1PI two-point function for massless vector field $a_\mu$}

The ${\cal N} =4$ $a_\mu$ 1PI two-point function follows the general formula \eqref{4.1Pi}, only with $n_f=4$ and $n_s=6$. One can immediately recognize clean cancelation of the quadratic IR divergences after substituting ${\cal N} =4\to n_f=4, n_s=6$ into \eqref{IRcancel}.

\subsection{The one-loop 1PI two-point function for the scalar $\phi_m$}

The one-loop 1PI two-point function, $\Sigma_{mn}(p)$, of the field $\phi_m$ is the sum of five diagrams 
Fig. \ref{fig:N= 4scalar-photon-tadpole}, Fig. \ref{fig:N=4scalar-photon-bubble}, Fig. \ref{fig:N=4scalar-scalar-tadpole} and 
Fig. \ref{fig:N=4scalar-photino-bubbles},  whose values read
\begin{figure}
\begin{center}
\includegraphics[width=5.5cm]{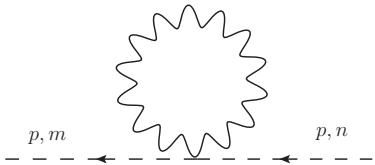}
\end{center}
\caption{$\cal N$=4 scalar-photon tadpole: $\Sigma_{mn}(p)_{\rm tad}$.}
\label{fig:N= 4scalar-photon-tadpole}
\end{figure}

\begin{figure}
\begin{center}
\includegraphics[width=5.5cm]{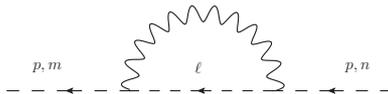}
\end{center}
\caption{$\cal N$=4 scalar-photon bubble: $\Sigma_{mn}(p)_{\rm bub}$.}
\label{fig:N=4scalar-photon-bubble}
\end{figure}

\begin{equation}
\Sigma_{mn}(p)_{\rm 4sc-tad} =\frac{e^2}{(4\pi)^2} 20 \delta_{mn}T_0,
\label{S4sc-tad}
\end{equation}

\begin{figure}
\begin{center}
\includegraphics[width=5cm]{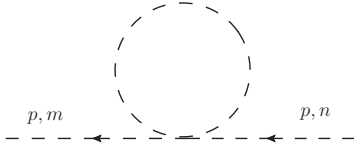}
\end{center}
\caption{$\cal N$=4 four-scalar tadpole: $\Sigma_{mn}(p)_{\rm 4sc-tad}$.}
\label{fig:N=4scalar-scalar-tadpole}
\end{figure}

\begin{equation}
\begin{split}
\Sigma&_{mn}(p)_{\rm f-bub}=32e^2\mu^{4-D}\delta_{mn} \int\; \frac{d^D\ell}{(2\pi)^D} \frac{p\ell-\ell^2}{\ell^2(\ell-p)^2}\sin^2\frac{\ell\theta p}{2}
\\=&\frac{8e^2}{(4\pi\mu^2)^2}\delta_{mn} \bigg[-4T_0+p^2\left((4\pi)^{2-\frac{D}{2}}(p^2)^{\frac{D}{2}-2}{\rm\Gamma}\bigg(2-\frac{D}{2}\right){\rm B}\left(\frac{D}{2}-1, \frac{D}{2}-1\right)\Bigg|_{D\to 4-\epsilon}
\\&-2I_K^0\bigg)\bigg],
\end{split}
\label{f-bN12mn}
\end{equation}
\begin{figure}
\begin{center}
\includegraphics[width=5cm]{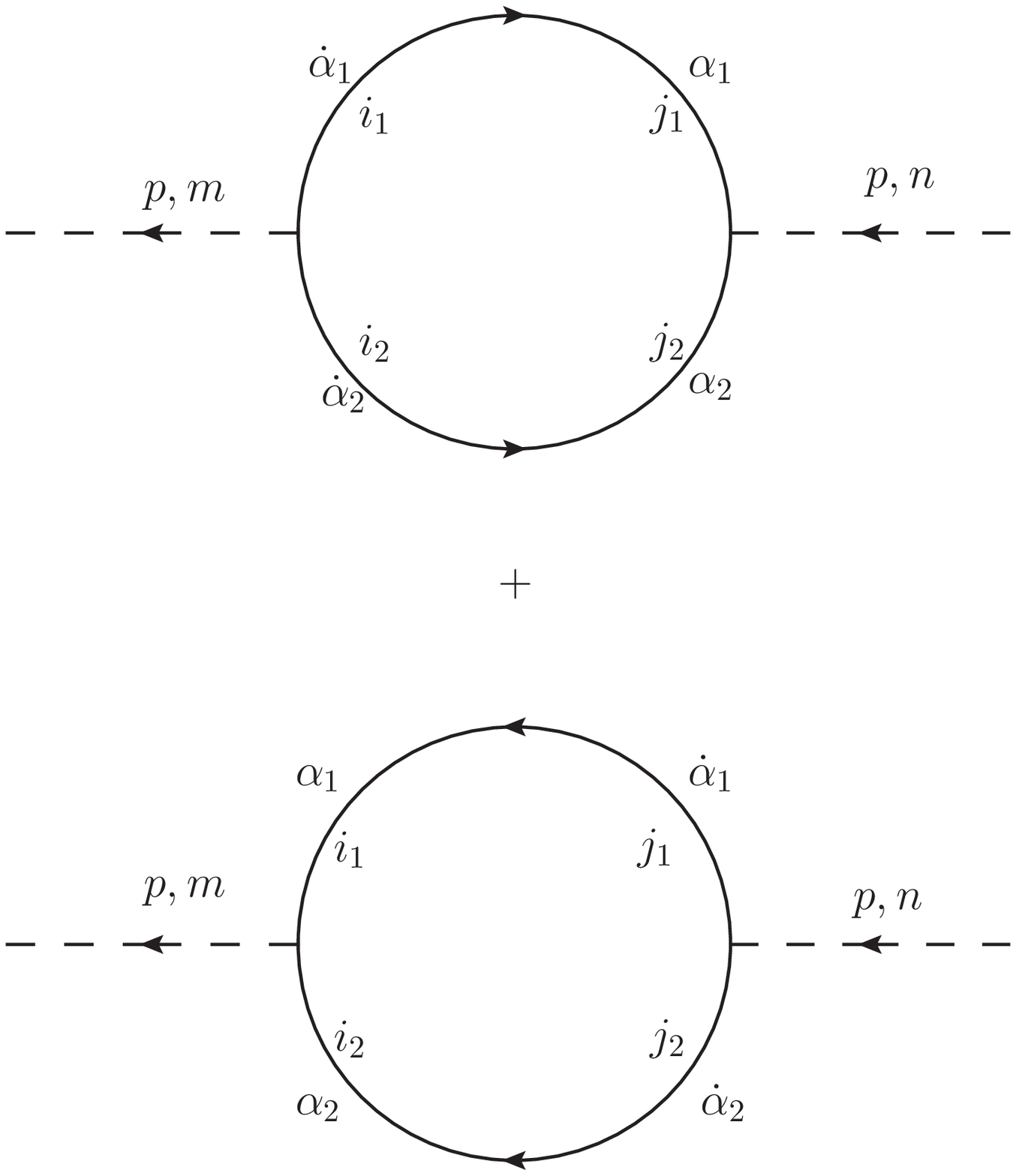}
\end{center}
\caption{$\cal N$=4 scalar-photino bubbles: $\Sigma_{mn}(p)_{\rm f-bub}$.}
\label{fig:N=4scalar-photino-bubbles}
\end{figure}
hence
\begin{equation}
\Sigma_{mn}(p)=\Sigma_{mn}(p)_{\rm tad}+\Sigma_{mn}(p)_{\rm bub}+\Sigma_{mn}(p)_{\rm 4sc-tad}+\Sigma_{mn}(p)_{\rm f-bub},
\label{N12mn}
\end{equation}
is again IR divergence free, only this time we have
\begin{equation}
\begin{split}
\Sigma_{mn}(p)|_{\rm IR}=&\left(\Sigma_{mn}(p)_{\rm tad}+\Sigma_{mn}(p)_{\rm bub}+\Sigma_{mn}(p)_{\rm 4sc-tad}+\Sigma_{mn}(p)_{\rm f-bub}\right)|_{\rm IR}
\\=&\frac{e^2}{(4\pi)^2}\delta_{mn}(12T_0-32T_0+20T_0)=0.
\end{split}
\label{5.6}
\end{equation}
The UV part reads
\begin{equation}
\Sigma_{mn}(p)|_{\rm UV}=\frac{e^2}{(4\pi)^2}\delta_{mn} p^2\left(4-2\tr\theta\theta\frac{p^2}{(\theta p)^2}-4(\theta\theta p)^2\frac{p^2}{(\theta p)^4}\right)\left(\frac{2}{\epsilon}+\ln(\mu^2(\theta p)^2)\right).
\label{5.7}
\end{equation}

\subsection{The one-loop 1PI two-point function for $\lambda_{\alpha\,i}$}

The one-loop 1PI two-point function, $\Sigma^{\dot{\alpha}\alpha\,i}_{\phantom{\dot{\alpha}\alpha}\,j}(p)$, of the field $\lambda_{\alpha\,i}$ is the sum of the three diagrams Fig. \ref{fig:N=4fermion-photon-tadpole}, Fig. \ref{fig:N=4fermion-photon-bubble} and Fig. \ref{fig:N=4fermion-scalar-bubble} whose values read
\begin{eqnarray}
\Sigma^{\dot{\alpha}\alpha\,i}_{\:\:\:\:\:\,j}(p)_{\rm tad}=0,
\label{Sigmaijtad0}
\end{eqnarray}
\begin{figure}
\begin{center}
\includegraphics[width=5cm]{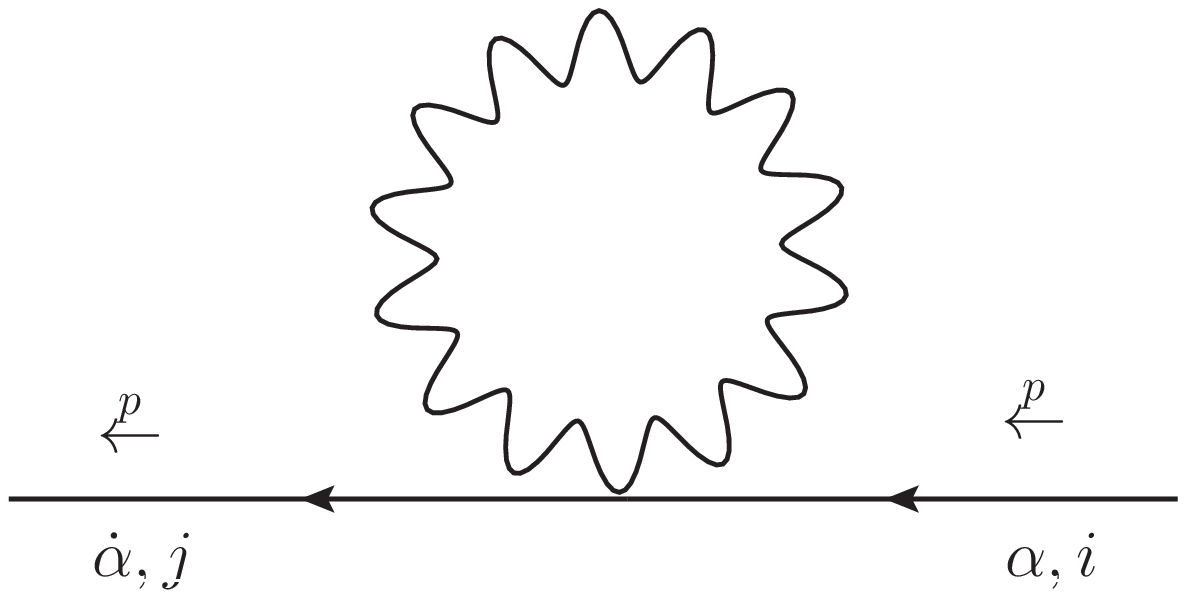}
\end{center}
\caption{$\cal N$=4 photino-photon tadpole: $ \Sigma^{\dot{\alpha}\alpha\,i}_{\:\:\:\:\:\,j}(p)_{\rm tad}$.}
\label{fig:N=4fermion-photon-tadpole}
\end{figure}
\begin{eqnarray}
\Sigma^{\dot{\alpha}\alpha\,i}_{\:\:\:\:\:\,j}(p)_{\rm bub}&=&
-\frac{e^2}{(4\pi)^2}\bar{\sigma}_{\mu}^{\dot{\alpha}\alpha}\,\delta^i_j\bigg[ p^\mu\; N_1(p)+(\theta\theta p)^\mu\; N_2(p)\bigg],
\label{Sigmaijbub}
\end{eqnarray}
with $N_{1,2}(p)$ being given in (\ref{N1}) and (\ref{N2}), respectively, and
\begin{figure}
\begin{center}
\includegraphics[width=5cm]{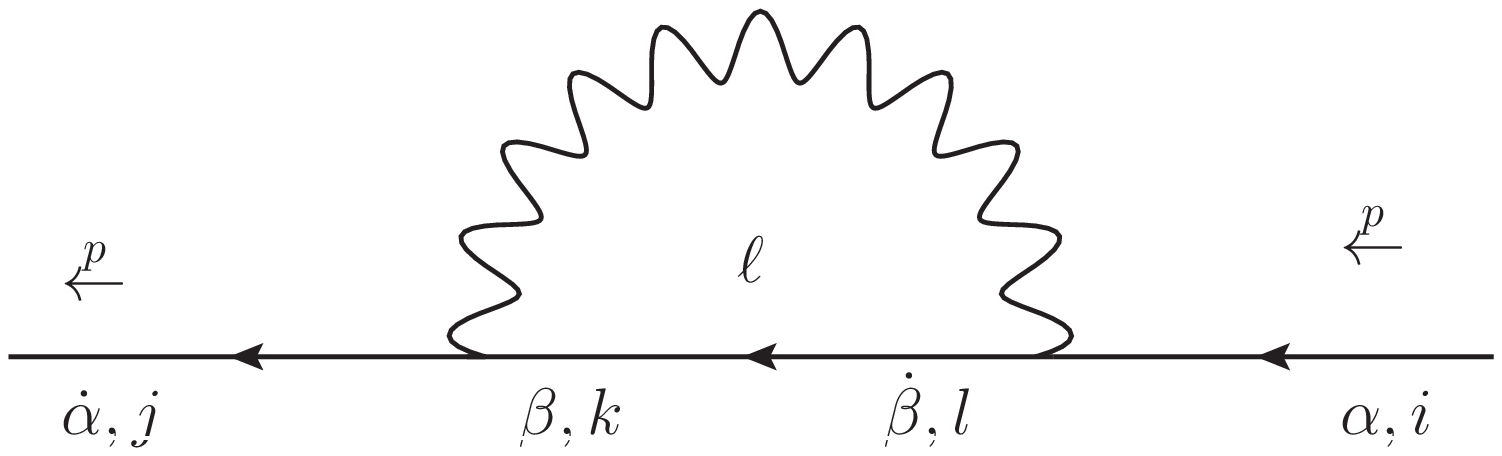}
\end{center}
\caption{$\cal N$=4 photino-photon bubble: $ \Sigma^{\dot{\alpha}\alpha\,i}_{\:\:\:\:\:\,j}(p)_{\rm bub}$.}
\label{fig:N=4fermion-photon-bubble}
\end{figure}
\begin{figure}
\begin{center}
\includegraphics[width=5cm]{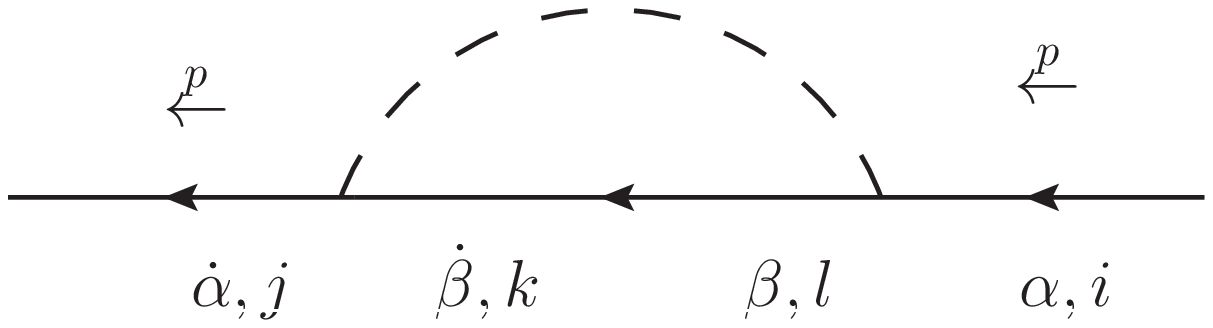}
\end{center}
\caption{$\cal N$=4 photino-scalar bubble: $ \Sigma^{\dot{\alpha}\alpha\,i}_{\:\:\:\:\:\,j}(p)_{\rm scal}$.}
\label{fig:N=4fermion-scalar-bubble}
\end{figure}
\begin{equation}
\begin{split}
\Sigma^{\dot{\alpha}\alpha\,i}_{\:\:\:\:\:\,j}(p)_{\rm scal}=&24e^2\mu^{4-D}\delta_j^i\int\frac{d^D\ell}{(4\pi)^D}\sin^2\frac{\ell\theta p}{2}\frac{\ell^\mu-p^\mu}{\ell^2(\ell-p)^2}\bar\sigma_\mu^{\dot\alpha\alpha}
\\=&-\frac{6e^2}{(4\pi)^2}\delta_j^ip^\mu\sigma_\mu^{\dot\alpha\alpha} \bigg((4\pi\mu^2)^{2-\frac{D}{2}}(p^2)^{\frac{D}{2}-2}{\rm\Gamma}\left(2-\frac{D}{2}\right){\rm B}\left(\frac{D}{2}-1, \frac{D}{2}-1\right)\Bigg|_{D\to 4-\epsilon}
\\&-2I_K^0\bigg).
\end{split}
\label{5.10}
\end{equation}
Hence,
\begin{equation}
\Sigma^{\dot{\alpha}\alpha\,i}_{\phantom{\dot{\alpha}\alpha}\,j}(p)= \Sigma^{\dot{\alpha}\alpha\,i}_{\:\:\:\:\:\,j}(p)_{\rm tad}+  \Sigma^{\dot{\alpha}\alpha\,i}_{\:\:\:\:\:\,j}(p)_{\rm bub}+ \Sigma^{\dot{\alpha}\alpha\,i}_{\:\:\:\:\:\,j}(p)_{\rm scal},
\label{Sigmaijsum}
\end{equation}
is quadratic IR divergence free, and the total UV divergences is presented below
\begin{equation}
\Sigma^{\dot{\alpha}\alpha\,i}_{\phantom{\dot{\alpha}\alpha}\,j}(p)|_{\rm UV}=-\frac{e^2}{(4\pi)^2}\delta_j^i\sigma^{\dot{\alpha}\alpha}_\mu p^\mu\left(6+\tr\theta\theta\frac{p^2}{(\theta p)^2}+2(\theta\theta p)^2\frac{p^2}{(\theta p)^4}\right)\left(\frac{2}{\epsilon} + \ln(\mu^2(\theta p)^2)\right).
\label{5.12}
\end{equation}


\section{Effect of gauge fixing on photon two point function}

In the prior sections we have shown that the quadratic IR divergent contribution to the photon two point function can be canceled by introducing supersymmetry. Yet we have still two unanswered question: First, all of our computations above are carried out in the commutative Feynman gauge, which, albeit convenient, is just one specific choice. We do not know whether the cancelation we found would be changed by a change of gauge fixing. Second, we have quite complicated UV divergence in the Feynman gauge in general, which may be modified by changing gauge fixing, as in the commutative gauge theories. To study these two issues we introduce in this section a new, non-local and nonlinear gauge fixing based on the Seiberg-Witten map then evaluate its effect to the photon two point function.

\subsection{The noncommutative Feynman gauge fixing action}

We introduce a new gauge fixing for non-local U(1) gauge theory via
the $\theta$-exact Seiberg-Witten map. In terms of BRST language, this gauge fixing contains BRST-auxiliary field $B$, and it is given by
\begin{gather}
\begin{split}
S&=S_{\rm U(1)}+S_{\rm gf}=S_{\rm U(1)}
+s\int \bar\omega\left(\partial_{\mu}A^{\mu}(a_\mu,\theta^{ij})+\frac{B}{2}\right),
\\
S_{\rm gf}&=\int B\left(\partial_{\mu}A^{\mu}+\frac{B}{2}\right)-
\bar\omega s(\partial_{\mu}A^{\mu})
=\int\frac{1}{2}(B+\partial_{\mu}A^{\mu})^2
-\frac{1}{2}(\partial_{\mu}A^{\mu})^2-\bar\omega \partial_{\mu}(sA^{\mu}),
\end{split}
\label{6.1}
\end{gather}
with $s$ being regular U(1) BRST transformation $s a_\mu=\partial_\mu\omega$, where $\omega$ is the U(1) ghost. Next we use consistency condition for SW map to get $sA_\mu(a_\mu,\theta^{ij})=D_\mu\Omega$, where $\Omega$ is $\rm U_{\star}(1)$ ghost and $D_\mu$ the $\rm U_{\star}(1)$ covariant derivative in the adjoint representation $D_\mu=\partial_\mu+i[A_\mu \stackrel{\star}{,} \phantom{\Phi}] $.

Since the SW map for $\Omega$ is actually the same as for the NC gauge parameter $\Lambda$ from \cite{Trampetic:2015zma}
we can derive the photon-photon and photon-ghost coupling in this gauge.
So by using the BRST transformations
\begin{equation}
\begin{split}
s\bar\omega&=B,\; \, sB=0,
\\ sA&^{\mu}(a_\mu,\theta^{ij})=s_{\rm NC}A^{\mu}= D^\mu \Omega,
\label{6.2}
\end{split}
\end{equation}
the  following gauge fixing action  is produced from (\ref{6.1})
\begin{equation}
S_{\rm gf}=
\int\frac{1}{2}(B+\partial_{\mu}A^{\mu})^2
-\frac{1}{2}(\partial_{\mu}A^{\mu})^2-\bar\omega \partial_{\mu}D^\mu \Omega,
\label{6.3}
\end{equation}
which after the application of the SW map resulting Feynman rules for the gauge fixing and ghost induced diagrams given in the appendix F. We name this gauge as ``the noncommutative Feynman gauge'' as it is formally identical to the Feynman gauge in the $\rm U_\star(1)$ gauge theory.

\subsection{One-loop contributions from the new NC gauge fixing action}

The new gauge fixing action \eqref{6.3} introduces additional terms to the three and four photon self-couplings as well as photon-ghost couplings, as summarized in \eqref{F.1}. Unlike the three and four photon couplings in the commutative Feynman gauge~\cite{Horvat:2013rga,Horvat:2015aca}, these new interaction terms are no longer transverse. It then becomes intriguing how the sum of the resulting loop integrals behave.

From Feynman rules \eqref{F.1} we find the following diagrams Fig.\ref{Gf3pb1}-\ref{GfGPt} contributions to the one loop photon two point function.
\begin{figure}
\begin{center}
\includegraphics[width=5cm]{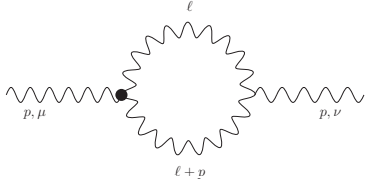}
\end{center}
\caption{Gauge fixing L:  3-photon bubble.}
\label{Gf3pb1}
\end{figure}
\begin{figure}
\begin{center}
\includegraphics[width=5cm]{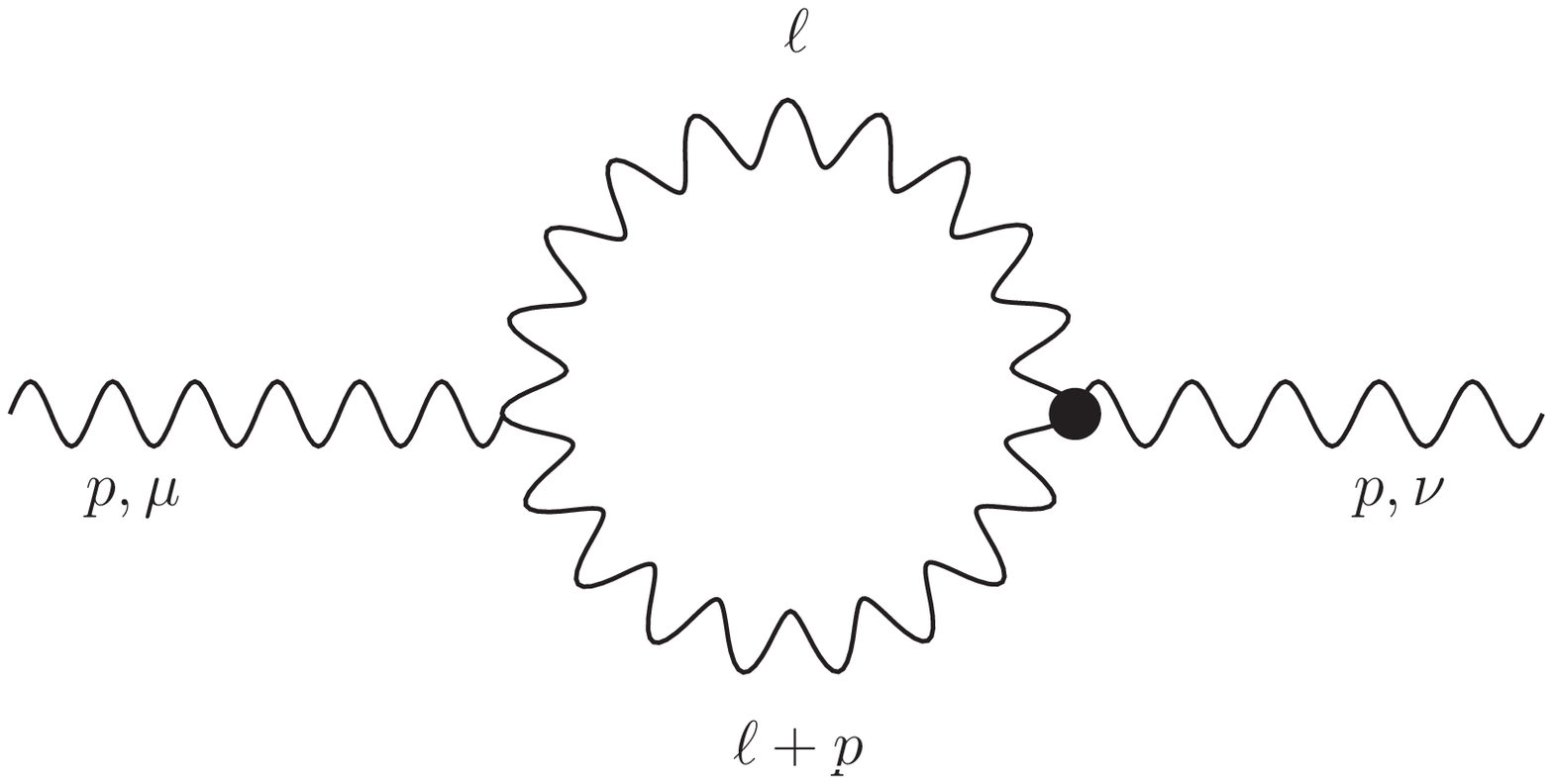}
\end{center}
\caption{Gauge fixing R:  3-photon bubble.}
\label{Gf3pb2}
\end{figure}
\begin{figure}
\begin{center}
\includegraphics[width=5cm]{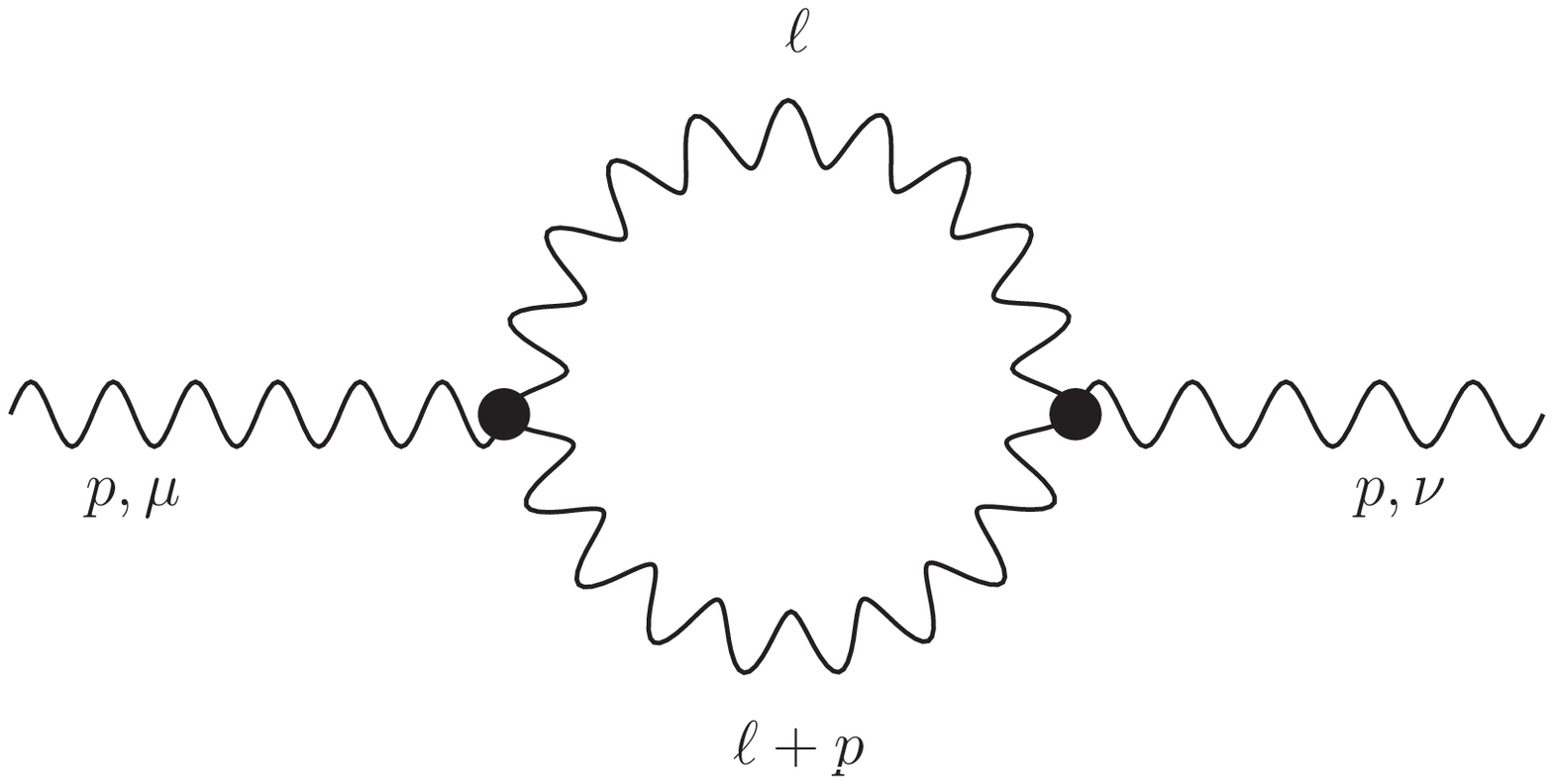}
\end{center}
\caption{Gauge fixing L\&R:  3-photon bubble.}
\label{Gf3pb3}
\end{figure}
\begin{figure}
\begin{center}
\includegraphics[width=4cm]{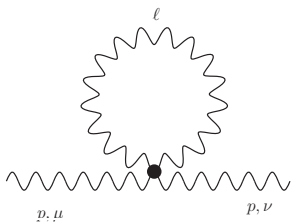}
\end{center}
\caption{Gauge fixing T:  4-photon tadpole.}
\label{Gf4pt}
\end{figure}
\begin{figure}
\begin{center}
\includegraphics[width=5cm]{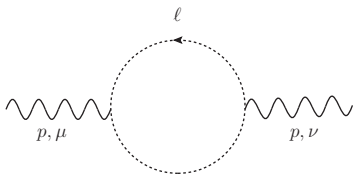}
\end{center}
\caption{Photon-ghost bubble.}
\label{GfGPb}
\end{figure}
\begin{figure}
\begin{center}
\includegraphics[width=5cm]{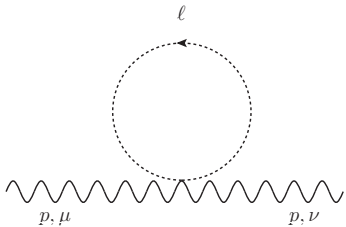}
\end{center}
\caption{2Photons-ghost tadpole.}
\label{GfGPt}
\end{figure}
Denoting the total sum of Fig's. \ref{Gf3pb1} to \ref{GfGPt} as $\Pi^{\mu\nu}_{\rm gf_{\rm total}}$, it turns out to be convenient to split it into two partial sums,
\begin{equation}
\Pi^{\mu\nu}_{\rm gf_{\rm total}}=\Pi^{\mu\nu}_{\rm gf_{\rm mix}}+\Pi^{\mu\nu}_{\rm gf}.
\label{6.4}
\end{equation}
Here $\Pi^{\mu\nu}_{\rm gf_{\rm mix}}$ presents the sum over Figures \ref{Gf3pb1} and \ref{Gf3pb2}, which contain one 3-photon vertex from the classical action and the other from gauge fixing action, while $\Pi^{\mu\nu}_{\rm gf}$ sums over the rest of them which are solely from the gauge fixing.\footnote{One more reason is that gauge fixing \eqref{6.1} and/or \eqref{6.3} can be added to {\em any} U(1) gauge invariant action, particularly the free U(1) action $S_{\rm free}=-\frac{1}{4}\int f_{\mu\nu}f^{\mu\nu}$. In this case $\Pi^{\mu\nu}_{\rm gf}$ would present the whole contribution to the one loop 1-PI photon two point function. Thus it is convenient to isolate it out.}

Evaluation of diagrams in Fig.\ref{Gf3pb1}-\ref{GfGPt} follows substantially the standard procedure used in the prior section, except the rising of the two new types of tadpole integrals. The first one takes the form of the second term in \eqref{vanishingtadpole} so it can be removed by our regularization prescription. The second one is a tadpole integral without any loop momenta in the numerator i.e. $\int\frac{d^D k}{(2\pi)^D}\,k^{-2}f_{\star_2}(k,p)^2$. This integral contains total effective loop momenta power $\ell^{-4}$ because of the additional denominator $(k\theta p)^{-2}$ from the nonlocal factor $f_{\star_2}(k,p)^2$, which is below the minimal power for the commutative tadpole integral to vanish~\cite{Leibbrandt:1975dj}. Consequently we observe unregularized UV divergence when computing this integral by transforming it into the bubble or applying the n-nested zero regulator method~\cite{Leibbrandt:1975dj}. We develop an alternative prescription \eqref{B.19} based on the parametrization \eqref{2.30} which is capable of dimensionally-regularizing this integral into a $1/\epsilon$ divergence plus the logarithmic UV/IR mixing term $\ln (\mu^2(\theta p)^2)$ at the $D\to 4-\epsilon$ limit \eqref{B.20}.

 We are able to express both $\Pi^{\mu\nu}_{\rm gf_{\rm mix}}$ and $\Pi^{\mu\nu}_{\rm gf}$ appropriately once $T_{-2}$ is added to the prior basis integral set $T_0$, $I_K^0$, $I_K^1$ and $I_H$. The outcome is listed as below
\begin{equation}
\begin{split}
\Pi^{\mu\nu}_{\rm gf_{\rm mix}}=&\frac{e^2}{(4\pi)^2}\left(2p^\mu p^\nu+p^{\{\mu}(\theta\theta p)^{\nu\}}\frac{p^2}{(\theta p)^2}\right)
\cdot\left(4+4 I_H\right),
\end{split}
\label{6.5}
\end{equation}
\begin{equation}
\Pi^{\mu\nu}_{\rm gf}=\frac{e^2}{(4\pi)^2}\Big(p^\mu p^\nu\cdot\Pi_A+p^{\{\mu}(\theta\theta p)^{\nu\}}\Pi_B\Big),
\label{6.6}
\end{equation}
\begin{equation}
\begin{split}
\Pi_A=&-\frac{1}{2}\bigg[(4\pi\mu^2)^{2-\frac{D}{2}}(p^2)^{\frac{D}{2}-2}\cdot 4\cdot{\rm\Gamma}\left(2-\frac{D}{2}\right){\rm B}\left(\frac{D}{2}-1, \frac{D}{2}-1\right)\Bigg|_{D\to 4-\epsilon}
\\&\cdot\left(1-\frac{p^2}{(\theta p)^2}\Big(\tr\theta\theta(\theta p)^2+(\theta\theta p)^2\Big)\right)+\frac{1}{2}p^2\tr\theta\theta T_{-2}-8 I^0_K
\\&+2\frac{p^2}{(\theta p)^2}\Big(\tr\theta\theta (\theta p)^2(4I^0_K+4I_H)+(\theta\theta p)^2(8I_K^0+4I_H)\Big)\bigg],
\end{split}
\label{6.7}
\end{equation}
\begin{equation}
\begin{split}
\Pi_B=&-\frac{1}{2}\bigg((4\pi\mu^2)^{2-\frac{D}{2}}(p^2)^{\frac{D}{2}-2}\cdot 2(D-1)\cdot{\rm\Gamma}\left(2-\frac{D}{2}\right){\rm B}\left(\frac{D}{2}-1, \frac{D}{2}-1\right)\Bigg|_{D\to 4-\epsilon}
\\&-\frac{1}{4}p^2 T_{-2}-\frac{p^2}{(\theta p)^2}\cdot 3 \cdot(4I^0_K+4I_H)\bigg).
\end{split}
\label{6.8}
\end{equation}
One can immediately notice that $\Pi^{\mu\nu}_{\rm gf_{\rm total}}$ contains only two tensor structures $p^\mu p^\nu$ and $p^{\{\mu}(\theta\theta p)^{\nu\}}$ which can not be combined into a transverse sum. The loss of transversality appears to be, of course, surprising. However we are going to develop reasoning/arguments for this seeming odd behavior in the next section and show that it is in fact understandable.

\section{Gauge fixing contribution without integrating out BRST-auxiliary field}


In order to achieve simple transversality we conclude that one has to keep BRST auxiliary field $B$ from being integrated out. Arguments for that are as follows. 

Starting with the action (\ref{6.1}) we write a generating functional
\begin{equation}
\begin{split}
Z\big[J^\mu,j, {\bar j}, h\big]&=\int Da_\mu D\omega D\bar\omega \;DB \cdot
\exp \Big[i\Big(S+\int\big(J^\mu a_\mu+{\bar j}\omega+j\bar\omega+hB\big)\Big)\Big]
\\ &=
\exp\Big[iW\big[J^\mu,j, {\bar j}, h\big]\Big],
\label{7.1}
\end{split}
\end{equation}
from where we have effective action in terms of "currents":
\begin{equation}
\Gamma\big[J^\mu,j, {\bar j}, h\big]=W\big[J^\mu,j, {\bar j}, h\big]+
\int\big(J^\mu a_\mu+{\bar j}\omega+j\bar\omega+hB\big).
\label{7.2}
\end{equation}
Regular BRST transformation $s$ acting on $Z$ vanishes, thus we have:
\begin{equation}
\begin{split}
sZ=0&=i\int Da_\mu D\omega D\bar\omega \;DB
\Big(J^\mu\partial_\mu\omega-j\cdot B\Big)\cdot
\exp \Big[i\Big(S+\int\big(J^\mu a_\mu+{\bar j}\omega+j\bar\omega+hB\big)\Big)\Big]
\\ &\Longrightarrow
\int\Big(\frac{\delta\Gamma}{\delta a_\mu}\partial_\mu\omega +
B\frac{\delta\Gamma}{\delta\bar\omega}\Big)
=\int\Big(-\omega\partial_\mu\frac{\delta\Gamma}{\delta a_\mu} +
B\frac{\delta\Gamma}{\delta\bar\omega}\Big)=0.
\label{7.3}
\end{split}
\end{equation}
Since the transversality condition means $\partial_\mu\frac{\delta\Gamma}{\delta a_\mu}=0$, which is satisfied in equation (\ref{7.3}) only for $B=0$. This however is not allowed if we do integrate out the $B$ field. Thus we do not perform  that, instead we construct a propagator from the following doublet combination
$\left(
    \begin{array}{c}
        a_\mu \\
       B
    \end{array}
\right)$.

\subsection{Formal analysis}
In order to compute the two point function(s) within the presence of the B-field, we must define the propagator(s) for the "kind of strange" vector-scalar field $\left(
    \begin{array}{c}
        a_\mu \\
       B
    \end{array}
\right)$ doublet. Starting with
\begin{equation}
S_{\rm U(1)}=\int-\frac{1}{4}f_{\mu\nu}f^{\mu\nu}=\frac{1}{2}\int
a_\mu\big(\partial^2 g^{\mu\nu}-\partial^\mu\partial^\nu\big)a_\nu,
\label{7.4}
\end{equation}
we get a quadratic part of $S$
\begin{equation}
S_{\rm quadratic}=\frac{1}{2}\int\Big(
a_\mu\big(\partial^2 g^{\mu\nu}-\partial^\mu\partial^\nu\big)a_\nu
+2B\cdot \partial_\mu a^\mu+B^2\Big),
\label{7.5}
\end{equation}
whose Fourier transform is as follows:
\begin{equation}
\begin{split}
{\tilde S}_{\rm quadratic}=\frac{1}{2}&\int\frac{d^4 k}{(2\pi)^4}\Big(
{\tilde a}^\mu(-k)\big(-k^2 g_{\mu\nu}+k_\mu k_\nu\big){\tilde a}^\nu(k)
\\&
+i{\tilde B}(-k) k_\mu{\tilde a}^\mu(k)-ik_\mu{\tilde a}^\mu(-k){\tilde B}(k)+
{\tilde B}(-k){\tilde B}(k)\Big)
\\
=\frac{1}{2}&\int\frac{d^4 k}{(2\pi)^4}
\left({\tilde a}_\mu(-k),{\tilde B}(-k)\right) T_0
\left(
    \begin{array}{c}
        {\tilde a}_\nu(k) \\
       {\tilde B}(k)
    \end{array}
\right).
\label{7.6}
\end{split}
\end{equation}
The Hermitian matrix
\begin{equation}
T_0=
\begin{pmatrix}
  T_{0_{11}}^{\mu\nu} & {} T_{0_{12}}^\mu
  \\
   T_{0_{21}}^\nu & {} T_{0_{22}}
\end{pmatrix}
=
\begin{pmatrix}
   -k^2 g^{\mu\nu}+k^\mu k^\nu & {} -i k^\mu
  \\
   i k^\nu & {}1
\end{pmatrix},
\label{7.7}
\end{equation}
is then the inverse of the propagator in the momentum space. Next we inverse $T_0$ to obtain 
\begin{equation}
T_0^{-1}=
\begin{pmatrix}
  G_{\rho\mu} & {}A_\rho
  \\
 B_\mu  & {}G
\end{pmatrix}
\Longrightarrow
T^{-1}\cdot T_0=1=
\begin{pmatrix}
  \delta_\rho^\nu & {}0
  \\
 0  & {}1
\end{pmatrix}.
\label{7.8}
\end{equation}
From (\ref{7.7}) and (\ref{7.8}) we have four equations:
\begin{gather}
\begin{split}
&
 \delta_\rho^\nu =-k^2 G_{\rho\nu}+k^\mu G_{\rho\mu} k^\nu+iA_\rho  k^\nu,\quad
0=-ik^\mu G_{\rho\mu}+A_\rho
\\&
0=-k^2 B^\nu+k^\mu B_\mu k^\nu+iG k^\nu,\quad
1=-iB_\mu k^\mu + 1,
\end{split}
\label{7.9}
\end{gather}
which we solve by using simple Ansatz: $A_\rho=A\cdot k_\rho$,
$B_\mu=B\cdot k_\mu$ $\Longrightarrow$
$G=0,\;B=\frac{i}{k^2}$  $\Longrightarrow$ $A=B^\dagger=\frac{-i}{k^2}$
$\Longrightarrow$ $G_{\rho\mu}=-\frac{g_{\rho\mu}}{k^2}$. Taking into account a text book convention $( -k^2 g^{\mu\nu}+k^\mu k^\nu)G_{\mu\nu}=i \delta_\rho^\mu$ for the phase factor we add overall factor $i$ and obtained correct photon propagator, as illustrated in Fig \ref{BaPp}.
\begin{figure}
\begin{center}
\includegraphics[width=8cm]{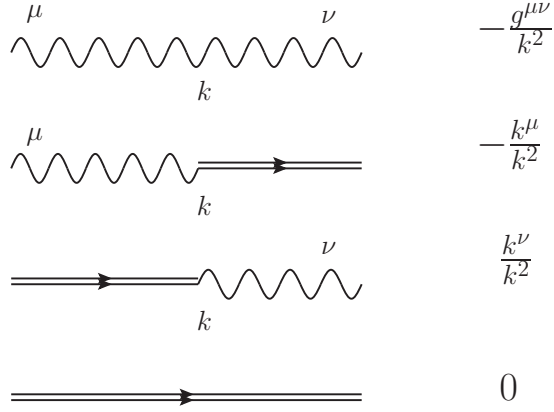}
\end{center}
\caption{Illustration of the above procedure in computing photon propagator. BRST-auxiliary field $B$ (Bauxiliary) is denoted by a double full line.}
\label{BaPp}
\end{figure}

One can write down the usual field redefinition $B'=B+\partial_\mu a^\mu$ in \eqref{6.3} in the Fourier transformed context as
\begin{equation}
\begin{pmatrix}
\tilde a_\mu(k)
\\
\tilde B'(k)
\end{pmatrix}
=A_0\cdot\begin{pmatrix}
\tilde a_\mu(k)
\\
\tilde B(k)
\end{pmatrix}
=\begin{pmatrix}
1 & 0
\\
i k^\mu & 1
\end{pmatrix}
\cdot\begin{pmatrix}
\tilde a_\mu(k)
\\
\tilde B(k)
\end{pmatrix}.
\label{7.10}
\end{equation}
The $A_0^{-1}$ then diagonalizes the bilinear form \eqref{7.7} into
\begin{equation}
T'_0={A_0^{-1}}^\dagger T_0 A_0^{-1}=\begin{pmatrix}
-k^2 g^{\mu\nu} & 0
\\
0 & 1
\end{pmatrix}.
\label{7.11}
\end{equation}
It is easy to see that inverting this $T'_0$ gives the expected Feynman propagator. From that viewpoint the B-integration can be achieved by diagonalization. 
One can formally generalize the tree level diagonalization procedure to the one loop. Consider in general the one loop corrections as another Hermitian matrix $T_1$ adding to the $T_0$ matrix, we write the quadratic part of the 1-loop corrected effective action in the momentum space as
\begin{equation}
\begin{split}
{\tilde \Gamma}^1_{\rm quadratic}
=\frac{1}{2}&\int\frac{d^4 k}{(2\pi)^4}
\left({\tilde a}_\mu(-k),{\tilde B}(-k)\right) \left(T_0+T_1\right)
\left(
    \begin{array}{c}
        {\tilde a}_\nu(k) \\
       {\tilde B}(k)
    \end{array}
\right).
\label{7.6}
\end{split}
\end{equation}
Next we express $T_1$ in terms of its components
\begin{equation}
T_1=
\begin{pmatrix}
  T_{1_{11}}^{\mu\nu} & iT_{1_{12}}^{\mu}=-iT_{1_{21}}^{\mu}
  \\
  iT_{1_{21}}^{\nu} & T_{1_{22}}
\end{pmatrix}.
\label{7.12}
\end{equation}
The Slavnov-Taylor identity \eqref{7.3} then requires that $T_{1_{11}}^{\mu\nu}=\frac{\delta\tilde \Gamma^1}{\delta\tilde a_\mu\delta\tilde a_\nu}$ has to be transverse, while others not. One can now replace $A_0$ by a new transformation
\begin{equation}
A=\begin{pmatrix}
1 & 0
\\
i (k^\mu+T_{1_{21}}^{\mu})  & (1+T_{1_{22}})
\end{pmatrix},
\label{7.13}
\end{equation}
with $A^{-1}$ diagonalizing the 1-loop corrected bilinear form $T_0+T_1$
\begin{equation}
T'={A^{-1}}^\dagger (T_0+T_1) A^{-1}=\begin{pmatrix}
-k^2g^{\mu\nu}+T_{1_{11}}^{\mu\nu}+\Pi'^{\mu\nu} & 0
\\
0 & \left(1+T_{1_{22}}\right)^{-1}
\end{pmatrix},
\label{7.14}
\end{equation}
where
\begin{equation}
\Pi'^{\mu\nu}=k^\mu k^\nu-\frac{\left(k^\mu+T_{1_{21}}^{\mu}\right)\left(k^\nu+T_{1_{21}}^{\nu}\right)}{1+T_{1_{22}}}.
\label{7.15}
\end{equation}
We then conjecture that the formal leading order expansion of $\Pi^{\mu\nu}$ with respect to the coupling constant $e$ corresponds to the gauge fixing corrections to the 1-loop 1-PI photon two point function, i.e.
\begin{equation}
\Pi^{\mu\nu}_{\rm gf_{\rm total}}=\Pi'^{\mu\nu}\left(k^\mu\to p^\mu\right)|_{e^2}=p^\mu p^\nu\cdot T_{1_{22}}-p^{\{\mu}T_{1_{21}}^{\nu\}}.
\label{7.16}
\end{equation}
And, as we shall see below, this relation/conjecture indeed holds.

\subsection{The action of the gauge and BRST-auxiliary fields and Feynman rules}

We define the noncommutative photon-auxiliary field action by using the first and the second order SW maps for the NC gauge field, $A^{(1)}_\mu$ and $A^{(2)}_\mu$, respectively:
\begin{equation}
\begin{split}
S_{B-a_\mu}&=\int -(\partial^\mu B) (A^{(1)}_\mu + A^{(2)}_\mu)
\\&
=\int -(\partial^\mu B)\frac{1}{2}\theta^{ij}a_i\star_2\left(\partial_j a_\mu+f_{j\mu}\right)+(\partial^\mu B)\frac{1}{8}\theta^{ij}\theta^{kl}\bigg(\Big[a_i\partial_j\big(a_k(\partial_l a_\mu+f_{l\mu})\big)\Big]_{\star_{3'}}
\\&{\phantom{xxxx}}-2\Big[a_i(f_{jk}f_{\mu l}-a_k\partial_l f_{j\mu})\Big]_{\star_{3'}}
+\Big[(\partial_j a_\mu+f_{j\mu})a_k(\partial_l a_i+f_{li})\Big]_{\star_{3'}}\bigg),
\label{7.17}
\end{split}
\end{equation}
from where we obtain the corresponding photon-auxiliary field interaction vertices. Corresponding Feynman rules from the above action generate one-loop correction to the quadratic effective action and are given in appendix G.

\begin{figure}
\begin{center}
\includegraphics[width=5cm]{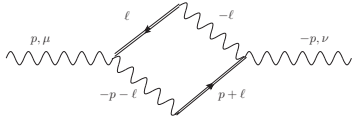}
\end{center}
\caption{Bubble contribution to photon polarization:  $B_{B-a_\mu}^{\mu\nu}(p)$, with $B-a_\mu$ propagators running in the loop in opposite directions.}
\label{Baabubb}
\end{figure}
\begin{figure}
\begin{center}
\includegraphics[width=4cm]{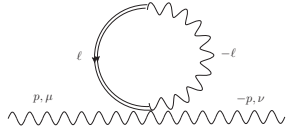}
\end{center}
\caption{The 3gauge-1Bauxiliary fields tadpole contribution to photon polarization: $T_{B-a_\mu}^{\mu\nu}(p)$.}
\label{Baaatad}
\end{figure}
\begin{figure}
\begin{center}
\includegraphics[width=4cm]{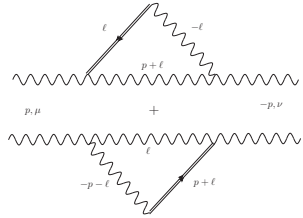}
\end{center}
\caption{Two bubbles with mixing $a_\mu$ and $B-a_\mu$ propagators emerging from SW-interacting U(1).}
\label{Baaabubb1+1}
\end{figure}

\begin{figure}
\begin{center}
\includegraphics[width=4.5cm]{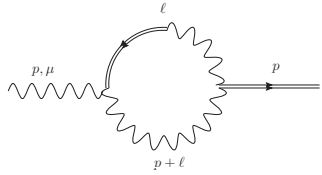}
\end{center}
\caption{Bubble contribution to $iT_{1_{21}}$, with $B-a_\mu$-propagator in the loop.}
\label{BabubbT12}
\end{figure}
\begin{figure}
\begin{center}
\includegraphics[width=3.5cm]{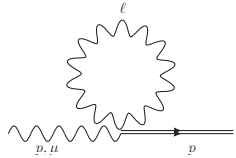}
\end{center}
\caption{Tadpole contribution to $iT_{1_{21}}$, with photon-loop.}
\label{TadpoleT12}
\end{figure}

\begin{figure}
\begin{center}
\includegraphics[width=4cm]{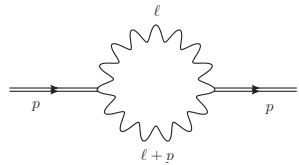}
\end{center}
\caption{Bubble contribution to $T_{1_{22}}$, with photon-loop.}
\label{BaabubbleT22}
\end{figure}
\begin{figure}
\begin{center}
\includegraphics[width=4cm]{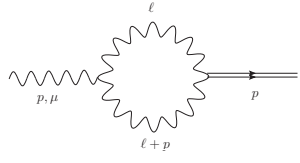}
\end{center}
\caption{Bubble contribution $i\Pi^{\mu}_{\rm mix}$ to $iT_{1_{21}}$, with photon-loop.}
\label{Bubb3photBmix}
\end{figure}

\subsection{One-loop contributions to the photon effective action up to the quadratic order, from the BRST auxiliary field $B$}

Based on our vertex read-out convention, $\partial_\mu=i p_\mu$, we obtain the following correspondence rule between the matrix elements of one loop correction $T_1$ and the 1-PI loop diagrams
\begin{equation}
\begin{split}
T_{1_{11}}^{\mu\nu}=&\Pi^{\mu\nu}+\Pi_{B-a_\mu}^{\mu\nu},
\\
\Pi_{B-a_\mu}^{\mu\nu}=&{\rm Fig.\ref{GfGPb}+Fig.\ref{GfGPt}+Fig.\ref{Baabubb}+Fig.\ref{Baaatad}+Fig.\ref{Baaabubb1+1}},
\\
i T_{1_{21}}^{\mu}=&i\Pi^\mu+i\Pi_{\rm mix}^\mu,
\quad i\Pi^\mu={\rm Fig.\ref{BabubbT12}+Fig.\ref{TadpoleT12}},
\\
T_{1_{22}}=&{\rm Fig.\ref{BaabubbleT22}},
\quad i\Pi_{\rm mix}^\mu={\rm Fig.\ref{Bubb3photBmix}}.
\end{split}
\label{7.26}
\end{equation}
Note that $\Pi^{\mu\nu}$ denotes all contributions from the classical action, which is the same as summing over all contributions to the photon two point function computed in sections 2-5 and thus transverse. Explicit computation first revolves that
\begin{equation}
\Pi_{B-a_\mu}^{\mu\nu}=0,
\label{7.28}
\end{equation}
thus the Slavnov-Taylor identity is actually trivially fulfilled. 

The rest of the matrix elements listed in \eqref{7.26} are nonzero and boil down to the following expressions
\begin{equation}
\Pi^\mu=\frac{e^2}{(4\pi)^2}\Big(p^\mu\Pi_{\rm I}+(\theta\theta p)^\mu\Pi_{\rm II}\Big),
\end{equation}
where
\begin{gather}
\Pi_{\rm I}=(4\pi\mu^2)^{2-\frac{D}{2}}(p^2)^{\frac{D}{2}-2}\cdot (D-2)\cdot{\rm\Gamma}\left(3-\frac{D}{2}\right){\rm B}\left(\frac{D}{2}-1, \frac{D}{2}-1\right)\Bigg|_{D\to 4-\epsilon}-(4I_K^0+6 I_H),
\label{7.29}\\
\begin{split}
\Pi_{\rm II}=\frac{p^2}{(\theta p)^2}\bigg(&(4\pi\mu^2)^{2-\frac{D}{2}}(p^2)^{\frac{D}{2}-2}\cdot (D-1)\cdot{\rm\Gamma}\left(3-\frac{D}{2}\right){\rm B}\left(\frac{D}{2}-1, \frac{D}{2}-1\right)\Bigg|_{D\to 4-\epsilon}
\\&-\frac{1}{8}(\theta p)^2 T_{-2}-(6I_K^0+6I_H)\bigg),
\end{split}
\label{7.30}
\end{gather}
and
\begin{gather}
\Pi_{\rm mix}^\mu=-\frac{e^2}{(4\pi)^2}\left(p^\mu+(\theta\theta p)^\mu\frac{p^2}{(\theta p)^2}\right)\left(4+4 I_H\right).
\label{7.32}
\end{gather}
Finally we have
\begin{equation}
\begin{split}
T_{1_{22}}=&\frac{1}{2}\frac{e^2}{(4\pi)^2}\Bigg((4\pi\mu^2)^{2-\frac{D}{2}}(p^2)^{\frac{D}{2}-2}{\rm\Gamma}\left(2-\frac{D}{2}\right){\rm B}\left(\frac{D}{2}-1, \frac{D}{2}-1\right)\Bigg|_{D\to 4-\epsilon}
\\&\cdot 4\cdot\left((D-3)+\frac{p^2}{(\theta p)^2}\Big(\tr\theta\theta+2(\theta\theta p)^2\Big)\right)-\frac{1}{2}p^2\tr\theta\theta T_{-2}
\\&-8I_K^0-24 I_H-\frac{p^2}{(\theta p)^2}\cdot 2\cdot\Big(\tr\theta\theta (\theta p)^2(4I_K^0+4I_H)
+(\theta\theta p)^2(8I_K^0+4I_H)\Big)\Bigg).
\end{split}
\label{7.33}
\end{equation}

Next we start to verify our conjecture \eqref{7.16}. First we derive the following relations from it, 
\begin{equation}
\Pi_A=(4\pi)^2e^{-2}T_{1_{22}}-2\Pi_{\rm I},\;\;
\Pi_B=-\Pi_{\rm II},\;\;
\Pi^{\mu\nu}_{\rm gf_{\rm mix}}=-p^{\{\mu}\Pi_{\rm mix}^{\nu\}}.
\label{7.36}
\end{equation}
One then immediately observes that the second and third relations do fulfill. As for the first one we can compute its right hand side
\begin{equation}
\begin{split}
(4\pi)^2e^{-2}T_{1_{22}}&-2\Pi_{\rm I}=\frac{1}{2}\Bigg((4\pi\mu^2)^{2-\frac{D}{2}}(p^2)^{\frac{D}{2}-2}{\rm\Gamma}\left(2-\frac{D}{2}\right){\rm B}\left(\frac{D}{2}-1, \frac{D}{2}-1\right)\Bigg|_{D\to 4-\epsilon}
\\&\cdot 4\cdot\left((D-3)+\frac{p^2}{(\theta p)^2}\Big(\tr\theta\theta+2(\theta\theta p)^2\Big)\right)-\frac{1}{2}p^2\tr\theta\theta T_{-2}
\\&-8I_K^0-24 I_H-\frac{p^2}{(\theta p)^2}\cdot 2\cdot\Big(\tr\theta\theta (\theta p)^2(4I_K^0+4I_H)
+(\theta\theta p)^2(8I_K^0+4I_H)\Big)\bigg)
\\&-2\Bigg((4\pi\mu^2)^{2-\frac{D}{2}}(p^2)^{\frac{D}{2}-2}\cdot (D-2)\cdot{\rm\Gamma}\left(3-\frac{D}{2}\right){\rm B}\left(\frac{D}{2}-1, \frac{D}{2}-1\right)\Bigg|_{D\to 4-\epsilon}
\\&-(4I_K^0+6 I_H)\Bigg)
\\=&\frac{1}{2}\Bigg((4\pi\mu^2)^{2-\frac{D}{2}}(p^2)^{\frac{D}{2}-2}{\rm\Gamma}\left(2-\frac{D}{2}\right){\rm B}\left(\frac{D}{2}-1, \frac{D}{2}-1\right)\Bigg|_{D\to 4-\epsilon}
\\&\cdot 4\cdot\left(-1+\frac{p^2}{(\theta p)^2}\Big(\tr\theta\theta+2(\theta\theta p)^2\Big)\right)-\frac{1}{2}p^2\tr\theta\theta T_{-2}
\\&+8I_K^0-\frac{p^2}{(\theta p)^2}\cdot 2\cdot\Big(\tr\theta\theta (\theta p)^2(4I_K^0+4I_H)
+(\theta\theta p)^2(8I_K^0+4I_H)\Big)\Bigg)=\Pi_A,
\end{split}
\label{7.37}
\end{equation}
which is in agreement with (\ref{6.7}). Thus the conjectured relation \eqref{7.16} is proven.

In fact  the gauge fixing contribution to the 1-loop correction is a shift out of the on-shell point of the gauge fixing functional. Therefore it does not need to be transverse. The fact that the pure gauge fixing and mixing contributions satisfies \eqref{7.16} independently is because the former can be considered as a gauge fixing to the free $\rm U(1)$ gauge theory. 

Finally let's briefly discuss the divergences in the gauge fixing configuration(s). Using the results from appendix B we can see that $\Pi^{\mu\nu}_{\rm gf_{\rm total}}$ contains no quadratic IR divergent term, therefore the quadratic IR divergence cancelation we found in the prior sections are also preserved under this gauge fixing choice point. We can also extract the UV \& logarithmic divergence at the $D\to 4-\epsilon$ limit
\begin{gather}
\Pi_A|_{\rm UV}=-\left(2-\frac{p^2}{(\theta p)^2}
\Big(3\tr\theta\theta(\theta p)^2+4(\theta\theta p)^2\Big)\right)\left(\frac{2}{\epsilon}+\ln(\mu^2(\theta p)^2)\right),
\label{7.38}\\
\Pi_B|_{\rm UV}=-\frac{7}{2}\frac{p^2}{(\theta p)^2}\left(\frac{2}{\epsilon}+\ln(\mu^2(\theta p)^2)\right),
\label{7.39}
\end{gather}
while the $\Pi^{\mu\nu}_{\rm gf_{\rm mix}}$ is finite at this limit. We conclude our analysis by listing of all UV plus  logarithmic divergences in $\Pi^{\mu\nu}_{\rm total}=\Pi^{\mu\nu}+\Pi^{\mu\nu}_{\rm gf_{\rm total}}$, which can be decomposed into seven symmetric tensor structures
\begin{equation}
\begin{split}
\Pi^{\mu\nu}_{\rm total}|_{\rm UV}=&
\frac{e^2}{(4\pi)^2}\Big(g^{\mu\nu}\cdot\Xi_1+p^\mu p^\nu\cdot\Xi_2+(\theta p)^\mu(\theta p)^\nu\cdot\Xi_3
\\&+\Big[g^{\mu\nu}(\theta p)^2-(\theta\theta)^{\mu\nu}p^2
+ p^{\{\mu}(\theta\theta p)^{\nu\}}\Big]\Xi_4
\\&+\Big[(\theta\theta)^{\mu\nu}(\theta p)^2+(\theta\theta p)^\mu(\theta\theta p)^\nu\Big]\Xi_5
+ (\theta p)^{\{\mu} (\theta\theta\theta p)^{\nu\}}\cdot \Xi_6
\\&+p^{\{\mu}(\theta\theta p)^{\nu\}}\cdot\Xi_7\Big).
\label{7.40}
\end{split}
\end{equation}
Explicit computation then yields
\begin{gather}
\begin{split}
\Xi_1|_{\rm UV}= p^2\bigg(\frac{4}{3}-\frac{4}{3}{n_f}-\frac{1}{3}{n_s}+\frac{p^2}{(\theta p)^4}\Big(3\tr\theta\theta(\theta p)^2+4(\theta\theta p)^2\Big)\bigg)\left(\frac{2}{\epsilon}+\ln(\mu^2(\theta p)^2)\right),
\end{split}
\label{7.41}\\
\Xi_2|_{\rm UV}=\left(-2-\frac{4}{3}+\frac{4}{3}{n_f}+\frac{1}{3}{n_s}\right)\left(\frac{2}{\epsilon}+\ln(\mu^2(\theta p)^2)\right),
\label{7.42}\\
\Xi_3|_{\rm UV}=B_2|_{\rm UV}=2\frac{p^2}{(\theta p)^2}\bigg(2-\frac{p^2(\tr\theta\theta)}{(\theta p)^2}
\bigg)\left(\frac{2}{\epsilon}+\ln(\mu^2(\theta p)^2)\right),
\label{7.43}\\
\Xi_4|_{\rm UV}=B_3|_{\rm UV}=2\frac{p^2}{(\theta p)^2}\left(\frac{2}{\epsilon}+\ln(\mu^2(\theta p)^2)\right),
\label{7.44}\\
\Xi_5|_{\rm UV}=B_4|_{\rm UV}=-4\frac{p^4}{(\theta p)^4}\left(\frac{2}{\epsilon}+\ln(\mu^2(\theta p)^2)\right),
\label{7.45}\\
\Xi_6|_{\rm UV}=B_5|_{\rm UV}=4\frac{p^4}{(\theta p)^4}
\left(\frac{2}{\epsilon} + \ln(\mu^2(\theta p)^2)\right),
\label{7.46}\\
\Xi_7|_{\rm UV}=\Pi_B|_{\rm UV}=-\frac{7}{2}\frac{p^2}{(\theta p)^2}\left(\frac{2}{\epsilon}
+\ln(\mu^2(\theta p)^2)\right).
\label{7.47}
\end{gather}

\section{Summary and discussion}

In this paper we have computed the one-loop contributions to all propagators of the noncommutative super Yang-Mills U(1) theory with ${\cal N}$=1, 2 and 4 supersymmetry and defined by the means of the $\theta$-exact Seiberg-Witten map. We have shown that for ${\cal N}$=1, 2 and 4 the quadratic noncommutative IR divergence,
\begin{equation*}
\frac{(\theta p)^\mu(\theta p)^\nu}{(\theta p)^4},
\end{equation*}
--a trade-mark of the noncommutative gauge theories-- which occur in the bosonic, fermionic and scalar loop-contributions to the photon propagator cancel each other, rendering photon propagator free of them as befits of the Supersymmetry. Indeed, from \eqref{IRcancel} for ${\cal N}=1,2,4$, one gets:
\begin{eqnarray}
{\cal N} &=& 1: n_f=1, n_s=0,
\nonumber\\
{\cal N} &=& 2: n_f=2, n_s=2,
\nonumber\\
{\cal N} &=& 4: n_f=4, n_s=6,
\nonumber\\
\Pi_{\rm total}^{\mu\nu}(p)|_{\rm IR}&=&\frac{e^2}{(4\pi)^2}\bigg[\Big(\frac{32}{3}+\frac{64}{3}\Big)-32\cdot n_f + 16\cdot n_s\bigg]\frac{(\theta p)^\mu(\theta p)^\nu}{(\theta p)^4}=0.
\label{N=1,2,4sumIR}
\end{eqnarray}
This cancellation, occuring in the case at hand, is nontrivial since Supersymmetry acts nonlinearly -see (\ref{2.14})-- on the ordinary fields. Let us recall that the cancellation of quadratic noncommutative IR divergences is also a feature of noncommutative super Yang-Mills theories when formulated in terms of noncommutative fields~\cite{Zanon:2000nq, Ruiz:2000hu, Ferrari:2003vs, AlvarezGaume:2003mb,Ferrari:2004ex}. Hence, our result concerning the cancellation of the quadratic noncommutative IR divergences really points into direction that the $\theta$-exact Seiberg-Witten map really provides  quantum duals of the same underlying theory.

We have shown --see (\ref{FullSc2point}) and (\ref{5.6})-- that the  characteristic quadratic noncommutative IR divergences,
\begin{equation*}
\frac{1}{(\theta p)^2},
\end{equation*}
which arise in the individual contributions to the one-loop propagators  of the scalar fields in the ${\cal N}$=2 and 4 Supersymmetry, also cancel each other at the end of the day. The same holds for the photino field as well. 

Since the previous cancellations occur both in the ordinary Feynman gauge and in the noncommutative Feynman gauge --see section 7--, our computations further indicate that the cancellation is robust against changing the gauge fixing and may have real {\it physical}, and therefore gauge invariant, content. Let us recall that independence of gauge-fixing parameter of the cancellation of noncommutative IR divergences in the dual theory, i.e., in ${\cal N}$=1 U(1) super Yang-Mills theory formulated in terms of the noncommutative fields, has been shown to hold --see Ref.~\cite{Ruiz:2000hu}.

In this paper we have also worked out explicitly the one-loop UV divergent contributions --which show as poles at $D=4$-- to all propagators of the theory: see (\ref{2.42}), (\ref{UVterm1})--(\ref{UVterm2}), (\ref{4.14}),  (\ref{4.17}), (\ref{5.7}), (\ref{5.12}) and (\ref{7.40})--(\ref{7.47}). It is noticeable that the pole parts displayed in the equations we have just quoted contain non-polynomial, i.e., non-local, terms whose denominator is a power of $\theta p$. With respect to this we would like to point out that, in keeping with Weinberg's power counting theorem~\cite{Weinberg:1959nj},  Feynman integrals whose degree of UV divergence are not the same along all directions are liable to give rise to the pole contributions which are non-polynomial. This is exactly our situation since our integrands contain factors of the type
\begin{equation*}
\frac{1}{(q^2)^n\,((q+p)^2)^m\,(q\theta p)^s},\quad s=1,2,
\end{equation*}
and these factors approach to zero as $\Lambda^{(-2n-2m-s)}$  along the direction parallel to $\theta p$, and as $\Lambda^{(-2n-2m)}$ along any direction orthogonal to $\theta p$. Hence UV divergences with a non-polynomial
dependence on the momenta may occur and our computations show that  indeed they do occur. We would like to recall that a similar situation, --i.e. the non-polynomial UV divergences-- happens in ordinary Yang-Mills theories in the light-cone gauge~\cite{Leibbrandt:1987qv, Leibbrandt:1993np}.

Now, the UV divergences of two-point functions are in general gauge dependent quantities. We have verified that this is so in our case by computing the one-loop propagator of the gauge field both in the ordinary Feynman gauge and in the noncommutative Feynman gauge --see Section 6. The result for the first type of gauge is in (\ref{UVterm1})--(\ref{UVterm2}) and in (\ref{7.40})--(\ref{7.47}) for the
second type of gauge fixing term: their differences stand out.  Hence, extracting gauge invariant information from the UV divergences is our next challenge along this line of research and it will require the computation of three and higher point functions.

Let us finally remark that UV/IR mixing effects also work for the non-polynomial UV divergent contributions we have obtained. Indeed, as seen in (\ref{2.42}), (\ref{UVterm1})--(\ref{UVterm2}), (\ref{4.14}),  (\ref{4.17}), (\ref{5.7}), (\ref{5.12}) and (\ref{7.40})--(\ref{7.47}) every pole in $2/\epsilon$ comes hand in hand with the logarithmic noncommutative IR divergence $\ln(\mu^2(\theta p)^2)$. The reader is referred to the final part of Appendix B for further information regarding this issue.

\section{Acknowledgments}
The work by C.P. Martin has been financially supported in part by the Spanish MINECO through grant FPA2014-54154-P. J.Y. has been fully supported by Croatian Science Foundation under Project No. IP-2014-09-9582. The work  J.T. is conducted under the European Commission and the Croatian Ministry of Science, Education and Sports Co-Financing Agreement No. 291823. In particular, J.T. acknowledges project financing by the Marie Curie FP7-PEOPLE-2011-COFUND program NEWFELPRO: Grant Agreement No. 69, and  Max-Planck-Institute for Physics, and W. Hollik for hospitality.  J.T. would also like to acknowledge L. Alvarez-Gaume for fruitful discussions and CERN Theory Division, where part of this work was conducted, for hospitality. We would like to acknowledge the COST Action MP1405  (QSPACE). We would also like to thank J. Erdmenger, W. Hollik and A. Ilakovac, for fruitful discussions. A great deal of computation was done by using MATHEMATICA 8.0Mathematica \cite{mathematica} plus the tensor algebra package xACT~\cite{xAct}. Special thanks to A. Ilakovac and D. Kekez for the computer software and hardware support.

\appendix

\section{Seiberg-Witten differential equations for the SYM U(1)}

Let $\Phi$ be a noncommutative field, either boson or fermion, in $d$ dimensions, which gauge transforms under the adjoint of U(N). Then its NC BRST transformation reads
\begin{equation}
s_{\rm NC}\Phi=-i[\Phi\stackrel{\star}{,} \Omega],
\label{A.1}
\end{equation}
where $\Omega$ is the noncommutative U(N) ghost field in $d$ dimensions that parametrizes the noncommutative BRST transformations of the U(N) gauge field $A_\mu$ in $d$ dimensions:
\begin{equation}
s_{\rm NC}A_{\mu}=\partial_{\mu}\Omega-i[A_\mu\stackrel{\star}{,} \Omega],\quad s_{\rm NC}\Omega=i\Omega\star\Omega.
\label{A.2}
\end{equation}

Let $\phi$ and $a_\mu$ be an ordinary matter and gauge fields in $d$ dimensions which take values in the  Lie algebra of U(N) in the fundamental representation. Let the BRST transformations of $\phi$ and $a_\mu$  be
\begin{equation}
s\phi=-i[\phi,\omega],\quad sa_{\mu}=\partial_{\mu}\omega-i[a_\mu,\omega],\quad s\omega=i\omega\cdot\omega,
\label{A.3}
\end{equation}
where $\omega$ is the ordinary ghost field in $d$ dimensions which also takes values in Lie algebra of U(N) in the fundamental representation. Then,  the SW map $\{\Omega[e\cdot a,\omega;\theta^{\mu\nu}],A_\mu[e\cdot a;\theta^{\mu\nu}], \Phi[e\cdot a,\phi;\theta^{\mu\nu}]\}$ is a solution to the problem
\begin{equation}
\begin{split}
&s\Omega=i\Omega\star\Omega,\, \quad \Omega[e\cdot a,\omega;\theta^{ij}=0]=\omega,\\
&sA_{\mu}=\partial_{\mu}\Omega-i[A_\mu\stackrel{\star}{,} \Omega],\quad A_\mu[e\cdot a;\theta^{ij}=0]=e \cdot a_\mu,\\
&s\Phi=-i[\Phi\stackrel{\star}{,} \Omega],\quad \Phi[e\cdot a,\phi;\theta^{ij}=0]
=\phi.
\end{split}
\label{sweq}
\end{equation}
It is known that  the following set of differential equations---called the Seiberg-Witten differential equations~\cite{Seiberg:1999vs,Barnich:2003wq}---furnish a solution to the problem in the system of equations \eqref{sweq}:
 \begin{equation}
 \begin{split}
&\frac{d}{dt} \Omega=-\frac{1}{4}\theta^{ij}\Big\{A_i\stackrel{\star_t}{,} \partial_j\Omega\Big\},\quad \Omega[t=0]=\omega,\\
&\frac{d}{dt} A_\mu=-\frac{1}{4}\theta^{ij}\Big\{A_i\stackrel{\star_t}{,}\partial_j A_\mu+F_{j\mu}\Big\},\quad A_\mu[t=0]=e\cdot a_\mu,\\
&\frac{d}{dt} \Phi=-\frac{1}{4}\theta^{ij}\Big\{A_i\stackrel{\star_t}{,}\partial_j \Phi+{\cal D}_j\Phi\Big\},\quad \Phi[t=0]=\phi.
\end{split}
\label{swphi}
\end{equation}
Note that $\mu$ runs from $0$ to $d-1$, while $i$ runs from $1$ to $d-1$, respectively.

Now we show how  a solution to the previous problem can be obtained by solving Seiberg-Witten differential equations for a U(N) gauge field in $d+1$ dimensions. Let $A_{M}=(A_{\mu},A_d)$ be a noncommutative gauge field in $d+1$ dimensions and in the fundamental representation of U(N) and let $\hat{\Omega}$ denote the corresponding noncommutative ghost field. Then the Seiberg-Witten differntial equations for $A_{M}$ and $\hat{\Omega}$ read
\begin{equation}
 \begin{split}
&\frac{d}{dt} \hat{\Omega}=-\frac{1}{4}\theta^{IJ}\Big\{A_I\stackrel{\star_t}{,}\partial_J\hat{\Omega}\Big\},\quad \hat{\Omega}[t=0]=\hat{\omega},\\
&\frac{d}{dt} A_M=-\frac{1}{4}\theta^{IJ}\Big\{A_I\stackrel{\star_t}{,}\partial_J A_\mu+F_{J M}\Big\},\quad A_M[t=0]=e\cdot a_M,
\end{split}
\label{swdeq}
\end{equation}
where $I$ and $J$ run from $1$ to $d$, and $a_M=(a_\mu,a_{d+1})$ and $\hat{\omega}$ are the corresponding ordinary fields in $d+1$ dimensions.

Let us assume that the coordinate $X^d$ commutes with all the others, i.e., $\theta^{IJ}$ is such that $\theta^{I d}=0$. Now, let $A_M[e\cdot a_M', ;\theta^{IJ}]$ and $\hat{\Omega}[e\cdot a_M',\hat{\omega};\theta^{IJ}]$ be the solution to \eqref{swdeq}  and let us take now $a_{M}$ and $\hat{\omega}$ to be independent of $x^d$, so that $A_M[e\cdot a_M', ;\theta^{IJ}]$ and $\hat{\Omega}[e\cdot a_M',\hat{\omega};\theta^{IJ}]$ become independent of $x^d$. Now, for these $A_M[e\cdot a_M', ;\theta^{ij}]$ and $\hat{\Omega}[e\cdot a_M',\hat{\omega};\theta^{ij}]$ the SW differential equations in \eqref{swdeq} boil down to
\begin{equation}
 \begin{split}
&\frac{d}{dt} \hat{\Omega}=-\frac{1}{4}\theta^{ij}\Big\{A_i\stackrel{\star_t}{,}\partial_j\hat{\Omega}\Big\},\quad \hat{\Omega}[t=0]=\hat{\omega},\\
&\frac{d}{dt} A_\mu=-\frac{1}{4}\theta^{ij}\Big\{A_i\stackrel{\star_t}{,}\partial_j A_\mu+F_{j \mu}\Big\},\quad A_\mu[t=0]=e\cdot a_\mu,\\
&\frac{d}{dt} A_d=-\frac{1}{4}\theta^{ij}\Big\{A_i\stackrel{\star_t}{,}\partial_j A_d+{\cal D}_j A_d\Big\},\quad A_d[t=0]=e\cdot a_d,
\end{split}
\label{swdeqred}
\end{equation}
where we have taken into account that $F_{\mu d}=\partial_\mu A_d-i[A_\mu,A_d]={\cal D}_\mu A_d$, for $A_d$ does not depend on $x^d$. It is plain that if we replace $A_d$ with $\Phi$ and $e\cdot a_d$ with $\phi$ in \eqref{swdeqred}, one obtains \eqref{swphi}. We thus conclude that the SW map for $\Phi$ can be obtained from the SW map  $A_d[e\cdot a_\mu,e\cdot a_d;\theta^{ij}]$ that solves \eqref{swdeq} by replacing $e\cdot a_d$ with $\phi$. We also deduce the following relation between the gauge field strength $F_{\mu d}$ and the covariant derivative $\cal D_{\mu}$
\begin{equation}
{\cal D}_{\mu}[A_\nu]\Phi\,=\,F_{\mu d}[e\cdot a_\mu,e\cdot a_d;\theta^{ij}]\vert_{e\cdot a_d\rightarrow\, \phi}.
\label{A.8}
\end{equation}

Next, let $A_{\mu}[a_\nu;\theta^{ij}]$, $\Lambda_{\alpha}[a_\mu,\lambda_{\alpha};\theta^{ij}]$ and $D^{(nc)}[a_\mu,D;\theta^{ij}]$ be $\theta$-exact Seiberg-Witten maps given by the following expansions in terms of the coupling constant $e$:
\begin{equation}
\begin{split}
&A_{\mu}[a_\nu;\theta^{ij}]=e\,(a_{\mu}+ e\, A^{(1)}_{\mu}[a_\nu;\theta^{ij}]+e^2\,A^{(2)}_{\mu}[a_\nu;\theta^{ij}])+\mathcal O\left(e^4\right),
\\&\Lambda_{\alpha}[a_\nu,\lambda_{\alpha};\theta^{ij}]=\lambda_{\alpha}+ e\, \Lambda_{\alpha}^{(1)}[a_\nu,\lambda_{\alpha};\theta^{ij}]+e^2\,\Lambda^{(2)}_{\alpha}[a_\nu,\lambda_{\alpha};\theta^{ij}]+\mathcal O\left(e^3\right),
\\&D^{(nc)}[a_\nu,D;\theta^{ij}]=D+ e\, D^{(1)}[a_\nu,D;\theta^{ij}]+e^2\,D^{(2)}[a_\nu,D;\theta^{ij}]+\mathcal O\left(e^3\right).
\end{split}
\label{A.9}
\end{equation}
Taking the $n$-th variations, of the $m$-th order of the NC gauge field $A_{\mu}$, its supersymmetric fermion partner $\Lambda_{\alpha}$ and of the NC auxiliary field $D^{(nc)}$, we obtain the following expressions
\begin{equation}
\begin{split}
&\delta^{n}A_{\mu}^{(m)}=A^{(m)}_{\mu}[a_\nu+\delta^{n}a^\nu;\theta^{ij}]-A^{(m)}_{\mu}[a_\nu;\theta^{ij}]+\mathcal O\left(\xi^2\right),
\\&\delta^{n}\Lambda_{\alpha}^{(m)}[a_\nu,\lambda_{\alpha};\theta^{ij}]=\Lambda_{\alpha}^{(m)}[a_\nu+\delta^{n}a^\nu,\lambda_{\alpha}+\delta^{n}\lambda_{\alpha};\theta^{ij}]-\Lambda_{\alpha}^{(m)}[a_\nu,\lambda_{\alpha};\theta^{ij}]+\mathcal O\left(\xi^2\right),
\\&\delta^{n}D^{(m)}[a_\nu,D;\theta^{ij}]=D^{(m)}[a_\nu+\delta^{n}a^\nu,D+\delta^{n}D;\theta^{ij}]-D^{(m)}[a_\nu,D;\theta^{ij}]+\mathcal O\left(\xi^2x\right).
\end{split}
\label{A.10}
\end{equation}
Here $m=1,2$, while $\delta^{n}a^\nu$, $\delta^{n}\lambda_{\alpha}$ and $\delta^{n}D$ for $n=0,1$,  have been given in \eqref{2.18}.

\section{Integrals}

A fairly large number of special function integrals occur in studying NCQFT, with or without SW map. Some description of the integrals relevant to this work was given in~\cite{Horvat:2013rga}, where we used a set of seven special function integrals to present the nonplanar part of the bubble integrals at $D=4$. During this work and our prior study on NC tadpole integrals~\cite{Horvat:2015aca} we studied additional new integrals and found some new relations among all of them. Here we present a new list of five integrals which are used to present all loop integral results in the main text.

The original set of seven integrals include four Bessel K-function integrals, and three integrals over a function $H[z]$ which can be expressed in terms of hypergeometric functions
\begin{equation}
\begin{split}
H[z]
=\lim\limits_{D\to 4}\bigg[&\left(\frac{z}{2}\right)^{D-2}{\rm\Gamma}\left(1-\frac{D}{2}\right)
\;_1F_2\left(\frac{1}{2};\frac{3}{2},\frac{D}{2};\left(\frac{z}{2}\right)^2\right)
\\&+\frac{1}{3-D}
\cdot{\rm\Gamma}\left(\frac{D}{2}-1\right)\;_1F_2\left(\frac{3-D}{2};\frac{4-D}{2},\frac{5-D}{2};\left(\frac{z}{2}\right)^2\right)\bigg].
\end{split}
\label{B.1}
\end{equation}
Below is a list of these integrals 
\begin{gather}
I_1=\int\limits_0^1 dx\, (x(1-x)p^2)^{\frac{1}{2}}((\theta p)^2)^{-\frac{1}{2}}K_1\left[(x(1-x)p^2(\theta p)^2)^{\frac{1}{2}}\right],
\label{B.2}\\
I_2=\int\limits_0^1 dx\, K_0\left[(x(1-x)p^2(\theta p)^2)^{\frac{1}{2}}\right],
\label{B.3}\\
I_3=\int\limits_0^1 dx\, x K_0\left[(x(1-x)p^2(\theta p)^2)^{\frac{1}{2}}\right],
\label{B.4}\\
I_4=\int\limits_0^1 dx\, x^2 K_0\left[(x(1-x)p^2(\theta p)^2)^{\frac{1}{2}}\right],
\label{B.5}\\
I_5=\int\limits_0^1 dx\, H\left[(x(1-x)p^2(\theta p)^2)^{\frac{1}{2}}\right],
\label{B.6}\\
I_6=\int\limits_0^1 dx\, x H\left[(x(1-x)p^2(\theta p)^2)^{\frac{1}{2}}\right],
\label{B.7}\\
I_7=\int\limits_0^1 dx\, x^2 H\left[(x(1-x)p^2(\theta p)^2)^{\frac{1}{2}}\right].
\label{B.8}
\end{gather}
Later it is revolved that $I_1$ is directly related to the tadpole integrals
\begin{equation}
\begin{split}
A_1=&\int\frac{d^D k}{(2\pi)^D}\,\frac{(k\theta p)^2}{k^2}f_{\star_2}(k,p)^2=-8\frac{1}{(4\pi)^2}\frac{1}{(\theta p)^2}
\\
=&-\frac{1}{(4\pi)^2}\Big(8I_1+p^2(4I_2-12I_3+8I_4)\Big),
\end{split}
\label{B.9}
\end{equation}
and
\begin{equation}
\begin{split}
A_2=&\int\frac{d^D k}{(2\pi)^D}\,\frac{(k\cdot p)^2}{k^2}f_{\star_2}(k,p)^2=\frac{8}{3}\frac{1}{(4\pi)^2}\frac{p^2}{(\theta p)^4}
\\
=&\frac{1}{3}\frac{1}{(4\pi)^2}\frac{p^2}{(\theta p)^4}\Big(8I_1+p^2(12I_2-92I_3+104I_4)+4p^2(3I_5-26I_6+32I_7)\Big).
\end{split}
\label{B.10}
\end{equation}
It is convenient to use the tadpole integral in lieu of the integral $I_1$ since the tadpole integral is quadratic IR divergent only. We select
\begin{equation}
\begin{split}
T_0=&\frac{1}{4}(4\pi)^2 A_1=(4\pi)^2\int\frac{d^D k}{(2\pi)^D}\, \frac{1}{k^2}\sin^2\frac{k\theta p}{2}=-2\frac{1}{(\theta p)^2}
=-2I_1-p^2(I_2-3I_3+2I_4).
\end{split}
\label{B.11}
\end{equation}
to fulfill this task. We can extract an identity
\begin{equation}
(4I_2-12I_4)+(5I_5-16I_7)=0,
\label{B.12}
\end{equation}
from the relation $A_1=-3p^{-2}(\theta p)^2A_2$. The other two useful relations are:
\begin{equation}
I_2=2I_3,\:\:I_5=2I_6.
\label{B.13}
\end{equation}
Using above three relations we can reduce the rest six integrals $I_{2-7}$ to three, which we choose to be
\begin{gather}
I_K^0=I_2=\int\limits_0^1 dx\,K_0\left[(x(1-x)p^2(\theta p)^2)^{\frac{1}{2}}\right],
\label{B.14}\\
I_K^1=I_3-I_4=\int\limits_0^1 dx\,x(1-x)K_0\left[(x(1-x)p^2(\theta p)^2)^{\frac{1}{2}}\right],
\label{B.15}\\
I_H=I_5=\int\limits_0^1 dx\,H\left[(x(1-x)p^2(\theta p)^2)^{\frac{1}{2}}\right].
\label{B.16}
\end{gather}
Using the generalized power series expansions in the vicinity of $z=0$, \footnote{$\psi(z)=\frac{d}{dz}\ln\Gamma(z)$ denotes the zeroth order polygamma function.}
\begin{equation}
K_0[z]=-\sum\limits_{k=0}^\infty\frac{1}{\Gamma[k+1]^2}\left(\frac{z}{2}\right)^{2k}\left(\ln\frac{z}{2}-\psi(k+1)\right),
\label{B.161}
\end{equation}
and
\begin{equation}
\begin{split}
H[z]=-1+&\sum\limits_{k=0}^\infty\frac{\Gamma\left(k+\frac{3}{2}\right)}{\Gamma\left(k+\frac{5}{2}\right)\Gamma\left(k+1\right)\Gamma\left(k+2\right)}\left(\frac{z}{2}\right)^{2k+2}
\\&\cdot\left(\ln\frac{z}{2}+\frac{1}{2}\psi\left(k+\frac{1}{2}\right)-\frac{1}{2}\psi\left(k+1\right)-\frac{1}{2}\psi\left(k+\frac{3}{2}\right)-\frac{1}{2}\psi\left(k+2\right)\right),
\end{split}
\label{B.162}
\end{equation}
it is not difficult to see that the integrals $T_0$, $I_K^0$, $I_K^1$ and $I_H$ bear distinctive asymptotic behavior in the IR regime. The $T_0$ is quadratically IR divergent by definition, while $I_K^0$ and $I_K^1$ carry the  dual logarithmic noncommutative IR divergence (logarithmic UV/IR mixing) $\ln(p^2(\theta p)^2)$, with coefficients $-1/2$ and $-1/12$, respectively. The last integral $I_H$ is finite at the IR limit.

A new type of tadpole integral, which is UV divergent at the $D\to 4-\epsilon$ limit occurs repeatedly in the NC Feynman gauge computation part of this work. Here we provide an account of its evaluation. This new tadpole, denoted as $T_{-2}$, bears a very simple form
\begin{equation}
T_{-2}=(4\pi)^2\mu^{4-D}\int\frac{d^D k}{(2\pi)^D}\,\frac{1}{k^2}f_{\star_2}(k,p)^2.
\label{B.17}
\end{equation}
On the other hand, it turns out that $T_{-2}$ is not that simple to evaluate. Two usual regularization methods used before, turning tadpole to bubble or using the $n$-nested zero regulator, do not function here. The first one produces divergent special function integrals while the second contains unfavorable powers of the regulator. The parametrization discussed in the first section of this note offers us an alternative way to handle this problem. Using that parametrization we can express $T_{-2}$ as
\begin{equation}
\begin{split}
T_{-2}&=(4\pi)^2\mu^{4-D}\int\frac{d^{D-1} \ell}{(2\pi)^{D-1}}\,\int\limits_{-\infty}^{+\infty}\,\frac{dx}{2\pi}\frac{1}{\ell^2+x^2}\frac{4\sin^2\frac{|\theta p|}{2}x}{x^2(\theta p)^2}
\\
&=(4\pi)^2\mu^{4-D}\int\frac{d^{D-1} \ell}{(2\pi)^{D-1}}\,\frac{1}{(\theta p)^2}\left(-\frac{1}{|\ell|^3}+\frac{2|\theta p|}{|\ell|^2}+\frac{e^{-|\ell||\theta p|}}{|\ell|^3}\right).
\label{B.18}
\end{split}
\end{equation}
Unlike $A_2$, here we can only neglect the second term in the last parenthesis because the first and last exceed the minimal value of the loop momenta power $m=-2$ in the dimensional regularization prescription. Then one can introduce one more integrand $y$ to make the first and last terms into one
\begin{equation}
\begin{split}
T_{-2}&=(4\pi)^2\mu^{4-D}\int\frac{d^{D-1} \ell}{(2\pi)^{D-1}}\,\frac{1}{(\theta p)^2}\left(-\frac{1}{|\ell|^3}+\frac{e^{-|\ell||\theta p|}}{|\ell|^3}\right)
\\&=-(4\pi)^2\mu^{4-D}\int\limits_0^1 dy\int\frac{d^{D-1} \ell}{(2\pi)^{D-1}}\,\frac{1}{|\theta p|}\frac{e^{-y|\ell||\theta p|}}{|\ell|^2}
\\&=-(4\pi)^{\frac{5-D}{2}}\mu^{4-D}\frac{2}{\Gamma\left(\frac{D-1}{2}\right)}\int\limits_0^1 dy\,\int\limits_0^\infty dl\,|\theta p|^{-1}l^{D-4}e^{-ly|\theta p|}
\\&=-(4\pi)^{\frac{5-D}{2}}\mu^{4-D}\frac{2}{\Gamma\left(\frac{D-1}{2}\right)}\int\limits_0^1 dy\,|\theta p|^{2-D}y^{3-D}\Gamma\left(D-3\right)
\\&=-(4\pi)^{\frac{5-D}{2}}\mu^{4-D}\frac{2}{\Gamma\left(\frac{D-1}{2}\right)}|\theta p|^{2-D}\frac{\Gamma\left(D-3\right)}{4-D}
\\&=(4\pi\mu^2)^{\frac{4-D}{2}}\left(\frac{(\theta p)^2}{4}\right)^{1-\frac{D}{2}}\frac{\Gamma\left(\frac{D}{2}-2\right)}{D-3}.
\end{split}
\label{B.19}
\end{equation}
Finally, a familiar pattern emerges once we compute the $D\to 4$ limit
\begin{equation}
T_{-2}=-\frac{4}{(\theta p)^2}\left(\frac{2}{4-D}+\ln(\mu^2(\theta p)^2)+\ln\pi+\gamma_E+2\right)+\mathcal O(4-D).
\label{B.20}
\end{equation}
Here we see the logarithmic UV/IR mixing taking place via a single integral.

In the end all loop integrals are expressed via usual planar integrals plus nonplanar integrals $T_{-2}$, $T_0$, $I_K^0$, $I_K^1$ and $I_H$.

\section{Photino-photon Feynman rules}

The $\cal N$=1 photino action $S^{\rm photino}$, (\ref{2.17}), in the momentum space reads
\begin{equation}
\begin{split}
&S^{\rm photino}=\int\,\frac{d^4 p}{(2\pi)^4}\,\bar{\lambda}_{\dot{\alpha}}(p)p_{\mu}\bar{\sigma}^{\mu\,\dot{\alpha}\alpha}\lambda_{\alpha}(p)
\\&+\int\,\prod_{i=1}^4
\frac{d^4 p_i}{(2\pi)^4}\,(2\pi)^4\delta(p_1-p_2-p_3)\,
\bar{\lambda}_{\dot{\alpha}}(p_1)a_\mu(p_2)\lambda_{\alpha}(p_3)\bar{\sigma}^{\rho\,\dot{\alpha}\alpha}\,
{{V^{e^1}}_{\rho}^{\mu}}\big[p_1,-p_2,-p_3\big]
\\&+\int\,\prod_{i=1}^4
\frac{1}{2}\frac{d^4 p_i}{(2\pi)^4}\,(2\pi)^4\delta(p_1-\sum_{j=2}^4 p_j)\,
\\&{\hspace{2cm}}
\cdot\bar{\lambda}_{\dot{\alpha}}(p_1)a_\mu(p_2)a_\nu(p_3)\lambda_{\alpha}(p_4)\bar{\sigma}^{\rho\, \dot{\alpha}\alpha}\,{{V^{e^2}}_{\rho}^{\mu\nu}}\big[p_1,-p_2,-p_3,-p_4\big]+\mathcal O\left(e^3\right),
\end{split}
\label{C.1}
\end{equation}
where all three terms above are represented by Figs \ref{fig:photino-propagator}, \ref{fig:photino-photon}, and \ref{fig:2photinos-2photons}. For photino propagator in particular, see \cite{Dreiner:2008tw}:
\begin{figure}
\begin{center}
\includegraphics[width=5cm]{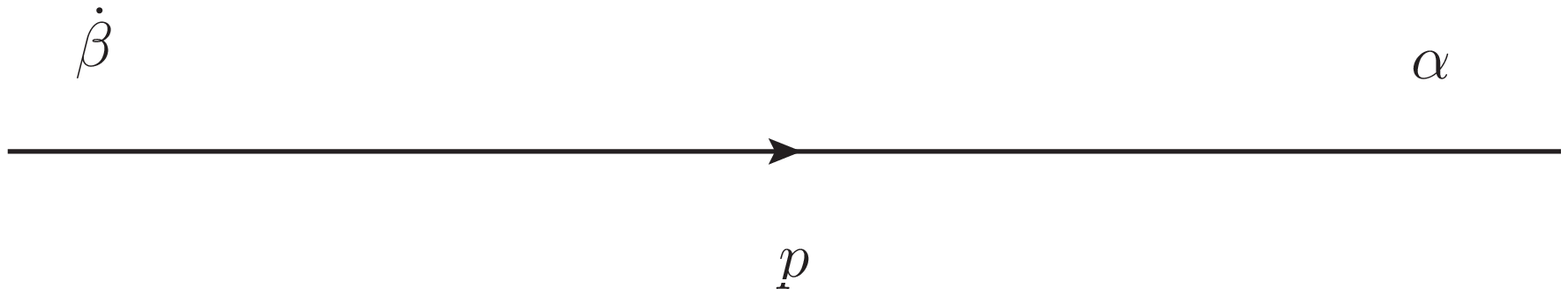}
\end{center}
\caption{ $\cal N$=1 photino propagator, Eq. (\ref{B.2}).}
\label{fig:photino-propagator}
\end{figure}
\begin{equation}
\langle 0|T\lambda_{\alpha}(x)\bar{\lambda}_{\dot{\beta}}(y)|0\rangle=\int\frac{d^4 p}{(2\pi)^4}\,\frac{i\,p_\mu\sigma^{\mu}_{\alpha\dot{\beta}}}{p^2+i\epsilon}\,e^{-ip(x-y)}.
\label{C.2}
\end{equation}
\begin{figure}
\begin{center}
\includegraphics[width=5cm]{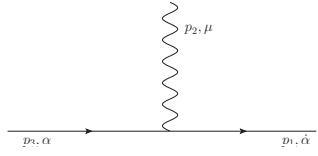}
\end{center}
\caption{ $\cal N$=1 photino-photon vertex: ${V^{e^1}}_{\rho}^{\mu}(p_1,p_2)$; $p_2+p_3-p_1=0$.}
\label{fig:photino-photon}
\end{figure}
\begin{figure}
\begin{center}
\includegraphics[width=5cm]{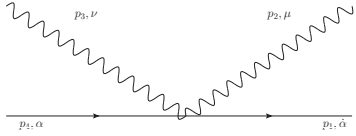}
\end{center}
\caption{ $\cal N$=1 photino-2photons vertex: ${V^{e^2}}_{\rho}^{\mu\nu}(p_2,p_3,p_4)$; $p_2+p_3+p_4-p_1=0$.}
\label{fig:2photinos-2photons}
\end{figure}


From the second line in (\ref{C.1}) and Fig \ref{fig:photino-photon} the photino-photon vertex reads as follows:
\begin{equation}
\bar{\sigma}^{\rho\,\dot{\alpha}\alpha}\,{{V^{e^1}}_{\rho}^{\mu}}(p_1,p_2)
=f_{\star_2}\left(p_1,p_2\right)\bigg(\bar{\sigma}^{\rho\,\dot{\alpha}\alpha} p_{2_\rho}(\theta p_1)^\mu-\bar{\sigma}^{\mu\,\dot{\alpha}\alpha}(p_2\theta p_1)-(\theta p_2)^\mu{\bar{\sigma}^{\rho\,\dot{\alpha}\alpha} p_{1_\rho}}\bigg),
\label{C.3}
\end{equation}
where $(p_2,\mu)$  is the photon incoming (momenta, index) and the fermion momentum $p_1$ flows through the vertex, as it should.


From the third line in (\ref{C.1}) and Fig \ref{fig:2photinos-2photons} the photino-2photons vertex reads as follows:
\begin{gather}
\begin{split}
&\bar{\sigma}^{\rho\,\dot{\alpha}\alpha}\,{V^{e^2}}_{\rho}^{\mu\nu}(p_2,p_3,p_4)=
\\&-i\bigg(\big(\theta p_2)^{\mu}\Big(\bar{\sigma}^{\rho\,\dot{\alpha}\alpha}p_{3_\rho}\big(\theta p_4)^{\nu}-\bar{\sigma}^{\nu\,\dot{\alpha}\alpha}(p_3\theta p_4)\Big)
f_{\star_2}\left(p_1,p_2\right)f_{\star_2}\left(p_3,p_4\right)
\\&+\big(\theta p_3)^{\nu}\Big(\bar{\sigma}^{\rho\,\dot{\alpha}\alpha}p_{2_\rho}\big(\theta p_4)^{\mu}-\bar{\sigma}^{\mu\,\dot{\alpha}\alpha}(p_2\theta p_4)\Big) 
f_{\star_2}\left(p_1,p_3\right)f_{\star_2}\left(p_2,p_4\right)
\\&+\frac{i}{2}\bigg(\bar{\sigma}^{\rho\,\dot{\alpha}\alpha}p_{2_\rho}\Big(\big(\theta p_3\big)^{\mu}
\big(\theta p_4\big)^{\nu}-\big(p_3\theta p_4\big)\theta^{\mu\nu}\Big)-\bar{\sigma}^{\nu\,\dot{\alpha}\alpha}\Big(\big(p_2\theta p_3\big)\big(\theta p_4\big)^{\mu}+\big(p_2\theta p_4\big)\big(\theta p_3\big)^{\mu}\Big)\bigg)
\\&\cdot \Big(f_{\star_{3'}}\left(p_2,p_4,p_3\right)+f_{\star_{3'}}\left(p_4,p_2,p_3\right)\Big)
\\&+\frac{i}{2}\bigg(\bar{\sigma}^{\rho\,\dot{\alpha}\alpha}p_{3_\rho}\Big((\theta p_4)^{\mu}(\theta p_2)^{\nu}+(p_2\theta p_4)\theta^{\mu\nu}\Big) +\bar{\sigma}^{\mu\,\dot{\alpha}\alpha}\Big((p_2\theta p_3)(\theta p_4)^{\nu}-(p_3\theta p_4)(\theta p_2)^{\nu}\Big)\bigg)
\\&\cdot \Big(f_{\star_{3'}}\left(p_3,p_4,p_2\right)+f_{\star_{3'}}\left(p_4,p_3,p_2\right)\Big)
\\&+\frac{i}{2}(\theta p_3)^{\mu}\Big(\bar{\sigma}^{\rho\,\dot{\alpha}\alpha}p_{3_\rho}(\theta p_4)^{\nu}-\bar{\sigma}^{\nu\,\dot{\alpha}\alpha}(p_3\theta p_4)\Big)
\\&\cdot\Big(f_{\star_{3'}}\left(p_4,p_2,p_3\right)+f_{\star_{3'}}\left(p_2,p_3,p_4\right)-2f_{\star_2}\left(p_1,p_2\right)f_{\star_2}\left(p_3,p_4\right)\Big)
\\&+
\frac{i}{2}\big(\theta p_4)^{\mu}\Big(\bar{\sigma}^{\rho\,\dot{\alpha}\alpha}p_{3_\rho}\big(\theta p_4)^{\nu}-\bar{\sigma}^{\nu\,\dot{\alpha}\alpha}(p_3\theta p_4)\Big)
\\&\cdot\Big(f_{\star_{3'}}\left(p_3,p_2,p_4\right)+f_{\star_{3'}}\left(p_2,p_3,p_4\right)-2f_{\star_2}\left(p_1,p_2\right)f_{\star_2}\left(p_3,p_4\right)\Big)
\\&+\frac{i}{2}(\theta p_2)^{\nu}\Big(\bar{\sigma}^{\rho\,\dot{\alpha}\alpha}p_{2_\rho}\big(\theta p_4)^{\mu}-\bar{\sigma}^{\mu\,\dot{\alpha}\alpha}(p_2\theta p_4)\Big)
\\&\cdot\Big(f_{\star_{3'}}\left(p_4,p_3,p_2\right)+f_{\star_{3'}}\left(p_3,p_2,p_4\right)-2f_{\star_2}\left(p_1,p_3\right)f_{\star_2}\left(p_2,p_4\right)\Big)
\\&+
\frac{i}{2}(\theta p_4)^{\nu}\Big(\bar{\sigma}^{\rho\,\dot{\alpha}\alpha}p_{2_\rho}\big(\theta p_4)^{\mu}-\bar{\sigma}^{\mu\,\dot{\alpha}\alpha}(p_2\theta p_4)\Big)
\\&\cdot\Big(f_{\star_{3'}}\left(p_2,p_3,p_4\right)+f_{\star_{3'}}\left(p_3,p_2,p_4\right)-2f_{\star_2}\left(p_1,p_3\right)f_{\star_2}\left(p_2,p_4\right)\Big)
\\&+\frac{i}{2}\bar{\sigma}^{\rho\,\dot{\alpha}\alpha}p_{1_\rho}\bigg(\Big(f_{\star_2}\left(p_1,p_2\right)f_{\star_2}\left(p_3,p_4\right)
+f_{\star_2}\left(p_1,p_3\right)f_{\star_2}\left(p_2,p_4\right)\Big)(\theta p_2)^\mu(\theta p_3)^\nu
\\&-f_{\star_3'}\left(p_4,p_2,p_3\right)\Big((p_2\theta p_3)\theta^{\mu\nu}+(\theta p_3)^\mu(\theta p_2)^\nu\Big)
\\&-(\theta p_2)^\mu\Big((p_2\theta p_3)(\theta p_4)^\nu+(\theta p_2)^\nu(p_3\theta p_4)\Big)f_{\rm (I)}\left(p_2,p_3,p_4\right)
\\&+(\theta p_3)^\nu\Big((p_2\theta p_3)(\theta p_4)^\mu-(\theta p_3)^\mu(p_2\theta p_4)\Big)f_{\rm (I)}\left(p_3,p_2,p_4\right)
\\&-\frac{1}{2}(\theta p_2)^\mu(\theta p_3)^\nu(p_2\theta p_3)\Big(f_{\rm (I)}\left(p_2,p_3,p_4\right)-f_{\rm (I)}\left(p_3,p_2,p_4\right)\Big)\bigg),
\end{split}
\label{C.4}
\end{gather}
with the photon momenta $(p_2,\mu)$, $(p_3,\nu)$ and the photino momentum $p_1$ being incoming.

\section{Scalar-photon Feynman rules}

From the scalar action (\ref{AcSreal}) we obtain the following scalar-photon Feynman rule corresponding to the Fig. \ref{fig:scalar-1photon} :
\begin{gather}
\begin{split}
S^\mu(p_1,p_2,p_3)= -e f_{\star_2}(p_1,p_3)&\Big((p_1 p_2)(\theta p_3)^\mu-(p_1p_3)(\theta p_2)^\mu
+ (p_2p_3)(\theta p_1)^\mu
\\&- p_1^\mu(p_2\theta p_3)-p_3^\mu(p_2\theta p_1)\Big),
\end{split}
\label{D.1}
\end{gather}
\begin{figure}
\begin{center}
\includegraphics[width=5cm]{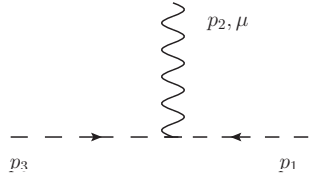}
\end{center}
\caption{Scalar-photon vertex: $S^\mu(p_1,p_2,p_3)$; $p_1+p_2+p_3=0$.}
\label{fig:scalar-1photon}
\end{figure}
and the following scalar-2photons Feynman rule corresponding to the Fig. \ref{fig:scalar-2photons}:
\begin{figure}
\begin{center}
\includegraphics[width=4.5cm]{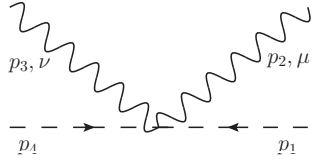}
\end{center}
\caption{Scalar-2photons vertex: $S^{\mu\nu}_{\rm rele}(p_1,p_2,p_3,p_4)$; $p_1+p_2+p_3+p_4=0$.}
\label{fig:scalar-2photons}
\end{figure}
\begin{equation}
S^{\mu\nu}_{\rm rele}(p_1,p_2,p_3,p_4)=S^{\mu\nu}_{1'} +S^{\mu\nu}_{2'} + S^{\mu\nu}_{3'}+S^{\mu\nu}_{4'} ,
\label{D.2}
\end{equation}
\begin{equation}
\begin{split}
&S^{\mu\nu}_{1'}=i e^2 \bigg\{f_{\star_2}(p_1,p_2) f_{\star_2}(p_4,p_3)
\cdot\Big((p_2 p_3)(\theta p_1)^\nu(\theta p_1)^\mu -
p_2^\nu(\theta p_4)^\mu(p_3\theta p_4)
\\&-p_3^\mu(\theta p_4)^\nu(p_2\theta p_1)+g^{\mu\nu}(p_3\theta p_4)(p_2\theta p_1) \Big)
+ f_{\star_2}(p_1,p_3) f_{\star_2}(p_4,p_2)
\\&
\cdot\Big((p_2 p_3)(\theta p_1)^\nu(\theta p_4)^\mu -p_2^\nu(\theta p_4)^\mu(p_3\theta p_1)
-p_3^\mu(\theta p_1)^\nu(p_2\theta p_4)+g^{\mu\nu}(p_2\theta p_4)(p_3\theta p_1) \Big)
\\&-
\Big((p_1 p_3)(\theta p_2)^\nu(\theta p_4)^\mu -p_1^\nu(\theta p_2)^\mu(p_3\theta p_4)
\\&-p_4^\mu(\theta p_3)^\nu(p_2\theta p_1)+(p_4 p_2)(\theta p_2)^{\nu}(\theta p_1)^{\mu} \Big)
+ f_{\star_2}(p_2,p_4) f_{\star_2}(p_3,p_1)
\\&
\cdot\Big((p_4 p_3)(\theta p_2)^\mu(\theta p_1)^\nu
-p_4^\nu(\theta p_2)^\mu(p_3\theta p_1)-p_1^\mu(\theta p_3)^\nu(p_2\theta p_4)+(p_1 p_2)(\theta p_3)^{\nu}(\theta p_4)^{\mu} \Big)\bigg\},
\end{split}
\end{equation}
\begin{equation}
\begin{split}
&S^{\mu\nu}_{2'}=\frac{i}{2} e^2 \bigg\{\Big[f_{\star_{3'}}(p_2,p_3,p_4) +f_{\star_{3'}}(p_4,p_2,p_3)\Big]
\cdot\Big((p_1 p_2)(\theta p_3)^\mu(\theta p_4)^\nu -(p_1 p_2)\theta^{\mu\nu}(p_3\theta p_4)
\\&-p_1^\mu(\theta p_4)^\nu(p_2\theta p_3)-p_1^\mu(\theta p_2)^\nu(p_3\theta p_4) \Big)
+ \Big[f_{\star_{3'}}(p_2,p_3,p_1) +f_{\star_{3'}}(p_1,p_2,p_3)\Big]
\\&
\cdot\Big((p_4 p_2)(\theta p_3)^\mu(\theta p_1)^\nu -(p_4 p_2)(p_3 \theta p_1) \theta^{\mu\nu}
-p_4^\mu(\theta p_1)^\nu(p_2\theta p_3)-p_4^\mu(\theta p_2)^\nu(p_3\theta p_1) \Big)
\\&
+ \Big[f_{\star_{3'}}(p_3,p_2,p_4) +f_{\star_{3'}}(p_4,p_3,p_2)\Big]
\cdot\Big((p_1 p_3)(\theta p_2)^\nu(\theta p_4)^\mu -(p_1 p_3)(p_2 \theta p_3) \theta^{\mu\nu}
\\&
-p_1^\nu(\theta p_3)^\mu(p_3\theta p_2)-p_1^\nu(\theta p_3)^\mu(p_2\theta p_4) \Big)
+ \Big[f_{\star_{3'}}(p_3,p_2,p_1) +f_{\star_{3'}}(p_1,p_3,p_2)\Big]
\\&
\cdot\Big((p_4 p_3)(\theta p_2)^\nu(\theta p_1)^\mu -(p_4 p_3)(p_2 \theta p_1) \theta^{\mu\nu}
-p_4^\nu(\theta p_1)^\mu(p_3\theta p_2)-p_4^\nu(\theta p_3)^\mu(p_2\theta p_1) \Big)\bigg\},
\end{split}
\end{equation}
\begin{equation}
\begin{split}
&S^{\mu\nu}_{3'}=\frac{i}{2} e^2 \bigg\{\Big[f_{\star_{3'}}(p_4,p_2,p_3) +f_{\star_{3'}}(p_2,p_3,p_4)-2 f_{\star_2}(p_2,p_1) f_{\star_2}(p_3,p_4)\Big]
\\&
\cdot\Big((p_1 p_3)(\theta p_3)^\mu(\theta p_4)^\nu -p_1^\nu(\theta p_3)^\mu (p_3\theta p_4)\Big)
+\Big[f_{\star_{3'}}(p_1,p_2,p_3) +f_{\star_{3'}}(p_2,p_3,p_1)
\\&
-2 f_{\star_2}(p_2,p_4) f_{\star_2}(p_3,p_1)\Big]
\cdot\Big((p_4 p_3)(\theta p_3)^\mu(\theta p_1)^\nu -p_4^\nu(\theta p_3)^\mu (p_3\theta p_1)\Big)
\\&
+\Big[f_{\star_{3'}}(p_4,p_3,p_2) +f_{\star_{3'}}(p_3,p_2,p_4)-2 f_{\star_2}(p_3,p_1) f_{\star_2}(p_2,p_4)\Big]
\\&
\cdot\Big((p_1 p_2)(\theta p_2)^\nu(\theta p_4)^\mu -p_1^\mu(\theta p_2)^\nu (p_2\theta p_4)\Big)
+\Big[f_{\star_{3'}}(p_1,p_3,p_2) +f_{\star_{3'}}(p_3,p_2,p_1)
\\&
-2 f_{\star_2}(p_3,p_4) f_{\star_2}(p_2,p_1)\Big]
\cdot\Big((p_4 p_2)(\theta p_2)^\nu(\theta p_1)^\mu -p_4^\mu(\theta p_2)^\nu (p_2\theta p_1)\Big)
\\&
+\Big[f_{\star_{3'}}(p_2,p_3,p_4) +f_{\star_{3'}}(p_3,p_2,p_4)-2 f_{\star_2}(p_2,p_1) f_{\star_2}(p_3,p_4)\Big]
\\&
\cdot\Big((p_1 p_3)(\theta p_4)^\mu(\theta p_4)^\nu -p_1^\nu(\theta p_4)^\mu (p_3\theta p_4)\Big)
+\Big[f_{\star_{3'}}(p_2,p_3,p_1) +f_{\star_{3'}}(p_3,p_2,p_1)
\\&
-2 f_{\star_2}(p_2,p_4) f_{\star_2}(p_3,p_1)\Big]
\cdot\Big((p_4 p_3)(\theta p_1)^\mu(\theta p_1)^\nu -p_4^\nu(\theta p_1)^\mu (p_3\theta p_1)\Big)
\\&
+\Big[f_{\star_{3'}}(p_3,p_2,p_4) +f_{\star_{3'}}(p_2,p_3,p_4)-2 f_{\star_2}(p_3,p_1) f_{\star_2}(p_2,p_4)\Big]
\\&
\cdot\Big((p_1 p_2)(\theta p_4)^\nu(\theta p_4)^\mu -p_1^\mu(\theta p_4)^\nu (p_2\theta p_4)\Big)
+\Big[f_{\star_{3'}}(p_3,p_2,p_1) +f_{\star_{3'}}(p_2,p_3,p_1)
\\&-2 f_{\star_2}(p_3,p_4) f_{\star_2}(p_2,p_1)\Big]
\cdot\Big((p_4 p_2)(\theta p_1)^\nu(\theta p_1)^\mu -p_4^\mu(\theta p_1)^\nu (p_2\theta p_1)\Big) \bigg\},
\end{split}
\end{equation}
\begin{equation}
\begin{split}
&S^{\mu\nu}_{4'}=\frac{i}{2}e^2(p_1\cdot p_4)\bigg(\Big(f_{\star_2}\left(p_1,p_2\right)f_{\star_2}\left(p_3,p_4\right)
+f_{\star_2}\left(p_1,p_3\right)f_{\star_2}\left(p_2,p_4\right)\Big)(\theta p_2)^\mu(\theta p_3)^\nu
\\&-f_{\star_3'}\left(p_4,p_2,p_3\right)\Big((p_2\theta p_3)\theta^{\mu\nu}+(\theta p_3)^\mu(\theta p_2)^\nu\Big)
\\&-(\theta p_2)^\mu\Big((p_2\theta p_3)(\theta p_4)^\nu+(\theta p_2)^\nu(p_3\theta p_4)\Big)f_{\rm (I)}\left(p_2,p_3,p_4\right)
\\&+(\theta p_3)^\nu\Big((p_2\theta p_3)(\theta p_4)^\mu-(\theta p_3)^\mu(p_2\theta p_4)\Big)f_{\rm (I)}\left(p_3,p_2,p_4\right)
\\&-\frac{1}{2}(\theta p_2)^\mu(\theta p_3)^\nu(p_2\theta p_3)\Big(f_{\rm (I)}\left(p_2,p_3,p_4\right)-f_{\rm (I)}\left(p_3,p_2,p_4\right)\Big)\bigg).
\end{split}
\end{equation}

\section{Scalar-fermion Feynman rules in the NC $\cal N$=2,4 SYM U(1)}

Feynman rules in  $\cal N$=2, described by Figs \ref{fig:propagator}-\ref{fig:4-scalarvertex}, are:
\begin{equation}
\begin{split}
\Sigma_{\alpha\dot\beta}(p)&=i\frac{\sigma^{\mu}_{\alpha\dot{\beta}}\:p_\mu}{p^2},
\\
\Gamma^{\alpha_1\alpha_2}(p_1,p_2)&=-2{\sqrt 2}ie\sin\frac{p_1\theta p_2}{2} \epsilon^{\alpha_1\alpha_2},
\\
\Gamma^{\dot\alpha_1\dot\alpha_2}(p_1,p_2)&=2{\sqrt 2}ie\sin\frac{p_1\theta p_2}{2} \epsilon^{{\dot\alpha}_1{\dot\alpha}_2},
\\
\Gamma(p_1,p_2,p_3,p_4)&=4ie^2\Big[\sin\frac{p_1\theta p_4}{2} \sin\frac{p_2\theta p_3}{2}+\{1\leftrightarrow2\}\Big].
\\
\end{split}
\label{E.1}
\end{equation}
Feynman rules in  $\cal N$=4 from Figs \ref{fig:Propagator}-\ref{fig:2,44-scalarvertex} are:
\begin{equation}
\begin{split}
\Sigma_i^j(p)&=i\frac{\sigma^{\mu}\:p_\mu\delta^j_i}{p^2},
\\
\Gamma_{i_1 i_2 m}^{\alpha_1\alpha_2}(p_1,p_2)&=2ie\big({\tilde\sigma}^{-1}_m\big)_{i_1 i_2}\sin\frac{p_1\theta p_2}{2} \epsilon^{\alpha_1\alpha_2},
\\
\Gamma_{i_1 i_2 m}^{\dot\alpha_1\dot\alpha_2}(p_1,p_2)&=2ie\big({\tilde\sigma}_m\big)_{i_1 i_2}\sin\frac{p_1\theta p_2}{2} \epsilon^{{\dot\alpha}_1{\dot\alpha}_2},
\\
\Gamma_{m_1 m_2 m_3 m_4}(p_1,p_2,p_3,p_4)&=
-4ie^2\Big[\sin\frac{p_1\theta p_2}{2} \sin\frac{p_3\theta p_4}{2}
\big(\delta_{m_1 m_3} \delta_{m_2 m_4} -\delta_{m_2 m_3}  \delta_{m_1 m_4}  \big)+ c.p. \Big].
\end{split}
\label{E.2}
\end{equation}

\begin{figure}
\begin{center}
\includegraphics[width=5cm]{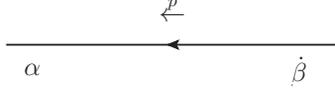}
\end{center}
\caption{$\cal N$=2 fermion propagator: $\Sigma_{\alpha\dot\beta}(p)$.}
\label{fig:propagator}
\end{figure}

\begin{figure}
\begin{center}
\includegraphics[width=5cm]{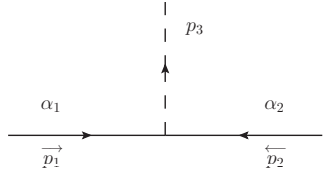}
\end{center}
\caption{$\cal N$=2 scalar-fermion vertex: $\Gamma^{\alpha_1\alpha_2}(p_1,p_2)$; $p_1+p_2-p_3=0$.}
\label{fig:Sfvertex}
\end{figure}

\newpage

\begin{figure}
\begin{center}
\includegraphics[width=5cm]{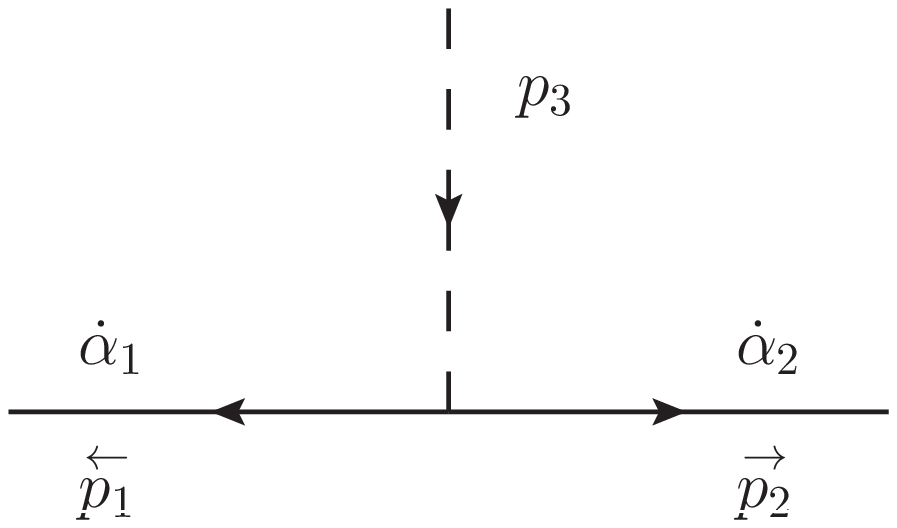}
\end{center}
\caption{$\cal N$=2 scalar-antifermion vertex: $\Gamma^{\dot\alpha_1\dot\alpha_2}(p_1,p_2)$;  $p_3-p_1-p_2=0$.}
\label{fig:Sfdotvertex}
\end{figure}

\begin{figure}
\begin{center}
\includegraphics[width=5cm]{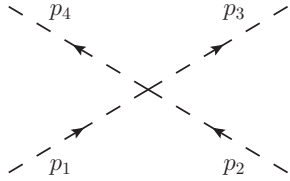}
\end{center}
\caption{$\cal N$=2 four-scalar vertex: $\Gamma(p_1,p_2,p_3,p_4)$; $p_1+p_2-p_3-p_4=0$.}
\label{fig:4-scalarvertex}
\end{figure}

\begin{figure}
\begin{center}
\includegraphics[width=5cm]{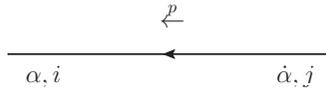}
\end{center}
\caption{$\cal N$=4 fermion propagator: $\Sigma_i^j(p)$.}
\label{fig:Propagator}
\end{figure}

\begin{figure}
\begin{center}
\includegraphics[width=5cm]{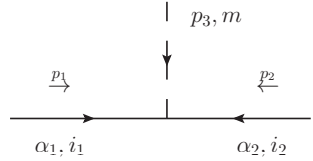}
\end{center}
\caption{$\cal N$=4 scalar-fermion vertex: $\Gamma_{i_1 i_2 m}^{\alpha_1\alpha_2}(p_1,p_2)$; $p_1+p_2+p_3=0$.}
\label{fig:Sfvertex}
\end{figure}

\begin{figure}
\begin{center}
\includegraphics[width=5cm]{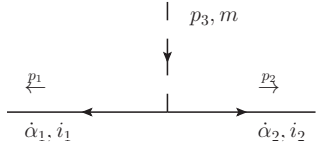}
\end{center}
\caption{$\cal N$=4 scalar-antifermion vertex: $\Gamma_{i_1 i_2 m}^{\dot\alpha_1\dot\alpha_2}(p_1,p_2)$; $p_3-p_1-p_2=0$.}
\label{fig:Sfvertex}
\end{figure}

\begin{figure}
\begin{center}
\includegraphics[width=5cm]{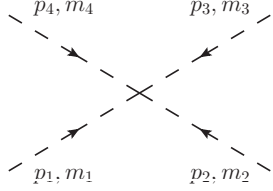}
\end{center}
\caption{$\cal N$=4 four-scalar vertex: $\Gamma_{m_1 m_2 m_3 m_4}(p_1,p_2,p_3,p_4)$; $p_1+p_2+p_3+p_4=0$.}
\label{fig:2,44-scalarvertex}
\end{figure}

\section{Feynman rules from the NC gauge fixing and ghost actions}

From Figs \ref{Gf3pv}-\ref{G2pv} we have the following Feynman rules:
\begin{gather}
\begin{split}
&
\Gamma_{\rm gf}^{\mu_1\mu_2\mu_3}(p_1,p_2,p_3)=-\frac{1}{2}f_{\star_2}(p_1,p_2)
\Big[2p_1^{\mu_1} p_1^{\mu_3}(\theta p_3)^{\mu_2}-
p_1^{\mu_1}(p_1p_3)\theta^{\mu_2\mu_3} +
2p_1^{\mu_1} p_1^{\mu_2}(\theta p_2)^{\mu_3}
\\&
-p_1^{\mu_1}(p_1p_2)\theta^{\mu_2\mu_3}
+ 2p_2^{\mu_2} p_2^{\mu_3}(\theta p_3)^{\mu_1}-
p_2^{\mu_2}(p_2p_3)\theta^{\mu_1\mu_3}
+ 2p_2^{\mu_2} p_2^{\mu_1}(\theta p_1)^{\mu_3}-
p_2^{\mu_2}(p_2p_1)\theta^{\mu_1\mu_3}
\\&
+2p_3^{\mu_3} p_3^{\mu_2}(\theta p_2)^{\mu_1}-
p_3^{\mu_3}(p_3p_2)\theta^{\mu_1\mu_2} +
 (p_3^{\mu_3} p_3^{\mu_1}(\theta p_1)^{\mu_2}-
p_3^{\mu_1}(p_1p_3)\theta^{\mu_1\mu_2}\Big],
\\&
\Gamma_{\rm gf}^{\mu_1\mu_2\mu_3\mu_4}(p_1,p_2,p_3,p_4)=-\frac{i}{8}f_{\star_2}(p_1,p_2)f_{\star_2}(p_3,p_4)
\Big[ (p_1+p_2)^{\mu_2}(\theta p_2)^{\mu_1}-
(p_1+p_2)\cdot p_1\theta^{\mu_1\mu_2}\Big]
\\&
\cdot\Big[ (p_3+p_4)^{\mu_4} (\theta p_4)^{\mu_3}-
(p_1+p_4)\cdot p_4\theta^{\mu_3\mu_4}\Big]
\\&
+\frac{i}{8}f_{\star_{3'}}[p_2,p_3,p_4]
\Big[ 2p_1^{\mu_1} p_1^{\mu_4} \big(2(\theta p_4)^{\mu_2}(\theta p_4)^{\mu_3}
-(p_3\theta p_4)\theta^{\mu_2\mu_3}\big)
\\&
+2p_1^{\mu_1} p_1^{\mu_2}\big(2(\theta p_2)^{\mu_4}(\theta p_4)^{\mu_3}
+(p_2\theta p_4)\theta^{\mu_3\mu_4}\big)
+p_1^{\mu_1}(p_1p_2)\big(2(\theta p_4)^{\mu_3}\theta^{\mu_2\mu_4}-(\theta p_4)^{\mu_2}\theta^{\mu_3\mu_4}\big)
\\&
-p_1^{\mu_1}(p_1p_4)\big(3(\theta p_3)^{\mu_2}\theta^{\mu_3\mu_4}+2(\theta p_3)^{\mu_4}\theta^{\mu_2\mu_3}
+2(\theta p_4)^{\mu_3}\theta^{\mu_2\mu_4}+(\theta p_4)^{\mu_2}\theta^{\mu_3\mu_4}
\big)
\Big]
\\&
+ \{S_4 \; {\rm permutations}\},
\\&
\Gamma_{\rm gh}^{\mu}(p_1,p_2)=f_{\star_2}(p_1,p_2)\Big[\frac{1}{2}(p_1+p_2)^2(\theta p_2)^\mu-(p_1+p_2)^\mu(p_1\theta p_2)\Big],
\\&
\Gamma_{\rm gh}^{\mu_1\mu_2}(p_1,p_2,p_3,p_4)=\Big\{\frac{1}{2}f_{\star_2}(p_1,p_2)f_{\star_2}(p_3,p_4)
\Big[2p_4^{\mu_2}(\theta p_2)^{\mu_1}-(p_4p_1)\theta^{\mu_1\mu_2}\Big](p_4\theta p_3)
\\&
\phantom{XXXXXXX}+\frac{1}{2}f_{\star_2}(p_2,p_3)f_{\star_2}(p_1,p_4)p_4^{\mu_1}(\theta p_3)^{\mu_2}(p_1\theta p_4)\Big\}
+\{1\leftrightarrow2\}.
\end{split}
\label{F.1}
\end{gather}

\begin{figure}
\begin{center}
\includegraphics[width=4.5cm]{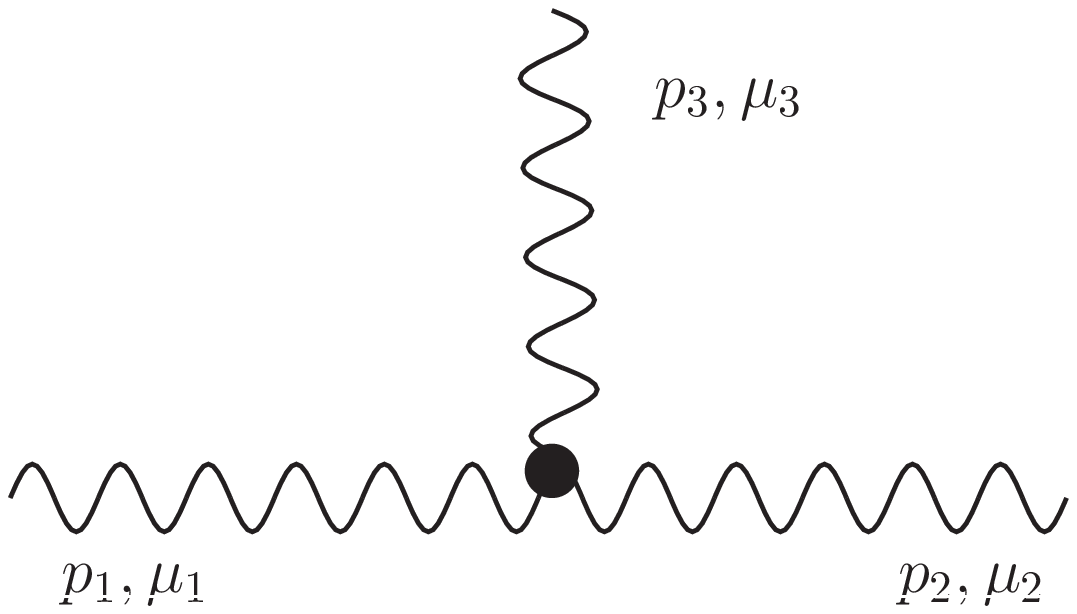}
\end{center}
\caption{Gauge fixing 3-photon vertex: $\Gamma_{\rm gf}^{\mu_1\mu_2\mu_3}(p_1,p_2,p_3)$; $p_1+p_2+p_3=0$.}
\label{Gf3pv}
\end{figure}

\begin{figure}
\begin{center}
\includegraphics[width=4.5cm]{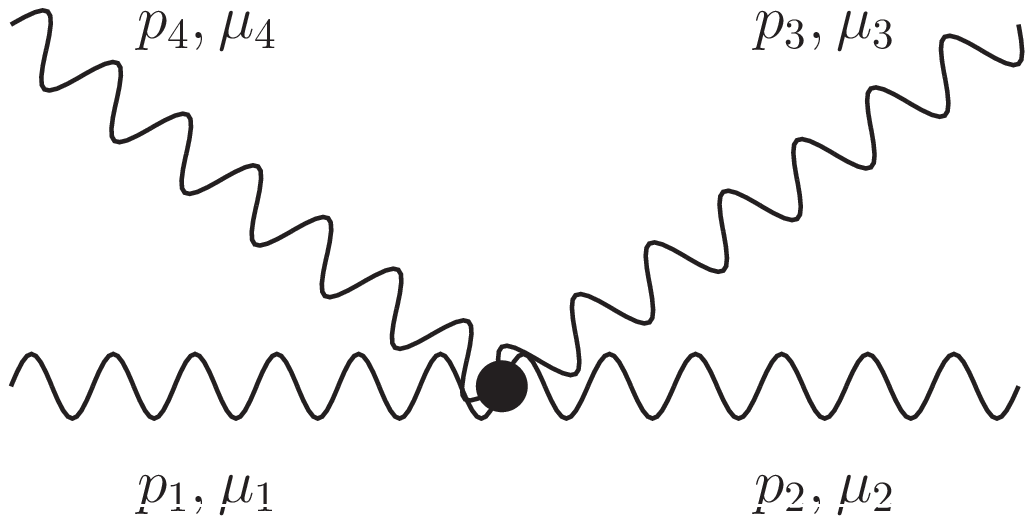}
\end{center}
\caption{Gauge fixing 4-photon vertex: $\Gamma_{\rm gf}^{\mu_1\mu_2\mu_3\mu_4}(p_1,p_2,p_3,p_4)$; $p_1+p_2+p_3+p_4=0$.}
\label{Gf4pv}
\end{figure}

\begin{figure}
\begin{center}
\includegraphics[width=5cm]{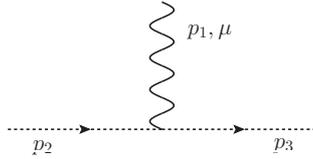}
\end{center}
\caption{Ghost-photon vertex: $\Gamma_{\rm gh}^{\mu}(p_1,p_2)$; $p_1+p_2-p_3=0$.}
\label{Gpv}
\end{figure}

\begin{figure}
\begin{center}
\includegraphics[width=5cm]{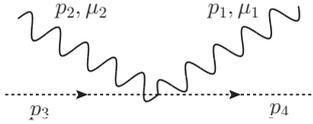}
\end{center}
\caption{Ghost-2photons vertex:  $\Gamma_{\rm gh}^{\mu_1\mu_2}(p_1,p_2,p_3,p_4)$; $p_1+p_2+p_3-p_4=0$.}
\label{G2pv}
\end{figure}

\section{Feynman rules from the gauge and BRST-auxiliary field interactions}

Feynman rule corresponding to Fig. \ref{Baa} is:
\begin{figure}
\begin{center}
\includegraphics[width=4cm]{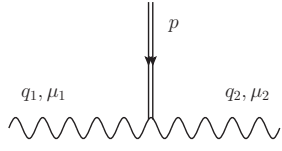}
\end{center}
\caption{The 2gauge-Bauxiliary field interactions: $\Gamma_{Baa}^{\mu_1\mu_2}(p,q_1,q_2)$; $p+q_1+q_2=0$.}
\label{Baa}
\end{figure}
\begin{equation}
\begin{split}
\Gamma_{Baa}^{\mu_1\mu_2}(p,q_1,q_2)=&-\frac{i}{2} f_{\star_2}(q_1,q_2)\Big(2p^{\mu_2}(\theta q_2)^{\mu_1}- (p q_2)\theta^{\mu_1\mu_2}+ 2p^{\mu_1}(\theta q_1)^{\mu_2}+
(p q_1)\theta^{\mu_1\mu_2}\Big),
\label{G.1}
\end{split}
\end{equation}
while the 3-gauge-B-auxiliary-field Feynman rule from Fig. \ref{Baaa} has the following form:
\begin{figure}
\begin{center}
\includegraphics[width=4cm]{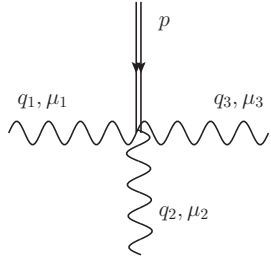}
\end{center}
\caption{The 3gauge-Bauxiliary field interactions: $\Gamma_{Baaa}^{\mu_1\mu_2\mu_3}(p,q_1,q_2,q_3)$; $p+q_1+q_2+q_3=0$.}
\label{Baaa}
\end{figure}

 \begin{equation}
\begin{split}
\Gamma^{\mu_1\mu_2\mu_3}_{Baaa}&(p,q_1,q_2,q_3)=\frac{1}{8} f_{\star_{3'}}(q_1,q_2,q_3)\Big(
3(p q_3) (\theta q_2)^{\mu_1}\theta^{\mu_2\mu_3}
+(p q_1) (\theta q_3)^{\mu_1}\theta^{\mu_2\mu_3}
\\&
+(p q_3) (\theta q_1)^{\mu_1}\theta^{\mu_2\mu_3}
-2(p q_1) (\theta q_2)^{\mu_2}\theta^{\mu_1\mu_3}
+2(p q_3) (\theta q_2)^{\mu_3}\theta^{\mu_1\mu_2}
\\&
+2(p q_3) (\theta q_2)^{\mu_2}\theta^{\mu_1\mu_3}
+2p^{\mu_3}(q_2\theta q_3)\theta^{\mu_1\mu_2}
-2p^{\mu_1}(q_1\theta q_2)\theta^{\mu_2\mu_3}
\\&
-4p^{\mu_3}(\theta q_2)^{\mu_1}(\theta q_3)^{\mu_2}
-4p^{\mu_1}(\theta q_3)^{\mu_2}(\theta q_1)^{\mu_3}\Big)
+ \{S_3 \; {\rm permutations}\}.
\label{G.2}
\end{split}
\end{equation}

\end{document}